\documentclass[sn-mathphys,Numbered]{sn-jnl}

\usepackage{graphicx}%
\usepackage{multirow}%
\usepackage{amsmath,amssymb,amsfonts}%
\usepackage{amsthm}%
\usepackage{mathrsfs}%
\usepackage[title]{appendix}%
\usepackage{xcolor}%
\usepackage{textcomp}%
\usepackage{manyfoot}%
\usepackage{booktabs}%
\usepackage{algorithm}%
\usepackage{algorithmicx}%
\usepackage{algpseudocode}%
\usepackage{listings}%
\usepackage{lscape}

\def\bng{\bngx}

\font\bngx=bang10

\def\*#1*#2{o\null{#2}{#1}}

\def\sh#1{\setbox0=\hbox{#1}%
     \kern-.02em\copy0\kern-\wd0
     \kern.04em\copy0\kern-\wd0
     \kern-.02em\raise.0433em\box0 }
\definecolor{awesome}{rgb}{0.98, 0.36, 0.51}

\newcommand{\hst}{\textit{HST}}
\newcommand{\jwst}{\textit{JWST}}

\newcommand{\db}{\texttt{DENSE BASIS}}

\newcommand{\galfit}{\texttt{Galfit}}

\newcommand{\grizli}{\texttt{Grizli}}
\newcommand{\astrodrizzle}{\texttt{Astrodrizzle}}

\newcommand{\macsj}{MACS~J1423.8$+$2404}
\newcommand{\macs}{MACS 1423}
\newcommand{\obj}{\texttt{Firefly Sparkle}}
\newcommand{\lya}{Lyman-$\alpha$$~$}

\newcommand{\Oiii}{[O~{\sc iii}]}

\newcommand{\Oiiia}{[O~{\sc iii}]~$\lambda$4959$~$}
\newcommand{\Oiiib}{[O~{\sc iii}]~$\lambda$5007$~$}
\newcommand{\zffs}{\rm z$_{spec}$ $=$ 8.304 $\pm$ 0.001}

\newcommand{\nclusters}{ten}
\newcommand{\ffsc}{Firefly Sparkle}
\newcommand{\ffs}{\ffsc$~$}

\theoremstyle{thmstyleone}%
\theoremstyle{thmstyletwo}%

\theoremstyle{thmstylethree}%

\begin{document}

\title[The \ffsc]{The \ffsc:
The Earliest Stages of the Assembly of A Milky Way-type Galaxy in a 600 Myr Old Universe}


\author*[1]{\fnm{Lamiya} \sur{Mowla ({\bng lamiiya mOla})}}\email{lmowla@wellesley.edu}
\equalcont{These authors contributed equally to this work.}

\author*[2]{\fnm{Kartheik} 
\sur{Iyer}}\email{kgi2103@columbia.edu}
\equalcont{These authors contributed equally to this work.}

\author[3,4]{\fnm{Yoshihisa} \sur{Asada}}
\author[3]{\fnm{Guillaume} \sur{Desprez}}
\author[5]{\fnm{Vivian Yun Yan} \sur{Tan}}
\author[8]{\fnm{Nicholas} \sur{Martis}}
\author[5]{\fnm{Ghassan} \sur{Sarrouh}}
\author[6,7]{\fnm{Victoria} \sur{Strait}}
\author[9]{\fnm{Roberto} \sur{Abraham}}
\author[8]{\fnm{Maru\v{s}a} \sur{Brada\v{c}}}
\author[6,7]{\fnm{Gabriel} \sur{Brammer}}
\author[5]{\fnm{Adam} \sur{Muzzin}}
\author[13]{\fnm{Camilla} \sur{Pacifici}}
\author[11]{\fnm{Swara} \sur{Ravindranath}}
\author[3]{\fnm{Marcin} \sur{Sawicki}}
\author[10]{\fnm{Chris} \sur{Willott}}
\author[3]{\fnm{Vince} \sur{Estrada-Carpenter}}
\author[12]{\fnm{Nusrath} \sur{Jahan}}
\author[3]{\fnm{Gaël} \sur{Noirot}}
\author[6,7]{\fnm{Jasleen} \sur{Matharu}}
\author[8]{\fnm{Gregor} \sur{Rihtar\v{s}i\v{c}}}
\author[3]{\fnm{Johannes} \sur{Zabl}}

\affil*[1]{\orgdiv{Whitin Observatory, Department of Physics and Astronomy}, \orgname{Wellesley College}, \orgaddress{\street{106 Central St.}, \city{Wellesley}, \postcode{02481}, \state{MA}, \country{USA}}}

\affil[2]{\orgdiv{Columbia Astrophysics Laboratory}, \orgname{Columbia University}, \orgaddress{\street{550 West 120th Street}, \city{New York}, \postcode{10027}, \state{NY}, \country{USA}}}

\affil[3]{\orgdiv{Department of Astronomy and Physics and Institute for Computational Astrophysics}, \orgname{Saint Mary's University}, \orgaddress{\street{923 Robie Street}, \city{Halifax}, \postcode{B3H 3C3}, \state{Nova Scotia}, \country{Canada}}}

\affil[4]{\orgdiv{Department of Astronomy}, \orgname{Kyoto University}, \orgaddress{\street{Sakyo-ku}, \city{Kyoto}, \postcode{606-8502}, \state{Kyoto}, \country{Japan}}}

\affil[5]{\orgdiv{Department of Physics and Astronomy}, \orgname{York University}, \orgaddress{\street{4700 Keele St.}, \city{Toronto}, \postcode{M3J 1P3}, \state{Ontario}, \country{Canada}}}

\affil[6]{\orgdiv{Cosmic Dawn Center (DAWN)},\country{Denmark}}

\affil[7]{\orgdiv{Niels Bohr Institute}, \orgname{University of Copenhagen}, \orgaddress{\street{Jagtvej 128}, \city{Copenhagen}, \postcode{DK-2200}, \country{Denmark}}}

\affil[8]{University of Ljubljana, Department of Mathematics and Physics, Jadranska ulica 19, SI-1000 Ljubljana, Slovenia}

\affil[10]{NRC Herzberg, 5071 West Saanich Rd, Victoria, BC V9E 2E7, Canada}

\affil[9]{David A. Dunlap Department of Astronomy and Astrophysics, University of Toronto, 50 St. George Street, Toronto, Ontario, M5S 3H4, Canada}

\affil[11]{Astrophysics Science Division, NASA Goddard Space Flight Center, 8800 Greenbelt Road, Greenbelt, MD 20771, USA}

\affil[12]{Shahjalal University of Science and Technology, University Ave, Sylhet 3114, Bangladesh}

\affil[13]{Space Telescope Science Institute, 3700 San Martin Drive, Baltimore, MD 21218 USA}




\maketitle


\textbf{ 
The most distant galaxies detected by JWST are assembling in a Universe that is less than 5\% of its present age. At these times, the progenitors of galaxies like the Milky Way are expected to be about 10,000 times less massive than they are now, with masses quite comparable to that of massive globular clusters seen in the local Universe. Composed today primarily of old stars and correlating with the properties of their parent dark matter halos, the first globular clusters are thought to have formed during the earliest stages of galaxy assembly. In this article we explore the connection between star clusters and galaxy assembly by showing JWST observations of a galaxy at {\zffs}, strongly lensed by a massive foreground cluster of galaxies, exhibiting a network of massive star clusters (the `\ffsc') cocooned in a diffuse arc. The \ffsc\ exhibits the hallmarks expected of a future Milky Way-type galaxy captured during its earliest and most gas-rich stage of formation. The mass distribution of the galaxy seems to be concentrated in ten distinct clusters ($\sim 49-57\%$ of the total mass), with individual cluster masses (M$_{*,\mathrm{cluster}} \sim 10^{5.3} - 10^{5.8}$M$_\odot$) that straddle the boundary between low-mass galaxies and high-mass globular clusters. The cluster ages suggest that they are gravitationally bound with star formation histories showing a recent starburst possibly triggered by the interaction with a companion galaxy at the same redshift at a projected distance of $\sim$2 kpc away from the \ffsc. The central star cluster shows nebular-dominated spectra consistent with high temperatures ($\mathrm{T_e}\sim 40,000$K) and a top-heavy initial mass function ($0.78<\alpha<2.0$), the product of formation in a very metal poor environment. Combined with abundance matching that suggests that this is likely to be a progenitor of galaxies like our own, the \ffs provides an unprecedented case study of a Milky Way-like galaxy in the earliest stages of its assembly in only a 600 million year old Universe.
}

\begin{figure}[t]%
\centering
\includegraphics[width=\textwidth]{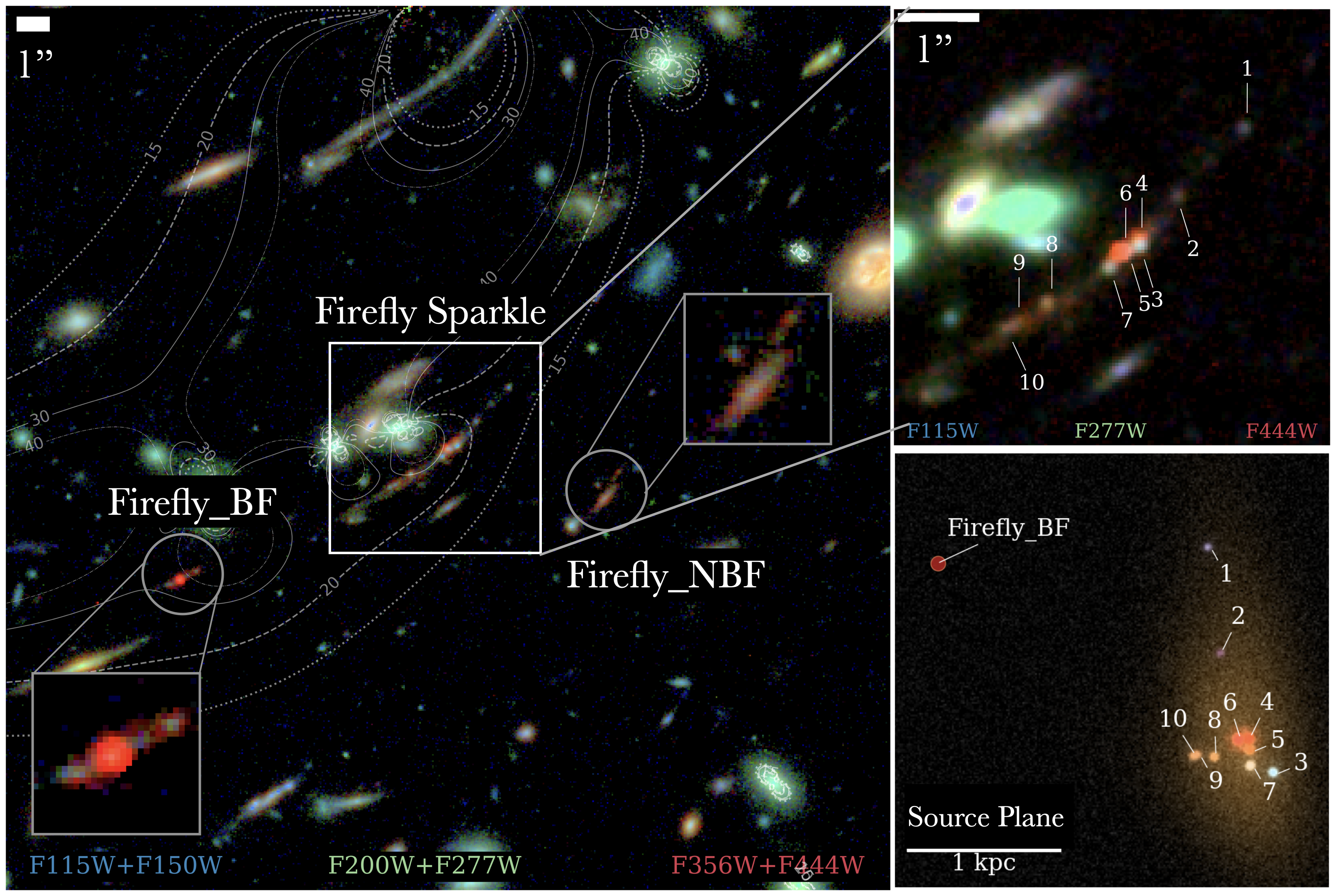}
\caption{The \ffs is a redshift \zffs\ gravitationally magnified arc lensed by the \macsj\ cluster. \textbf{Left}: Full field with the three objects of interest - \ffs (center), Firefly-Best Friend (FF-BF; bottom left), and Firefly-New Best Friend (FF-NBF; bottom right) - shown in boxes and circles. The contours show the lines of lensing magnifications ($\mu=$ 15, 20, 30, and 40). \textbf{Top right}: zoom in on the \ffs with the clusters labelled. \textbf{Bottom right}: source plane reconstruction of the \ffs with the position of neighbor FF-BF shown in red circle.}\label{fig:field}
\end{figure}

The \ffs is a gravitationally lensed arc first identified with the Hubble Space Telescope (HST) in the Cluster Lensing And Supernova survey (CLASH) survey of the galaxy cluster \macsj\ and reported as a $z>7$ candidate \citep{postman12}.
Follow-up ground based spectroscopy of the galaxy using MOSFIRE on  the Keck telescope resulted in a tentative redshift of $z=7.6$, based on a possible Ly$\alpha$ detection \citep{hoag17}. 
Recently, the CAnadian NIRISS Unbiased Cluster Survey (CANUCS) revisited the cluster field with JWST, observing it with the NIRISS, NIRCam and NIRSpec instruments \citep{willott22}.
NIRCam and NIRISS imaging was done in 11 bands covering 0.8 - 5 $\mu$m. As shown in Figure~\ref{fig:field}, the galaxy presents itself as a long magnified arc with \nclusters\ distinct clusters embedded within a low surface brightness component (the diffuse arc) that extends up to 5" in length, extending much further than seen in previous HST imaging.  Based on the unambiguous identification of \Oiiia and \Oiiib emission lines in NIRSpec Prism spectroscopy, we confirm that it is a high-redshift system (\zffs), albeit at a slightly higher redshift than reported by ground-based spectroscopy. There is no evidence for \lya emission in the NIRSpec spectrum. 
The arc is completely invisible in the bluest observed NIRCam and NIRISS bands (F090W), consistent with the arc and all clusters being at $z>7.5$. 

We created a magnification model using {\em Lenstool} \citep{Kneib:2004fx,jullo07} which is constrained with three multiple image systems \citep{hoag17} for which we provide spectroscopic redshifts in the CANUCS dataset. The magnification model of the \ffs arc shows that the system is magnified by factors between 16-26.
One of the highest magnification region ($\mu=$24) of the image is at the center of the arc and has strong \Oiii\ emission, which dominates the F444W flux, making it appear as red in the composite image of Figure \ref{fig:field}. The projected half-light size of the arc in the source plane is only 0.3 $\pm$ 0.1 kpc as most of the bright clusters are concentrated near the center of the galaxy. While eight out of the ten unresolved clusters are gathered towards the center of the galaxy in the source plane, clusters FF-1 and FF-2 lie along an elongated arm and the projected distance between FF-1 and the central cluster FF-5 is 1.4 physical kpc. The \ffs has two neighbors with z$_{\rm spec}$ $=$ 8.2996 $\pm$ 0.0008 and z$_{\rm spec}$ $=$ 8.2967 $\pm$ 0.0016. The closest neighbor, dubbed Firefly-BF (`best friend', henceforth called BF) ($\mu=$40), is within 2 kpc of FF-1. The BF also appears red at the center and has an \Oiii\ emission region that has been observed with with NIRSpec/Prism. Due to its high \Oiii\ equivalent width (EW) contributing to the F444W broadband flux, this source can appear to be a double-break source similar to those in \citealt{labbe23} and \citealt{desprez23}. Faint low-surface brightness features are visible at the corners of the arc close to the neighbor, hinting at a possible interaction between the two galaxies which may have triggered a burst of star formation in both of them (see Fig. 3). The second neighbor, Firefly-NBF (`new best friend'), is at a projected physical distance of 13 kpc in the source plane, and has a very faint \Oiii\ emission in NIRSpec. In this {\em Article}, we focus on the \ffs and its star clusters, touching only briefly on the spectroscopy and spectral energy distribution (SED) of the two neighboring galaxies when it is needed to complete the picture. 

We use NIRCam + NIRISS imaging to study the resolved \ffs in order to understand the distribution of stellar mass in the individual clusters vs the diffuse arc. Our photometry is derived by joint modelling of the {\nclusters} clusters, and the diffuse arc of the \ffs  in the NIRISS and NIRCam images using GALFIT \citep{Peng2010}. Nine of the {\nclusters} clusters are consistent with being unresolved, even in the highest resolution F115W NIRCam images. Only one central cluster (FF-4) exhibits an elongated component that we fit with a Gaussian ellipse (half-light radius, r$_{50}$=0.01"). The diffuse arc is also fitted with a Gaussian ellipse with a bending mode. All 11 components are fitted simultaneously for every filter, to derive the total flux of each component (for details, see Section \ref{sec:photometry}). An upper limit on the half-light radii (R$_{\rm eff}$) of the nine unresolved clusters is derived assuming 0.5 times the full width half maximum (FWHM) of the stellar point spread function (PSF) in the F115W image and is 0."02 (0.5 $\times$ 0."04). Since the best-fit major axis size of FF-4 (0."01) is smaller than the PSF size, we place an upper limit on FF-4's size as well. The tangential magnifications of the clusters range between $\mu_{\rm tan}=$ 12-21 resulting in an upper limit on the half-light sizes of merely 4-7 pc. Therefore, the {\ffs} may be offering an unprecedented glimpse into the building blocks of a galaxy as it undergoes its assembly at z $>$ 8.  Resolved photometry is also necessary for estimating the total mass of the {\ffsc}, as global spectral energy distribution (SED) fitting can bias stellar masses when a young stellar population outshines the first episodes of star formation (e.g., \citealp{sorba18, gimenez-arteaga23, narayanan23})

\begin{figure}[t]%
\centering
\includegraphics[width=\textwidth]{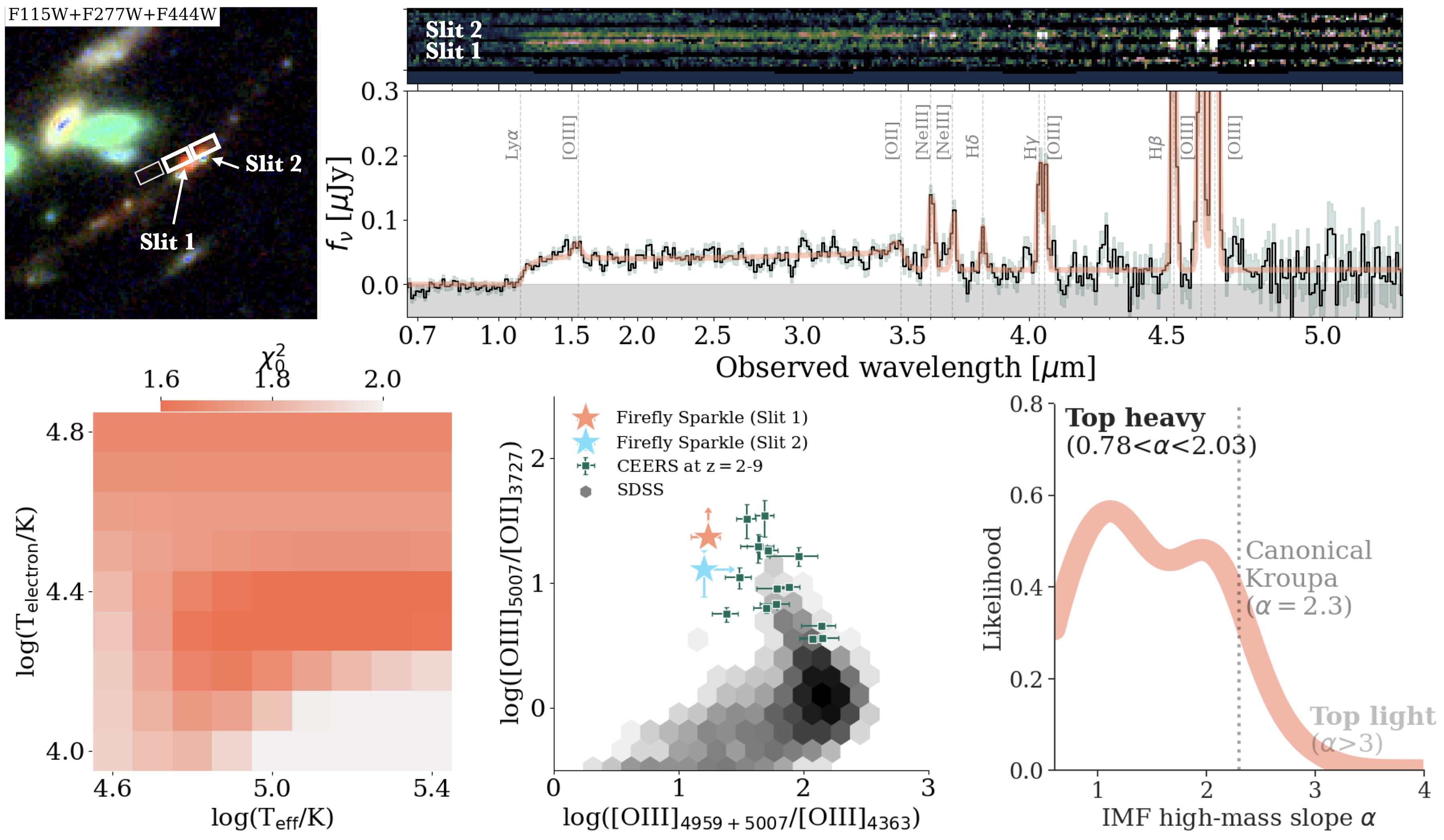}
\caption{Physical properties of the \ffs measured from the NIRSpec/Prism spectra of Slit 1 (which includes light from cluster 6 as well as fractional contributions from clusters 3,4, and 5; see Figure \ref{fig:field}) of the \ffsc. \textbf{Top Left:} Positions of the slitlets on the arc. \textbf{Top Right:} The 2D NIRSpec spectrum for Slit 1 (bottom line) and Slit 2 (top line) and the 1D spectrum of Slit 1 only. \textbf{Bottom Left:} Posteriors of the effective black-body temperature and electron temperature from CLOUDY modeling of the nebular continuum. The Slit 1 region exhibits electron temperature of T$_{\rm electron}\sim$ 20,000K, and ionizing source effective temperature of T$_{\rm effective} \sim$ 10$^5$K.  \textbf{Bottom middle:} Emission line diagnostics estimated from the fitted line ratios for RO3 and O32 hints at a metal poor stellar population. \textbf{Bottom right:} Marginal likelihood for the high-mass IMF slope from joint spectrophotometric fitting with \db suggest a top heavy IMF ($\alpha<$2), full posteriors shown in Figure \ref{fig:spectra}. }\label{fig:spec_prop}
\end{figure}

\begin{figure}[t]%
\centering
\includegraphics[width=\textwidth]{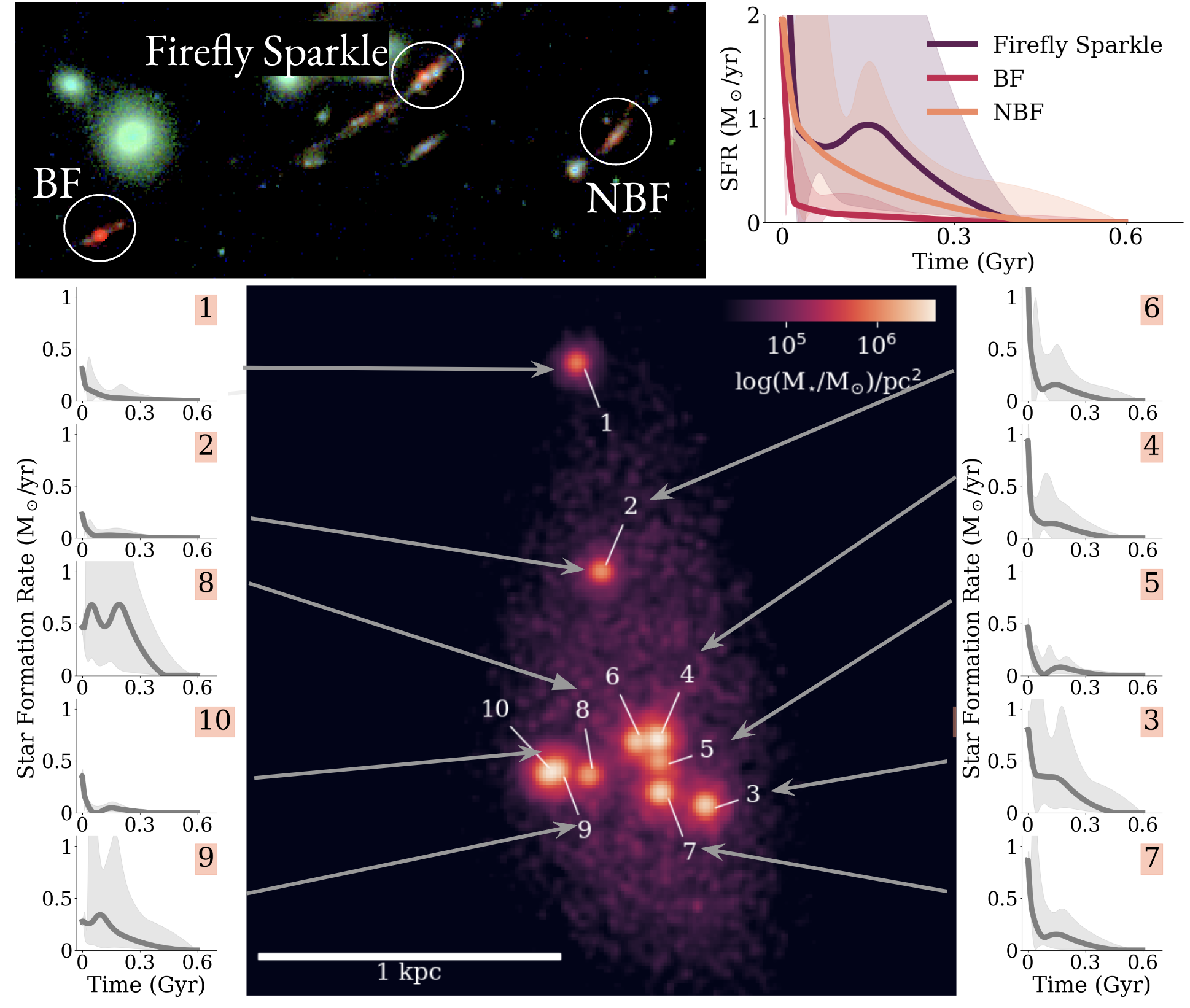}
\caption{\textbf{Top Left:} Image of the {\ffs}, FF-BF, and FF-NBF. \textbf{Top Right:} Non-parametric integrated star formation histories of these three galaxies reconstructed from SED fitting. The \ffs and FF-BF both show a recent burst of star formation in the last $\sim 50$ Myr indicative of recent interactions. \textbf{Bottom Middle:} The source plane stellar mass map of the {\ffs} reconstructed from the individual SED fits of the components and its source plane model. \textbf{Bottom Left and Right:} Star formation histories of the individual clusters inferred from SED fitting. Most of the clusters show ongoing star formation activity or starbursts.}\label{fig:sSFRage}
\end{figure}

\begin{figure}[h]%
\centering
\includegraphics[width=\textwidth]{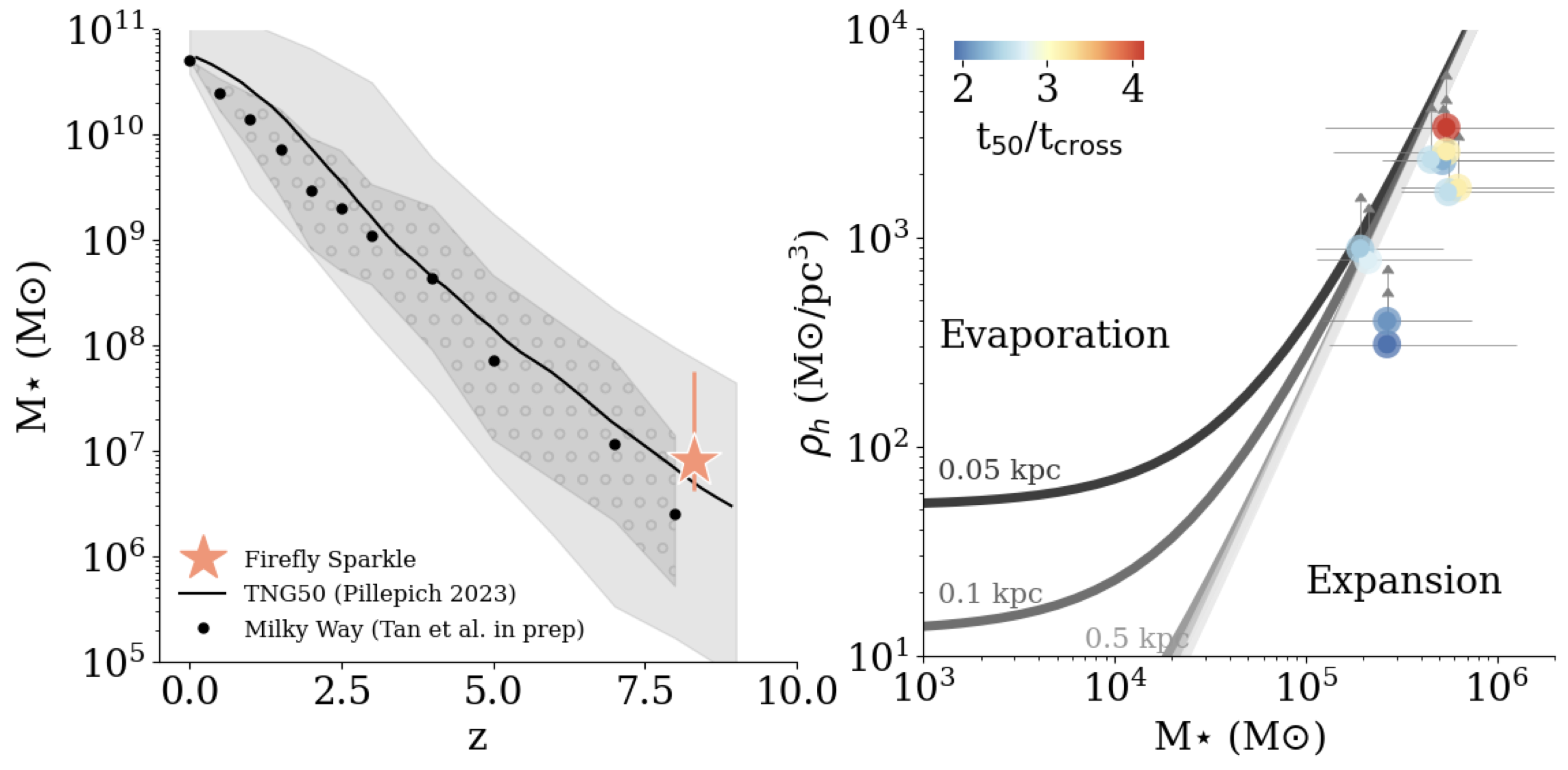}
\caption{The \ffs in context. \textbf{Left:} Progenitors of Milky Way galaxy analogues in the TNG50 simulation \citep{pillepich23} and from observations (abundance matching \citep{Behroozi:2013} applied to observed stellar mass functions \citep{grazian15,McLeod:2021,stefanon19}; Tan et al. in prep.). Given its stellar mass, the \ffs is within $1\sigma$ of the median mass of a progenitor at $z\sim 8.3$. \textbf{Right:} Stability of the individual star clusters in a tidal field given their stellar mass and mean density within a half-mass radius. The grey lines represent isochrones based on the balanced evolution model in \citealt{gieles2011} for different galactrocentric radii. The majority of the clusters are close to the boundary between expansion and evaporation dominated phase. The color indicate the age (t$_{50}$)/crossing time of the star clusters. Majority of these star clusters are not expected to survive to the present, but will instead expand and tidally stripped apart to form the stellar disk and the halo. However, star clusters kicked out to large galactocentric radii may survive to become present day globular clusters.}\label{fig:progen}
\end{figure}

We acquired NIRSpec Prism spectroscopy with two slits (hereafter referred to as Slit1 and Slit2) covering the central brightest region of the \ffsc, a third slit on the FF-BF, and a fourth slit on the FF-NBF. The spectra and the inferred properties from Slit 1 of {\ffs} are shown in Figure \ref{fig:spec_prop} while the extracted spectra of all four slits are shown in Figure \ref{fig:allspectra}. The spectrum of Slit 1 (which includes light from FF-6 as well as fractional contributions from FF-3, FF-4, and FF-5) exhibits characteristic nebular continuum features including a clear discontinuity at $\lambda_{\rm obs}\sim3.5\ \mu$m corresponding to a Balmer jump \citep{2024MNRAS.527.7965W}, and a smooth turnover at $\lambda_{\rm obs}<1.3\ \mu$m, possibly due to two-photon continuum \citep{Cameron2023}.
These features are quite similar to those observed in \citep{Cameron2023, 2023ApJ...952...74T, 2023MNRAS.518L..45C, 2023arXiv230600487U, 2023arXiv230600647H}, suggesting that the nebular continuum is a significant contributor to the overall SED for 
this star-forming cluster. 
The ionizing source effective temperature of $T_{\rm eff}=$10$^5$ K is obtained in two ways: from modeling the nebular continuum with CLOUDY and from line ratios after checking that the Balmer decrement is consistent with being dust-free (see \ref{subsec:specfit}). The temperature is much hotter than typical massive type O stars, and is suggestive of this star-forming cluster having a different ionizing source to produce the harder ionizing radiation leading to a higher upper mass limit for the IMF or a top heavy IMF \citep{Cameron2023}. 


We perform SED fitting in two stages (described in Section \ref{subsec:sedfit}) - we initially perform a joint spectrophotometric fit to the NIRSpec Prism spectrum along with the HST + NIRISS + NIRCam photometry in the two Slits where both exist (seen in Figure \ref{fig:spectra}), followed by photometric modeling of the star formation histories of the individual clusters folding in constraints from the spectrum. 
All ten clusters have intrinsic (corrected for magnification) stellar masses of $\sim$ 10$^5$ to 10$^6$ M$_{\odot}$ and specific star formation rates (sSFR) of $\sim$ 10$^{-7}$ year$^{-1}$. The smooth component of the arc has the highest stellar mass of log(M$_{\star}$/M$_{\odot})=$ 6.6$^{+0.8}_{-0.3}$ and a specific star formation rate similar to that of the star clusters (Fig.~\ref{fig:sSFRage}). The surface density of the star clusters range between 10$^3$-10$^4$ M$_{\odot}$/pc$^2$, lower than even higher redshift star clusters \citet{adamo24} but similar to the surface denisty of MW- globular cluster \citep{1979ARA&A..17..241H}. 



From our joint analysis of the spectrophotometric data, we find some evidence of a top-heavy IMF in Slit 1, which is the only spectrum with SNR high enough to constrain the slope. Varying the power-law slope of the Kroupa IMF in FSPS results in an excess of high mass stars, which (i) increase the nebular continuum from the two-photon component \citep[and therefore the strength of the Balmer jump;][]{Cameron2023}, and (ii) increases the equivalent width of the emission lines. The change in the strength of the Balmer jump and the contribution to the various emission lines is distinct from SFR and allows us to obtain joint posteriors on the two quantities. The resulting fits rule out models with IMFs that are more top-light than the canonical Kroupa value ($\alpha = 2.3$) and show a preference for lower, more top-heavy, slopes ($\alpha_{\rm Slit~1} = 1.3_{-0.5}^{+0.9}$. This is consistent with theoretical expectations in the early universe when massive star clusters form in gas-rich environments \citep{2015ApJ...808...24C, Cameron2023}.

While the primary objective of this article is to highlight the \ffs object as a representative of high-z galaxy formation with both spatially resolved and spectrophotometric observations, with both the main galaxy and its companions having secure redshifts, it is important to note that the current analysis comes with several caveats. Changes in the magnification model as we further refine and add sources with future spectroscopic confirmations could change the estimated masses and magnifications of the clusters. However, the estimated t$_{50}$, sSFR, and surface densities are insensitive to this change and therefore do not change our overall picture. Additionally, given the extreme shear of this object it is unlikely that the magnification will significantly change from its current value. A second possible systematic is the modeling of the spectrophotometry, which could be affected by choices of stellar population and photoionisation models and assumed prior on parameters like the SFH, affecting estimates of stellar masses, IMF and star formation histories. The correct assumptions for these parameters and prior does not currently have consensus within the literature, but as JWST enables more observations at high redshifts we are gaining further insights into the possible systematics and accounting for them. In the current analysis, we have tried to mitigate the effects of these factors by adopting flexible non-parametric models where possible, and providing independent measurements where possible (e.g. for the electron temperature). We also find that these independent measurements tend to corroborate each other, e.g. the high electron temperature is consistent with scenarios with a top-heavy IMF \citep{Cameron2023}. Finally, a concern is that the absence of the rest-frame K-band makes estimating the stellar masses extremely fraught. While future JWST/MIRI observations will help definitively estimate the mass, our resolved analysis and non-parametric SFHs are expected to mitigate this effect \citep{sorba18,carnall19b, leja19} and provide a more robust estimate of the total mass of the system. 

The motivation for the \ffs came from the Sparkler \citep{mowlaiyer2022} - a strongly lensed galaxy at $z=1.38$ surrounded by old star clusters that could be resolved only with JWST. In contrast to the Sparkler, the \ffs represents one of JWST's first spectrophotometric observations of an extremely lensed galaxy assembling at high redshifts, with clusters that are in the process of formation instead of seen at later epochs. In addition to the star clusters, the \ffs is especially interesting in context as a possible Milky-Way like progenitor. Abundance matching methods allow us to estimate the masses of galaxies that might be MW progenitors at high redshifts \citep{Behroozi:2013, papovich15}, and have been used to study the evolution of systems at high-z in both simulations and observations that can come to resemble our galaxy at late times \citep{papovich15, graziani17}. By comparing this against galaxies from the TNG50 simulation \citep{pillepich23} or abundance matching using observed galaxy mass functions, we can estimate the range of stellar masses that a possible Milky Way or M31 progenitor can have at the redshift of the \ffs (see left panel of Figure \ref{fig:progen}; detailed further in Section \ref{subsec:projmatch}). We find that the \ffs has a total demagnified stellar mass of log M$_* \sim 6.8^{+0.68}_{-0.26}$M$_\odot$, which is consistent with the mass we expect for a MW or M31-like galaxy in the present day universe. The implications of this are far reaching, but primarily that the \ffs could represent the early stages of a galaxy much like our own, unlike the extremely massive systems that aren't magnified by lensing seen more commonly with JWST. 

We think of galaxies assembling at high-z in a very gas-rich environment \citep{papovich16} through accretion fueled starbursts, leading to disk formation \citep{2023BAAA...64..143I, 2015ApJ...808L..17F, 2024arXiv240200561N}. 
These gas-rich disks are subsequently disrupted or enhanced by mergers, instabilities, feedback, mis-aligned accretion and other events \citep{2016ASSL..418..355B, 2017MNRAS.464..635M, 2018ApJ...861..148M,  2022MNRAS.512.3806V}.  
However, the current observations indicate that galaxy assembly at high redshifts occurs in a fundamentally different mode, with a majority of the stellar mass and SFR happening in dense gravitationally bound star clusters that seem to be stable against disruption by feedback (see Figure \ref{fig:progen} and section \ref{subsec:density_bound} for more details), or occurring in a regime where feedback is less effective \citep{2023MNRAS.523.3201D, 2024MNRAS.527.5929Y, 2023MNRAS.525L.117R}. The extreme conditions in these clusters can also alter the star formation efficiency and IMF, which can be tested with the current observations \citep{lancaster21}. This scenario is also favored by statistical studies \citep{2024MNRAS.527.5929Y}, which find that efficient stellar-driven winds (such as those arising from a top-heavy IMF or bursty star formation) could account for the self-consistent evolution of the galaxy mass function at high redshifts.
These `billiard balls' are sites of rapid star formation that may be stripped by tidal forces in the galaxy's nascent disk to become nuclear star clusters \citep{2016ASSL..418..355B, 2017MNRAS.464..635M, 2022ApJ...928..106T}. In particular, \citep{bournaud14} find in simulations that clumps can survive for $>100$ Myr because although they are being tidally stripped and losing mass, the gas-rich turbulent environment allows them to reaccrete gas at rates high enough to compensate for this. 
Another possibility is that some of these are ejected from the galaxy by mergers and other high-energy events and form the first generation of the galaxy's globular cluster population (Cluster 1 in this observation is a prime candidate for this). \citep{2014ApJ...785...71G,  2014MNRAS.443.3675M, 2015llg..book..477E, 2018RSPSA.47470616F} 
The \ffs provides a rich source for future observations that can test these scenarios to better understand the earliest stages of formation for galaxies like our own.


\section{Methods}\label{sec11}

\subsection{Image preparation}

The cluster field \macsj\ was observed with JWST/NIRCam imaging using filters F090W, F115W, F150W, F200W, F277W, F356W, F410M, and F444W with exposure times of $6.4 \mbox{ks}$ each, reaching signal-to-noise between 5 and 10 for a $m_{\rm AB} = 29$ point source. It was also observed with JWST/NIRISS imaging using filters F115W, F150W, and F200W.
 
To reduce the imaging data we use the photometric pipeline that is presented in more detail in \citet{asada24}. Briefly, the raw data  has been reduced using the public grism redshift \& line analysis software {\grizli} \citep{grizli23}, which masks imaging artifacts, provides astrometric calibrations based on the Gaia Data Release 3 catalog \citep{gaia2023}, and shifts images using {\astrodrizzle}. The photometric zero-points are applied as described in \citet{grizliphot}. RGB image created using 6 filters of NIRCam observation of the \ffs is shown in Fig.\ref{fig:field}. We utilize background and bright cluster galaxy subtracted image to remove the effect of background in our photometry. The methodology for modeling and removing diffuse light from cluster galaxies and intracluster light (ICL) is presented in \citet{martis2024}. The NIRCam depths (0.3" diameter aperture) for F090W, F115W, F150W, F200W, F277W, F356W, F410M, and F444W are 7.2, 6.6, 5.2, 4.4, 3.0, 2.9, 5.5, and 4.3 nJy and the NIRISS depths for F115WN, F150WN, and F200WN are 3.6, 4.3, and 4.0 nJy respectively \citep{willott23}.

\subsection{Photometry of the {\ffsc}}
\label{sec:photometry}

We perform photometry in ten JWST bands (NIRISS: F115WN, F150WN and F200WN; NIRCam: F115W, F150W, F200W, F277W, F356W, F410M, and F444W) in which the \ffs is detected from their morphological fit with GALFIT.  In other {\jwst} and {\hst} filters the \ffs is not/barely detected) and hence we place upper limits for the entire source. Since the object is resolved into at least \nclusters\ distinct clusters and a diffuse galaxy component, we perform a morphological fit using {\galfit} \citep{peng10} to extract the photometric information. 

\begin{figure*}[t]
\begin{center}
\includegraphics[width=0.97\textwidth]{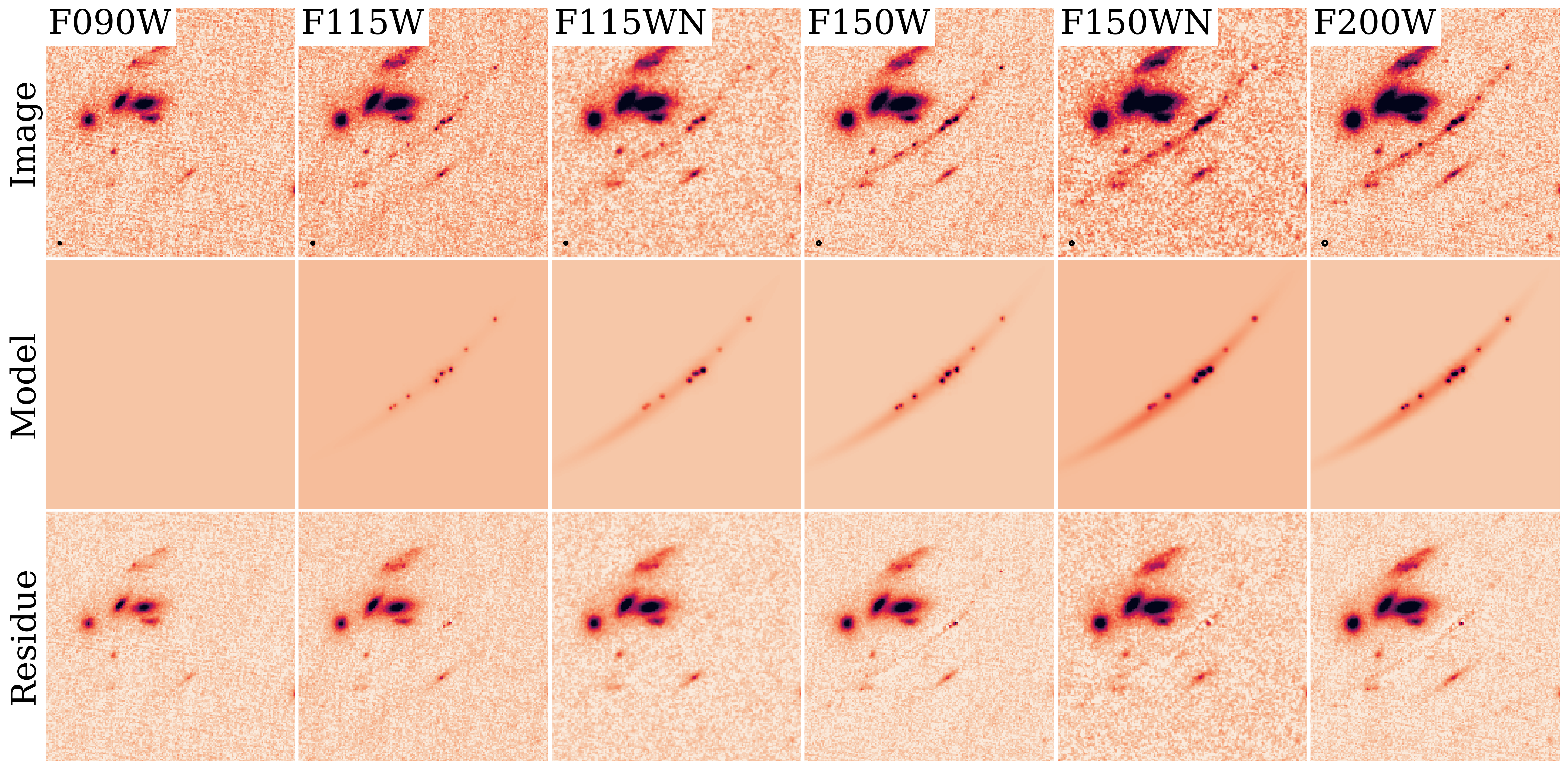}
\includegraphics[width=0.97\textwidth]{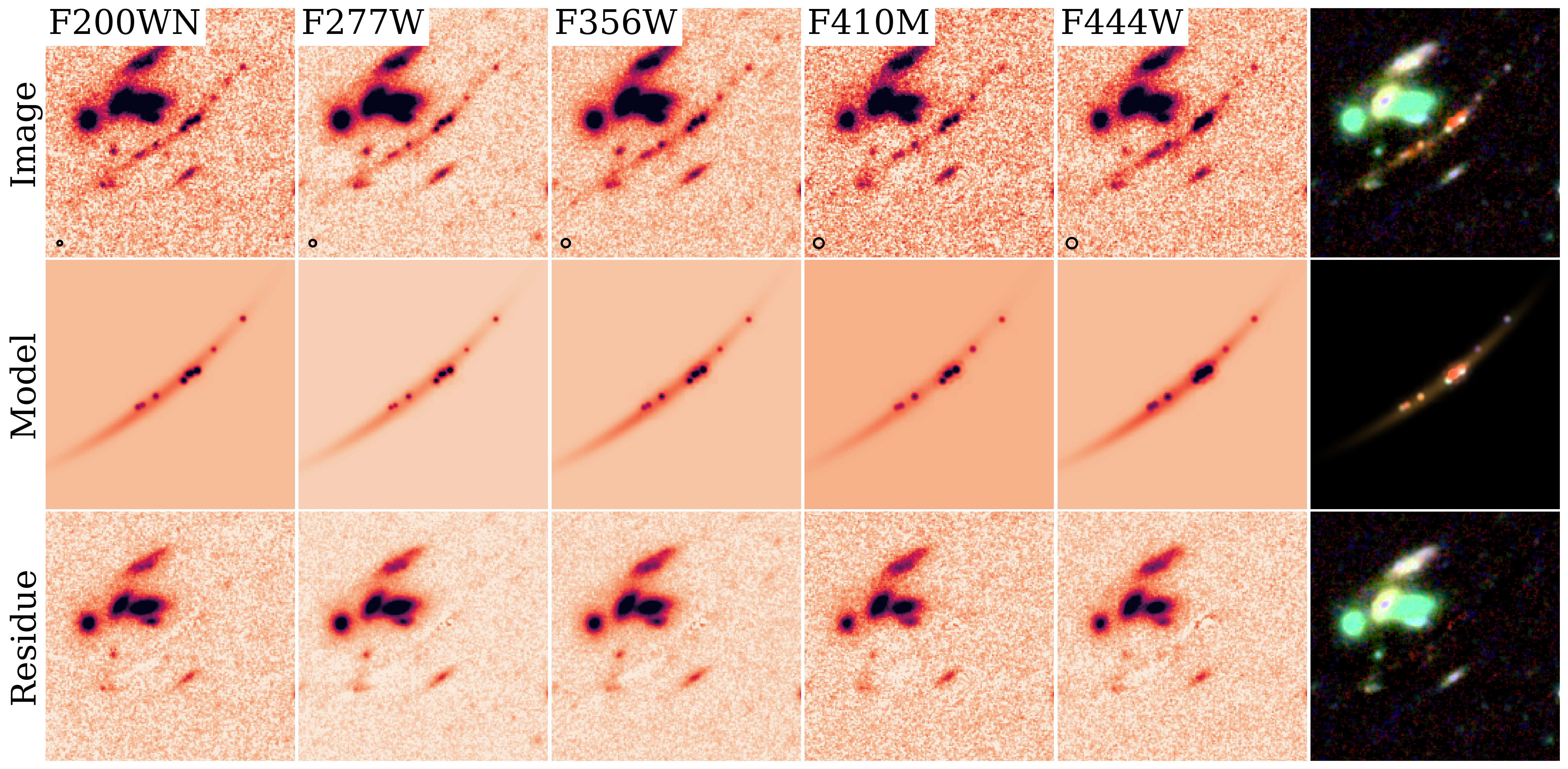}
\end{center}
\caption{Morphological fit with GALFIT of the \ffs in 11 JWST/NIRCam and NIRISS filters for photometric extraction and size determination. The images, their respective models, and residues for 11 filters and the RGB image (R: F444W, G: F277W, B: 115W) are shown. The \ffs is completely invisible in the bluest filter (F090W). Based on the reduced $\chi^2$ of the fits, nine out the ten clusters of the \ffs are consistent with being point sources in F115W. The full model consists of nine point sources, an elliptical Gaussian for a semi-resolved cluster (FF-4), and an elliptical Gaussian with a bending mode for the diffuse arc. The photometry is derived from the total model flux of the 11 components. The error of the photometry is estimated by injecting the full model in random locations in the \macs field, and refitting them them with GALFIT. The upper limit on the size of the nine unresolved clusters is determined from the HWHM of the F115W PSF (0."02).}
\label{fig:allgalfit}
\end{figure*}

We measure empirical point spread function (PSF) determined from the observations of stars. For the morphological fit, we create 10"$\times$10" postage stamps in all ten filters from the BCG subtracted images. We first model the 3 bright galaxies in the upper left corner of the postage stamps near the \ffs (see Fig. \ref{fig:allgalfit}) and subtract them from all ten filter images to ensure that light from the wings of these galaxies is not significantly affecting our photometry. We determine the priors for the centers of the ten clusters by visual inspection. We then use GALFIT to fit an elliptical Gaussian for FF-4, nine point sources for the other nine clusters, and another elliptical Gaussian with bending mode turned on for the diffuse arc, to the F115W image which has the highest resolution. The coordinates for the centers of the point sources (initial guess from visual inspection), the total magnitudes (initial guess M$=$27), the radii, and the ellipticity of the elliptical Gaussians, the bend parameter (\texttt{B2} on GALFIT) for the diffuse arc, and the sky background, are all left as free parameters. In the next stage, we use the best-fit center coordinates from F115W as initial guess and repeat the fit on the F444W image, which has the highest signal to noise for the arc. From this fit we determine the radii (4."0 and 0."2), and the ellipticity of the two Gaussians (b/a$=$0.08 for both), as well as the bending mode 2 amplitude \texttt{B2$=$2.14} for the arc, that we fix for the rest of the fits. We now fit all the components in all ten filters to determine their fluxes. We provide initial guesses for the coordinates of the clusters from the F115W fit, allowing them to vary by $\pm$0.5 pixel ($0"02$) in each direction to account for the centroiding uncertainty of the empirical PSF in each filter. The resulting models and residuals are shown in Fig.~\ref{fig:allgalfit}. Residuals from the fits are negligible, confirming the original visual impression that the \nclusters$~$ compact sources are unresolved and an additional smooth component is present. 

To derive the uncertainty in our flux estimation, we inject the full \ffs model in 100 random locations in our 10"$\times$10" postage stamps (avoiding the edge) and refit with the exact same setting of GALFIT. We find no significant systematic offset between the fitted flux and the injected flux for any of the eleven components, in any of the filters, showing that our photometric technique is robust to background variations across all filters. The uncertainty in the photometry is calculated from the biweight scale of the 100 refitted fluxes. The resulting photometry and the RGB image of the model and the residue are shown in Fig. \ref{fig:allgalfit}
The agreement between NIRISS and NIRCam fluxes in the three overlapping filters is another confirmation of the robustness of photometry. We have used updated zero-points \citep{grizliphot}, and corrected for Milky Way extinction using the value of color excess $E(B-V)=0.0272$ from \citealp{schlafly11} and assuming the extinction law by \citep{fitzpatrick99} using the factor between the extinction coefficient and color excess $R_{\rm V}=3.1$.

\begin{figure}[h]%
\centering
\includegraphics[width=\textwidth]{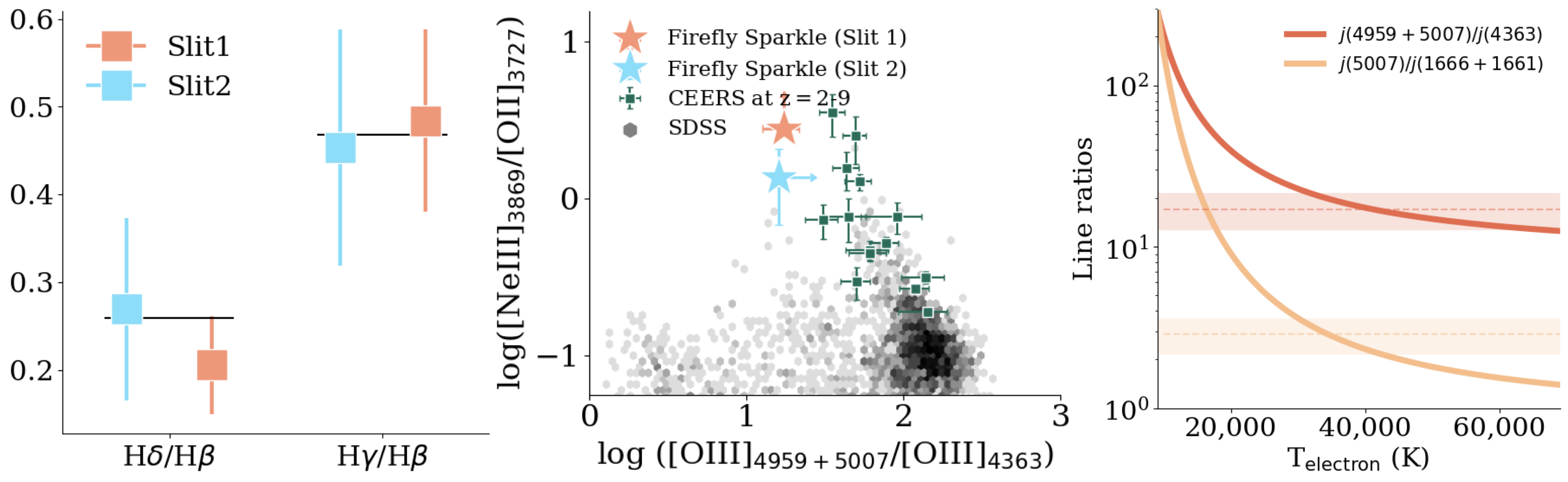}
\caption{Inferring properties from the NIRSpec Prism spectrum of the \ffsc. \textbf{Left:} Balmer decrements of H$\delta$/H$\beta$ and H$\gamma$/H$\beta$ in slit 1 (red) and slit 2 (blue) spectra. Black solid line denotes the line ratios under Case B recombination. The line ratios indicates the dust attenuation is not significant in both spectra. \textbf{Middle:} Similar to bottom middle in Fig.~\ref{fig:spec_prop}, but Ne3O2 ratio is used instead as an indicator of the ionization parameter. \textbf{Right:} Electron temperature measurements from [O {\sc iii}] emission lines in \ffs slit 1. Two different emission line ratios independently suggest a high electron temperature of $T_{\rm e, O^{++}}\sim40000$ K. }\label{fig:spectra_prism}
\end{figure}

\begin{figure}[h]%
\centering
\includegraphics[width=0.8\textwidth]{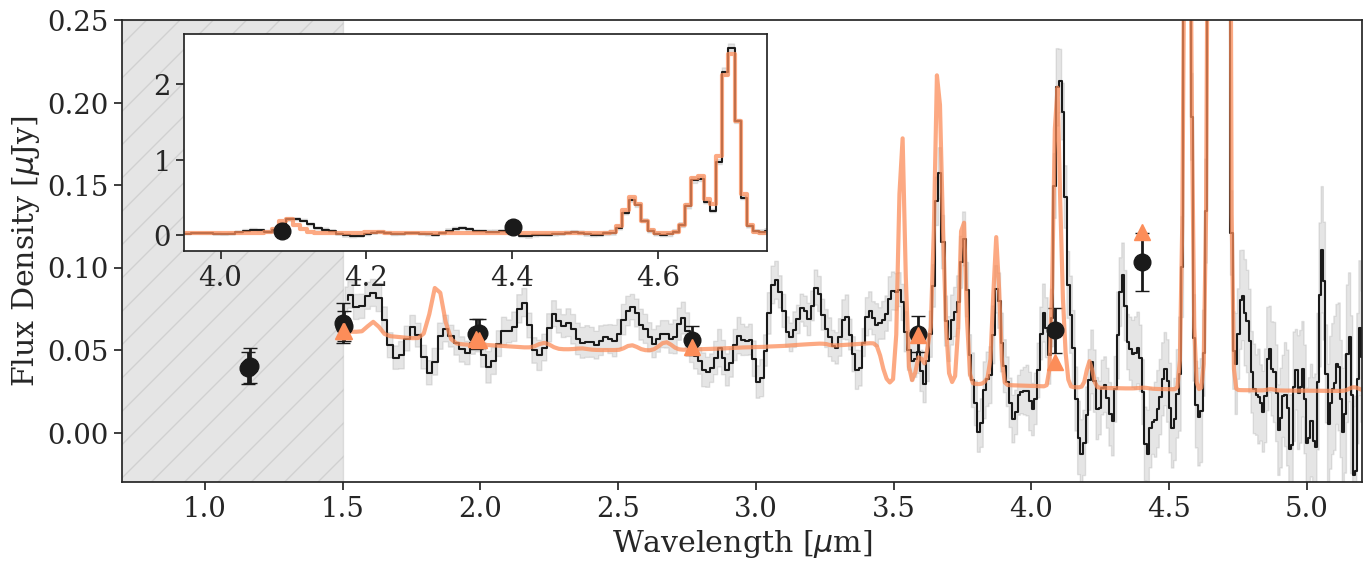}
\includegraphics[width=0.8\textwidth]{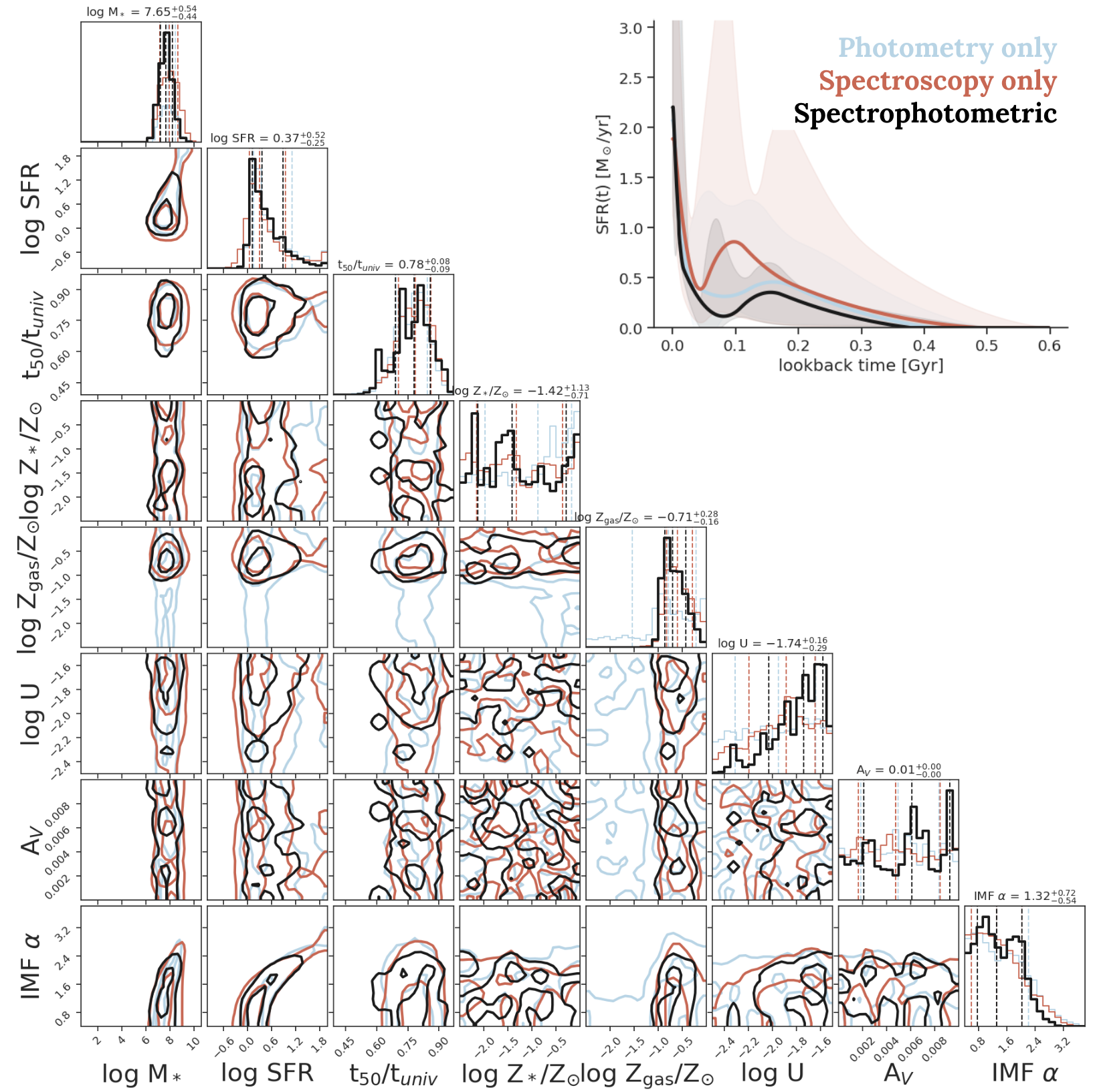}
\caption{Inferring physical properties from the spectrophotometric fits. A figure that shows the DB fit of the spectra for Slit 1 along with a corner plot showing the posteriors from the fitting.}\label{fig:spectra}
\end{figure}

\begin{figure}[h]%
\centering
\includegraphics[width=\textwidth]{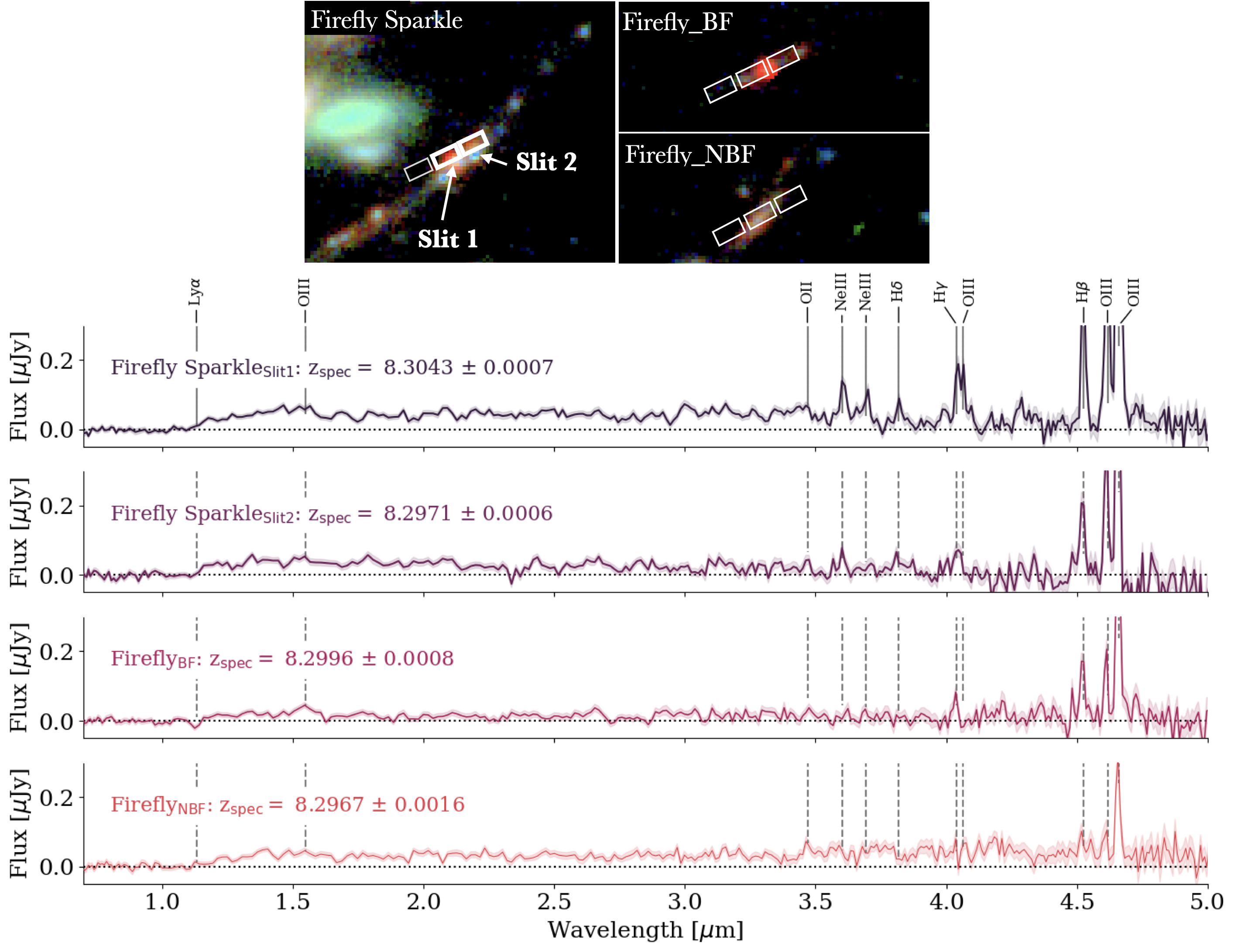}
\caption{NIRSpec Prism spectra for two slits of the \ffs, along with those of the nearby FF-BF and FF-NBF companions. Slit 1 contains contributions from FF-3, FF-4, FF-5, FF-6 in Figure 1, while Slit 2 contains contributions from FF-5, FF-6 and the diffuse arc. Strong emission lines and a Lyman break in all the spectra unambiguously determine the redshifts of all the components. There is a slight oversubtraction of background at $\lambda>$ 4$\mu$m for \ffs Slit2 and BF due to their locations close to the bar of the NIRSpec MSA shutter. Further analysis of these regions is left for follow-up observations.}\label{fig:allspectra}
\end{figure}

\subsection{Spectroscopy extraction and spectral fitting}

NIRSpec spectroscopy has been acquired for \macsj\ and spectra were obtained for the \ffsc, BF and NBF. The spectra are observed using the PRISM/CLEAR disperser and filter, through three Micro-Shutter Assembly (MSA) masks per cluster with total exposure time of 2.9 ks per MSA configuration.

The NIRSpec data was processed using the STScI {\it JWST} pipeline (software version 1.8.4 and {\tt jwst\_1030.pmap}) and the {\tt msaexp} package \citep{brammer2022msaexp}. The processing of the raw data is done in a way similar to Asada et al. (in prep). We used the standard {\it JWST} pipeline for the level 1 processing where we obtained the rate fits files from the raw data. We enabled the jump step option \texttt{expand\_large\_events} to mitigate  contamination by snowball residuals, and also used a custom persistence correction that masked out pixels which approach saturation within the following 1200 s for any readout groups. We then used \texttt{msaexp} to do level 2 processing where we performed the standard wavelength calibration, flat-fielding, path-loss correction, and photometric calibration, and obtained the 2D spectrum before background subtraction. Since central and upper shutter contains different clusters (see Fig.~\ref{fig:spec_prop} top left to find the shutter positions), we need custom background subtraction to avoid self subtraction as discussed in Asada et al. (in prep.). We thus build the background 2D spectrum by stacking and smoothing the sky spectrum in the empty pixels, and obtain the background subtracted 2D spectrum of \ffsc. Further details of the custom background subtraction method can be found in Asada et al. (in prep.). We finally extract the 1D spectrum separately in Slit 1 and Slit 2, by collapsing the 2D spectrum using an inverse-variance weighted kernel following the prescription by \citep{Horne_1986}.

\subsubsection{Spectral fitting in \ffs Slit1}\label{subsec:specfit}
The resulting 1D spectrum of \ffs in Slit 1 is shown in Fig.~\ref{fig:spec_prop}. 
The spectrum exhibits characteristic nebular continuum features including a clear discontinuity at $\lambda_{\rm obs}\sim3.5\ \mu$m corresponds to a Balmer jump, and a turnover at $\lambda_{\rm obs}<1.3\ \mu$m possibly due to the two-photon emission. These features suggest the nebular continuum should dominate over the stellar continuum in the rest-frame UV to optical spectrum within Slit 1 \citep{Cameron2023}. 
We thus model the continuum of the spectrum with nebular continuum using the photoionization code CLOUDY v23 \citep{Chatzikos2023}. To determine the dust attenuation value in the continuum model fitting, we first measure the H$\gamma$/H$\beta$ ratio by fitting Gaussian profiles. The ratio agrees well with the Case B recombination, and no significant dust attenuation is indicated. Therefore, in the continuum spectral modeling, we use pure-hydrogen gas irradiated by an ionizing source having blackbody SED without dust attenuation. We vary the effective temperature of the blackbody ($T_{\rm eff}$) and the electron temperature of the (ionized) hydrogen gas ($T_{\rm e, H^{+}}$), and search for the best-fitting model continuum by $\chi^2$ minimization. We mask out emission line regions in the fitting. The best-fit model has $\log(T_{\rm eff}/K)=5.0$ and $\log(T_{\rm e, H^{+}}/K)=4.3$, which is fully consistent with the results in \citep{Cameron2023}.
The result of continuum fitting does not change if we consider a slight dust attenuation ($A_V=0.1$ mag) in the fitting. 
As discussed in \citep{Cameron2023}, the effective temperature of $\log(T_{\rm eff}/K)=5.0$ is much hotter than typical massive type O stars, and is suggestive of this star-forming cluster having a top heavy IMF.
The IMF of this cluster is further discussed in Section \ref{subsec:sedfit}.

Having the model continuum, we subtract the underlying model continuum from the observed spectrum, and measure the spectroscopic redshift and emission line fluxes by fitting Gaussian profiles.
The best-fitting model spectrum with nebular continuum and Gaussian profiles is shown in Fig.~ \ref{fig:spec_prop} top-right by the red solid curve.
We detect emission lines of {\sc [O iii]}$\lambda\lambda4959,5007$, H$\beta$, {\sc [O iii]}$\lambda$4363, H$\gamma$, H$\delta$, [Ne {\sc iii}]$\lambda\lambda3869,3889$, and a blended line of [O {\sc iii}]$\lambda\lambda1661+1666$, but do not find significant detection of [{\sc O ii}]$\lambda3727$ and obtain the upper limit for the flux of this line.
We use these emission lines fluxes to estimate the physical parameters in the Slit 1.
We first estimate the dust attenuation based on Balmer decrements.
Both of H$\gamma$/H$\beta$ and H$\delta$/H$\beta$ ratio are consistent with theoretical predictions in Case B recombination \citep{Osterbrock2006} within the uncertainties, suggesting there is no significant dust attenuation (red squares in the left panel of Fig.~\ref{fig:spectra_prism}). This result is consistent with the initial measurement before the continuum fitting above, and supports the validity of the dust-free assumption in the nebular continuum fitting process.
Therefore, we do not correct for dust attenuation in the following measurements of emission line ratios and physical parameters in this section.

We next measure the electron temperature using temperature-sensitive emission line ratios: [O {\sc iii}]$_{4959+5007}$/[O {\sc iii}]$_{4363}$ and [O {\sc iii}]$_{5007}$/[O {\sc iii}]$_{1661+1666}$.
We assume the electron density to be $n_e=10^3\ {\rm cm^{-3}}$, and obtain consistent independent temperature measurements ($T_{\rm e,O^{++}}=4.1^{+2.7}_{-0.9}$ and $3.4^{+0.8}_{-0.4}\times10^4\ $ K, respectively, right panel in Fig.~\ref{fig:spectra_prism}).
The two line ratios do not depend on the electron density significantly, and the resulting temperatures from these ratios change only within the uncertainties even if we assume different electron  density.
Note that the [O {\sc iii}]$_{5007}$/[O {\sc iii}]$_{1661+1666}$ ratio could be largely affected by the dust attenuation, and also the [O {\sc iii}]$\lambda$1666 emission line can potentially be blended with He {\sc ii}$\lambda$1640 emission line. However, the Balmer decrement suggests the effect of dust is not significant and the absence of other He {\sc ii} lines such as He {\sc ii}$\lambda$4686 indicates the contribution by He {\sc ii}$\lambda$1640 be negligible to the blended line flux.
Given that these two possible systematic uncertainties are expected to be small and work in opposite directions, we expect the electron temperatures independently measured from the two emission line ratios to be unaffected by them.

Based on the electron temperature measurement, we obtained the oxygen abundance from [O {\sc iii}]$_{4959+5007}$/H$\beta$ and [O {\sc ii}]$_{3727}$/H$\beta$ ratios, following the prescription by \citep{Izotov2006}.
We adopt the mean value of the two electron temperature measurements ($T_{\rm e,O^{++}}=3.8\times10^4$ K), and assume the electron density to be $n_e=10^3\ {\rm cm^{-3}}$.
The total oxygen abundance is calculated from ${\rm O^{++}/H^{+}}$ and ${\rm O^{+}/H^{+}}$, and the higher ionizing state oxygen is ignored \citep{Berg2021}.
As the [O {\sc ii}]$\lambda3727$ emission line is undetected, we can only obtain an upper limit for ${\rm O^{+}/H^{+}}$, but the upper limit for the abundance of the singly ionized oxygen is negligibly small as compared to the doubly ionized oxygen.
We thus derived the total oxygen abundance from ${\rm O^{++}/H^{+}}$, yielding $12+\log({\rm O/H})=6.99^{+0.15}_{-0.33}$ ($Z_{\rm gas}/Z_\odot=0.02\pm0.01$ assuming the solar abundance to be 8.6).

We also derive the ionization parameters using the ionization sensitive emission line ratios: [O {\sc iii}]$_{5007}$/[O {\sc ii}]$_{3727}$ and [Ne {\sc iii}]$_{3869}$/[O {\sc ii}]$_{3727}$.
Following the prescription by \citep{Kewley2002,Levesque2014}, we obtain the lower limit for the ionization parameters from these two ratios.
Both ratios provide similar limit of $\log(U)>-2.1$.

All the emission line flux measurements and the derived physical parameters in \ffsc\ slit 1 are presented in Table \ref{tab:spectrum}. We also compare the diagnostic emission line ratios in \ffsc\ with those in other galaxy population in the bottom-middle panel of Fig.~\ref{fig:spec_prop}. We use the ionization-sensitive line ratio O32 ([O {\sc iii}]$_{5007}$/[O {\sc ii}]$_{3727}$) and the temperature-sensitive line ratio RO3 ([O {\sc iii}]$_{4959+5007}$/[O {\sc iii}]$_{4363}$), and compare these line ratios with other [O {\sc iii}]$\lambda$4363-detected galaxies at $z=2$-$9$ from previous JWST observations \citep{Sanders2023} and those in the local universe from SDSS observations \citep{Ahn2014}. The middle panel of Fig.~\ref{fig:spectra_prism} presents a similar comparison, but uses another ionization-sensitive line ratio Ne3O2 ([Ne {\sc iii}]$_{3869}$/[O {\sc ii}]$_{3727}$) instead of O32.

\begin{table}
\begin{tabular}{cc|cc}
\hline\hline
 Line & Flux & Properties & values \\
\hline
\lbrack O {\sc iii}\rbrack$\lambda$5007 & 645.3 $\pm$ 10.1 & $T_{\rm e,O^{++}}$ (from [O {\sc iii}]4363) & $4.1^{+2.7}_{-0.9}\times10^4$ K \rule{0pt}{1.2em} \\
\lbrack O {\sc iii}\rbrack$\lambda$4959 & 201.7 $\pm$ 9.0 & $T_{\rm e,O^{++}}$ (from [O {\sc iii}]1666) & $3.4^{+0.8}_{-0.4}\times10^4$ K \rule{0pt}{1.2em} \\
H$\beta$ & 124.9 $\pm$ 7.5 & $12+\log({\rm O^{++}/H^{+}})$ &  $6.99^{+0.15}_{-0.33}$ \rule{0pt}{1.2em} \\
\lbrack O {\sc iii}\rbrack$\lambda$4363 & 47.5 $\pm$ 10.2 & $12+\log({\rm O^{+}/H^{+}})$ & $<5.77$ \rule{0pt}{1.2em} \\
H$\gamma$ & 60.5 $\pm$ 12.4 & $12+\log({\rm O/H})$ &  $6.99^{+0.15}_{-0.33}$ \rule{0pt}{1.2em} \\
H$\delta$ & 25.7 $\pm$ 6.9 & $\log(U_{\rm ion, O32})$ &  $>-2.05$ \rule{0pt}{1.2em} \\
\lbrack Ne {\sc iii}\rbrack$\lambda$3968 & 43.8 $\pm$ 7.9 & $\log(U_{\rm ion, Ne3O2})$ & $>-2.10$ \rule{0pt}{1.2em} \\
\lbrack Ne {\sc iii}\rbrack$\lambda$3869 & 76.9 $\pm$ 8.5 &  &  \rule{0pt}{1.2em} \\
\lbrack O {\sc ii}\rbrack$\lambda$3727 & $<$27.8 &  &   \rule{0pt}{1.2em} \\
\lbrack O {\sc iii}\rbrack$\lambda\lambda$1661,1666 & 217.2 $\pm$ 55.6 &  &  \rule{0pt}{1.2em} \\
\hline\hline
\end{tabular}
\caption{Emission line flux measurements and the physical properties from the spectrum of \ffsc\ Slit 1. Fluxes are in units of $10^{-20}\ {\rm erg\ s^{-1}\ cm^{-2}}$.}
\label{tab:spectrum}
\end{table}

\subsubsection{Spectral fitting in \ffsc\ Slit 2}
On the contrary to Slit 1, the extracted 1D spectrum in \ffsc\ Slit 2 does not show nebular continuum features, and the blue continuum is rather smooth with a sharp drop-out due to the Lyman break at $\lambda_{\rm obs}\sim1.1\ \mu$m. We thus derive the emission line fluxes from the Slit 2 spectrum by fitting Gaussian profiles with the continuum being modeled by a constant offset around each emission line.
We detect [O {\sc iii}]$\lambda\lambda$4959,5007, H$\beta$, H$\gamma$, H$\delta$, [Ne {\sc iii}]$\lambda$3869, and [O {\sc ii}]$\lambda$3727 emission lines in the Slit 2 spectrum, but do not detect [O {\sc iii}]$\lambda$4363.

We then derive the physical properties in the same way as done for \ffsc\ Slit 1 spectrum.
We measure the dust attenuation from Balmer decrement, H$\gamma$/H$\beta$ and H$\delta$/H$\beta$, and find both line ratios agree well with the predicted ratios under Case B recombination (blue squares in Fig.~\ref{fig:spectra_prism} left). This suggests the dust attenuation is negligible in the Slit 2 spectrum as well, and we do not make a dust attenuation correction.

Since we do not detect the temperature-sensitive emission lines of [O {\sc iii}]$\lambda$1666 or [O {\sc iii}]$\lambda$4363 in the Slit 2 spectrum, we cannot measure the electron temperature and the metallicity from direct-temperature method. We thus only obtain the upper limit for the electron temperature ($T_{\rm e, O^{++}}$) from the non-detection of [O{\sc iii}]$\lambda$4363. The electron temperature in \ffsc\ Slit 2 is revealed to be $T_{\rm e, O^{++}} <1.8\times10^4$ K (1-sigma) or $<4.5\times10^4$ K (3-sigma).
To visualize the difference of physical properties in the Slit 1 and Slit 2, we show the diagnostic emission line ratios of \ffsc\ Slit 2 in Fig.~\ref{fig:spec_prop} bottom-middle and Fig.~\ref{fig:spectra_prism} middle panels as well.

\subsection{SED fitting analysis}\label{subsec:sedfit}

\begin{figure*}[t]
\begin{center}
\includegraphics[width=\textwidth]{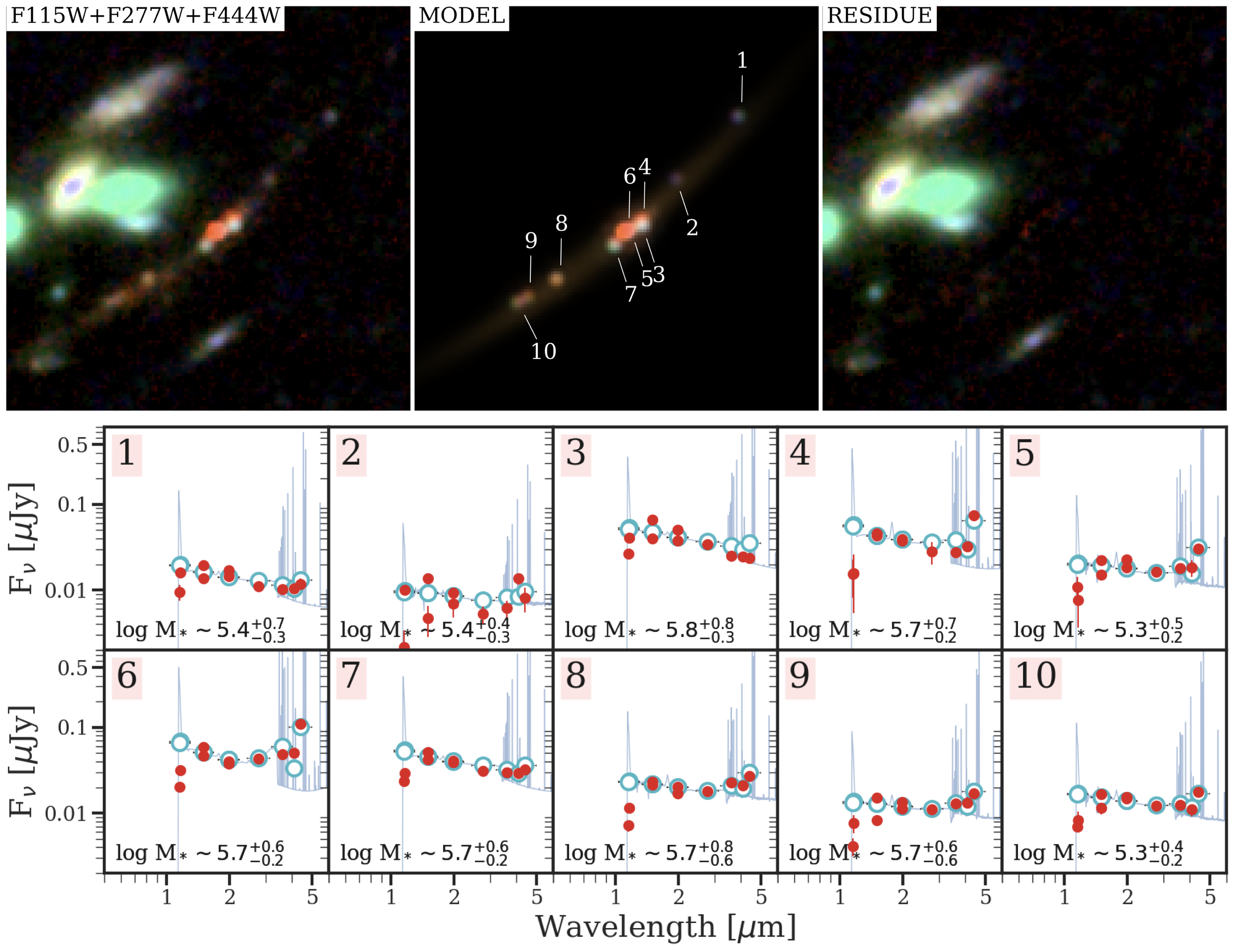}
\end{center}
\caption{\textbf{Top:} The multiwavelength 11-component model for the resolved structure in the \ffs consisting of 10 clusters and the diffuse arc shown in Figure \ref{fig:allgalfit}. The three panels show the observed image (left), the GALFIT model (middle) and the residuals (right) in composite F115W + F277W + F444W images. \textbf{Bottom:} Photometry for the 10 clusters are shown along with fits using \db$~$ and estimated stellar masses.}
\label{fig:phot_sed}
\end{figure*}

SEDs derived from our photometry were analysed using a slightly modified version of the {\db} method \citep{iyer17, iyer19} to determine nonparametric star formation histories (SFHs), masses, and ages for our sources in {\obj}. We adopt the Calzetti attenuation law \citep{calzetti01} and a Kroupa IMF \citep{Kroupa2001} with a flat prior for the high-mass slope $\alpha \in [0.]$. We fix the redshift to that found from the NIRSpec Prism Spectroscopy by the {\Oiiia} line at \zffs. All other parameters are left free. The primary advantage of using {\db} with nonparametric SFHs is that they allow us to account for flexible stellar populations, which is important at these redshifts \citep{tacchella23} since star formation is expected to be stochastic and may be underestimated if fit using traditional parametric assumptions \citep{carnall19b, lower20}. 

We perform our fitting in two stages - we initially perform a joint spectrophotometric fit to the NIRSpec Prism spectrum along with the HST + NIRISS + NIRCam photometry in the slits where both exist (seen in Figure \ref{fig:spectra}. We correct for slit loss considering two factors - the amount of light lost due to the changing PSF as a function of wavelength, and an overall multiplicative correction to match the spectrum against the photometric measurements. We modify the default {\db} method in this stage to additionally fit for the slope at the massive end of the IMF, the gas phase metallicity and the ionization parameter, using the relevant parameters from FSPS (\verb|'imf3', 'gas_logz', 'gas_logu'|).  Doing so allows us to significantly constrain priors on star formation rate, IMF, dust, ionization parameter and metallicity that we then use to fit the photometry. We find that the fits are consistent with negligible dust attenuation, consistent with our estimates from measuring the Balmer decrement. 
We also find that our fits significantly rule out the part of parameter space consistent with the canonical Chabrier-like or Kroupa-like IMF (with the high-mass slope $\approx 2.3$) in favour of more top-heavy slopes of $\approx 1.3^{+0.7}_{-0.5}$ for Slit 1, which contains portions of clusters 3,4,5,6. We find weaker constraints from the spectrum for Slit 2, which still skews toward top-heaviness albeit with large uncertainties $\approx 1.7^{+0.9}_{-0.7}$. Lastly, we find estimates of both stellar and gas-phase metallicities to be sub-solar, consistent with estimates from the line ratios. 


Using our photometry (Table \ref{tab:phottab}) we now determine the stellar properties of each individual component by running a second set of fits, which now use information from the spectra as priors to constrain a subset of the parameters - (i) we set dust to 0 consistent with the fits and Balmer decrement, (ii) we set a prior on the SFR based on the spectral posteriors (although changing this does not significantly affect our results), (iii), we set the ionisation parameter to the default value of $log U = -2$ that the spectrophotometric fits are consistent with, and (iv) we set the IMF high-mass slope to $\alpha \sim 1.0$ consistent with the spectrophotometric fits. We then fit for the stellar masses, SFR, non-parametric star formation histories, and metallicities of the individual star clusters. Both photometry and corresponding fits to the SED fit are shown in Fig.~\ref{fig:phot_sed}.
All ten components have intrinsic (corrected for magnification) stellar masses of $sim$10$^5$ M$_{\odot}$ and specific star formation rates (sSFR) of 10$^-7$ year$^{-1}$ . While the errorbars are large, the distinct colors of the clusters hint at different star formation histories. While the smooth component contains a large fraction of the stellar mass, the bulk ($\sim 57\%$) lies in the clusters. Table \ref{tab:physprop} lists the physical properties of the individual components as well as the full \ffs, BF and NBF galaxies.  



\subsection{Lens modelling}

We use \texttt{Lenstool} \cite{jullo07} to build a strong lensing model of the \macs\ cluster, to be fully presented in Desprez et al. (in prep.). This model is constrained with the three multiple image systems that were leveraged in \cite{hoag17}, for which we provide additional information obtained from the CANUCS data. The first two systems are those presented in \cite{limousin10}, one at $z=2.84$ for which we account for the two clusters visible in the four images of the objects, and the second one with three images at $z=1.779$ for which we identify another cluster in the two northernmost images for improved constraints. The last system is composed of five images \cite{zitrin15} for which we provide a new spectroscopic redshift measurement of $z=1.781$ that is in agreement with photometric and geometric redshifts previously measured.   

The different mass components are parameterised as double Pseudo-Isothermal Elliptical (dPIE) profiles \cite{eliasdottir07}. The model is composed of a large cluster scale mass halo, an independent galaxy scale one centered on the brightest cluster galaxy, and small galaxy scale mass components to account for the contribution all cluster members that follow a mass-luminosity scaling relation \cite{richard10}. For all galaxies, their positions, ellipticities, and orientations have been fixed to these measured from the images. The final best model manages to reproduce the position of the input multiple images with a distance rms of 0.46''.


Magnifications are obtained by generating convergence and shear maps around the \ffs with size 20'' and resolution 10\,milli-arcsec\,pixel$^{-1}$. These maps are created for the best optimised model, as well as for 100 randomly selection iterations of the Bayesian optimisation of \texttt{Lenstool} to compute associated errors. The source plane reconstruction is made using the best model and computing the source plane positions and magnification for the different point-like clusters. We use \texttt{Lenstool} to generate a source plane image reconstruction of the diffuse light of the galaxy using a smooth PSF deconvolved model of its light profile. 


\subsection{Size and Density of Star Clusters}
\label{subsec:density_bound}

We now investigate the spatial properties of the star clusters and their survavibility in the tidal field of the galaxy. Nine out of the ten star clusters are unresolved even in our highest resolution F115W NIRCam image. FF-4 has a slightly elongated shape visually, however have best-fit major axis size (0."01) smaller than the smallest PSF, making size estimate unreliable. Hence, we use the half-width half-max of the NIRCam F115W PSF ($0"02$) to set an upper limit on size of all ten star clusters. To determine the upper limits of the sizes of unresolved sources, we use tangential eigenvalue of  magnification $1/|\lambda_{\rm t}|$, which ranges between 14-24. This results in a size upper limit between 4-7 pc. The central star clusters have the highest magnification and the smallest upper limits, while the ones near the two ends of the arc have the lowest. 

We use the de-magnified sizes and stellar masses to estimate the density of the star clusters using methodology described in \citet{gieles2011}. We approximate an upper limit on the 3D half-mass radius from our projected upper limit on half-light size ($r_h=4/3r_{50}$). The density is calculated using $\rho_h\equiv3M/(8\pi r_h^3)$, where M is the total demagnified mass of the star clusters and the diffuse arc (assuming $\mu=$25). The resulting density vs. stellar mass plot is shown in left panel of Figure \ref{fig:progen}. The are color-coded by the age/crossing time, where the crossing time is calculated using $t_{50}=\sqrt{R^3 / GM}$. Here, the half-mass radius of the galaxy $R$ is calculated from the mass map (shown in Figure \ref{fig:sSFRage}. 

The isochrones on the left panel of Figure \ref{fig:progen} are based on the balanced evolution model in \citet{gieles2011}. In the local universe, globular clusters residing to the right of the isochrones are expected to be in an expansion dominated phase, while those on the left are expected to be in an evaporation dominated phase. To calculate the isochrones for $z=8.3$, we solve for Equation B6 in \citet{gieles2011}, assuming $\zeta=2$ for the early universe, and scaling the Milky Way rotation velocity to the mass of our galaxy. This figure is showing us that most of these star clusters are expected to survive to the present day universe, and will expand and then get ripped apart to form the stellar disk and the halo of the galaxy. The only way they survive is to get kicked out to large distances, away from the dense tidal field of the galaxy.

\begin{table}
\begin{tabular}{cccccc}
\hline\hline
Name & $\mu$ ($\mu_{\rm tan}$) & r$_{50}$ (pc) & log(M$_{\star}$/M$_{\odot}$) & log(sSFR/yr$^{-1}$) & t$_{50}$ (Myr) \vspace{0.2cm} \\ \hline \\
FF-1 & $15.6^{+ 3.1}_{- 2.0}$ ($13.9^{+ 0.9}_{- 2.8}$) & $<6.8^{+ 0.4}_{- 1.3}$ & $5.4^{+ 0.7}_{- 0.3}$ & $-7.1^{+ 0.7}_{- -0.1}$ & $72.5^{+ 106.1}_{- 72.5}$ \\
FF-2 & $17.1^{+ 3.7}_{- 2.2}$ ($15.2^{+ 1.0}_{- 3.2}$) & $<6.2^{+ 0.4}_{- 1.3}$ & $5.4^{+ 0.4}_{- 0.3}$ & $-7.5^{+ 0.7}_{- -0.4}$ & $78.6^{+ 101.2}_{- 78.6}$ \\
FF-3 & $20.6^{+ 4.6}_{- 3.4}$ ($18.6^{+ 1.5}_{- 4.3}$) & $<5.1^{+ 0.4}_{- 1.2}$ & $5.8^{+ 0.8}_{- 0.3}$ & $-7.1^{+ 0.8}_{- -0.0}$ & $120.9^{+ 62.5}_{- 120.9}$ \\
FF-4 & $21.1^{+ 4.8}_{- 3.5}$ ($19.0^{+ 1.6}_{- 4.5}$) & $<4.9^{+ 0.4}_{- 1.2}$ & $5.7^{+ 0.7}_{- 0.3}$ & $-7.1^{+ 0.7}_{- 0.0}$ & $96.7^{+ 81.9}_{- 96.7}$ \\
FF-5 & $22.5^{+ 5.4}_{- 3.9}$ ($20.5^{+ 1.8}_{- 5.0}$) & $<4.6^{+ 0.4}_{- 1.1}$ & $5.3^{+ 0.5}_{- 0.3}$ & $-7.1^{+ 0.6}_{- -0.0}$ & $102.7^{+ 77.0}_{- 102.7}$ \\
FF-6 & $23.7^{+ 6.0}_{- 4.3}$ ($21.8^{+ 2.0}_{- 5.5}$) & $<4.3^{+ 0.4}_{- 1.1}$ & $5.7^{+ 0.7}_{- 0.3}$ & $-7.1^{+ 0.7}_{- 0.0}$ & $84.6^{+ 98.8}_{- 84.6}$ \\
FF-7 & $24.7^{+ 6.5}_{- 4.6}$ ($23.0^{+ 2.3}_{- 5.9}$) & $<4.1^{+ 0.4}_{- 1.1}$ & $5.7^{+ 0.7}_{- 0.3}$ & $-7.1^{+ 0.7}_{- 0.0}$ & $96.7^{+ 79.5}_{- 96.7}$ \\
FF-8 & $26.1^{+ 8.0}_{- 4.6}$ ($24.3^{+ 3.0}_{- 6.3}$) & $<3.9^{+ 0.5}_{- 1.0}$ & $5.7^{+ 0.8}_{- 0.6}$ & $-7.6^{+ 0.9}_{- -0.1}$ & $157.1^{+ 21.5}_{- 157.1}$ \\
FF-9 & $24.2^{+ 7.5}_{- 3.9}$ ($22.2^{+ 2.9}_{- 5.4}$) & $<4.2^{+ 0.6}_{- 1.0}$ & $5.7^{+ 0.7}_{- 0.6}$ & $-7.7^{+ 0.9}_{- -0.1}$ & $120.9^{+ 57.7}_{- 120.9}$ \\
FF-10 & $24.0^{+ 7.4}_{- 3.9}$ ($22.0^{+ 3.0}_{- 5.3}$) & $<4.3^{+ 0.6}_{- 1.0}$ & $5.3^{+ 0.4}_{- 0.2}$ & $-7.2^{+ 0.5}_{- -0.0}$ & $90.6^{+ 88.5}_{- 90.6}$  \vspace{0.2cm} \\
\hline \hline
Name &  & r$_{50}$ (pc) & log($\mu$M$_{\star}$/M$_{\odot}$) & log(sSFR/yr$^{-1}$) & t$_{50}$ (Myr) \vspace{0.2cm} \\ \hline \\
FF-arc & - & 500$\pm$150 & $7.8^{+ 0.7}_{- 0.2}$ & $-7.0^{+ 0.7}_{- 0.0}$ & $60.4^{+ 109.2}_{- 60.4}$ \vspace{0.2cm} \\
\hline \\
\ffs & - & 300$\pm$150 & $7.8^{+ 0.6}_{- 0.3}$ & $-7.2^{+ 0.6}_{- 0.0}$ & $114.8^{+ 59.0}_{- 114.8}$ \\
BF & - & - & $7.1^{+ 0.4}_{- 0.0}$ & $-6.8^{+ 0.5}_{- 0.1}$ & $24.2^{+ 116.7}_{- 24.2}$ \\
NBF & - & - & $7.6^{+ 0.6}_{- 0.5}$ & $-7.3^{+ 0.6}_{- -0.1}$ & $84.6^{+ 98.2}_{- 84.6}$ \\
\hline\hline
\end{tabular}
\caption{Physical properties for individual GALFIT components of the main \ffs galaxy (top), the diffuse arc (middle) and the integrated photometry of the \ffs along with the of the BF and NBF companions.}
\label{tab:physprop}
\end{table}

\subsection{Abundance matching for MW and M31 Progenitors}
\label{subsec:projmatch}

To estimate the range of stellar masses of progenitors of both MW-mass and M31-mass galaxies at higher redshift, we adopt a semi-empirical approach combining both simulations and observations. We assume an evolving co-moving number density with redshift, as determined by the abundance matching code from \cite{Behroozi:2013}, with $z=0$ number densities of $\log(n/\text{Mpc}^{3}) =-2.95$ and $\log(n/\text{Mpc}^{3}) =-3.4$  respectively for MW and M31 mass analogues.  The code calculates a past median galaxy number density at $z_2$, given an initial number density at $z_1$, via peak halo mass functions. Since the merger rate per unit halo per unit $\Delta z$ is roughly constant, the evolution of cumulative number density of progenitors of any given galaxy is a power law, with the change described by $(0.16\Delta z)$ dex.

\cite{Behroozi:2013} chose to use peak halo mass functions because the resultant median number densities are less affected by the scatter in stellar mass and luminosity. However, this scatter does affect the $1\sigma$ errors in cumulative number density. The $1\sigma$ or 68 percentile range grows with increasing redshift, but this growth is also higher for more massive galaxies. 

As the code from \cite{Behroozi:2013} does not calculate stellar masses, we obtain the stellar mass ranges of the progenitor populations of MW and M31 analogues using stellar mass functions (SMFs) from various surveys \citep{Grazian:2015,McLeod:2021,Stefanon:2021}. We take the median cumulative number densities at each $\Delta z$, to find the stellar mass associated with that number density from the corresponding SMF. In addition , the $1\sigma$ errors on the given number density for each redshift are then used to determine the $1\sigma$ errors on the stellar mass of the progenitors.

\begin{landscape}
\begin{table}
\begin{tabular}{ccccccccccc}

\hline \hline 
ID & F115W & F115WN & F150W & F150WN & F200W & F200WN & F277W & F356W & F410M & F444W \\
\hline
FF-1 & $15.6 \pm {1.6}$ & $9.3 \pm {2.1}$ & $19.2 \pm {1.1}$ & $13.6 \pm {1.0}$ & $16.8 \pm {1.5}$ & $14.4 \pm {1.8}$ & $11.1 \pm {1.4}$ & $10.3 \pm {1.0}$ & $10.5 \pm {1.3}$ & $11.8 \pm {1.9}$ \\
FF-2 & $9.8 \pm {1.5}$ & $2.1 \pm {1.2}$ & $13.6 \pm {1.1}$ & $4.6 \pm {1.8}$ & $9.1 \pm {0.9}$ & $6.8 \pm {2.1}$ & $5.2 \pm {1.2}$ & $6.1 \pm {1.4}$ & $13.6 \pm {2.2}$ & $8.0 \pm {2.6}$ \\
FF-3 & $39.6 \pm {5.6}$ & $26.2 \pm {3.7}$ & $65.0 \pm {1.6}$ & $39.6 \pm {1.3}$ & $50.0 \pm {1.3}$ & $37.5 \pm {1.1}$ & $34.0 \pm {4.3}$ & $24.9 \pm {1.3}$ & $24.4 \pm {2.4}$ & $23.7 \pm {3.0}$ \\
FF-4 & $15.4 \pm {10.2}$ & $15.2 \pm {7.4}$ & $42.3 \pm {2.2}$ & $45.0 \pm {0.8}$ & $37.3 \pm {1.3}$ & $39.0 \pm {1.8}$ & $27.9 \pm {8.3}$ & $27.3 \pm {2.2}$ & $32.2 \pm {3.8}$ & $74.9 \pm {4.8}$ \\
FF-5 & $7.5 \pm {4.0}$ & $10.6 \pm {3.6}$ & $14.9 \pm {1.4}$ & $21.9 \pm {2.7}$ & $22.7 \pm {1.8}$ & $18.0 \pm {2.3}$ & $16.2 \pm {1.9}$ & $18.1 \pm {2.2}$ & $18.2 \pm {4.0}$ & $30.2 \pm {4.1}$ \\
FF-6 & $31.3 \pm {3.7}$ & $19.8 \pm {3.0}$ & $58.0 \pm {1.6}$ & $46.2 \pm {3.4}$ & $37.1 \pm {1.5}$ & $39.2 \pm {1.6}$ & $42.6 \pm {1.9}$ & $48.2 \pm {3.1}$ & $49.9 \pm {4.3}$ & $109.6 \pm {4.0}$ \\
FF-7 & $28.4 \pm {1.8}$ & $23.0 \pm {1.9}$ & $50.6 \pm {0.6}$ & $42.1 \pm {1.8}$ & $39.0 \pm {1.1}$ & $40.4 \pm {1.8}$ & $30.6 \pm {1.0}$ & $29.7 \pm {1.1}$ & $29.4 \pm {2.1}$ & $32.4 \pm {2.3}$ \\
FF-8 & $11.2 \pm {1.5}$ & $7.0 \pm {1.1}$ & $23.1 \pm {1.5}$ & $21.3 \pm {1.3}$ & $20.1 \pm {1.3}$ & $16.7 \pm {1.7}$ & $18.0 \pm {0.5}$ & $22.5 \pm {0.8}$ & $21.1 \pm {1.4}$ & $26.9 \pm {1.5}$ \\
FF-9 & $7.5 \pm {1.9}$ & $4.0 \pm {1.0}$ & $15.0 \pm {1.1}$ & $8.1 \pm {0.8}$ & $13.4 \pm {0.9}$ & $11.2 \pm {1.8}$ & $11.0 \pm {0.8}$ & $12.9 \pm {0.9}$ & $13.2 \pm {2.0}$ & $17.1 \pm {1.6}$ \\
FF-10 & $8.0 \pm {2.2}$ & $6.8 \pm {0.8}$ & $16.4 \pm {1.3}$ & $11.3 \pm {1.8}$ & $14.8 \pm {1.3}$ & $15.1 \pm {1.3}$ & $12.1 \pm {1.0}$ & $12.5 \pm {1.1}$ & $10.9 \pm {1.8}$ & $17.7 \pm {2.3}$ \\
FF-arc & $94.6 \pm {20.0}$ & $168.9 \pm {7.2}$ & $290.2 \pm {16.8}$ & $355.4 \pm {4.3}$ & $329.7 \pm {8.6}$ & $331.4 \pm {13.3}$ & $259.6 \pm {8.6}$ & $308.5 \pm {8.9}$ & $190.2 \pm {8.8}$ & $372.4 \pm {5.5}$ \vspace{0.2cm}\\
\hline \hline
\end{tabular}
\caption{Photometry of individual star clusters and the diffuse arc of the \ffs. Fluxes are in units of nJy.}
\label{tab:phottab}
\end{table}
\end{landscape}

\section*{Declarations}


\begin{itemize}
\item \textbf{Acknowledgements} We thank Norm Murray for the discussions about the formation of star clusters and their dynamic state, and Erica Nelson, Greg Bryan and Rachel Somerville for helpful discussions. We thank Noopur Gosavi for coining the name Firefly and Olivier Trottier for his help with Figure 4. 
\item \textbf{Funding} This research was enabled by grant 18JWST-GTO1 from the Canadian Space Agency, and funding from the Natural Sciences and Engineering Research Council of Canada.
Support for KI was provided by NASA through the NASA Hubble Fellowship grant HST-HF2-51508 awarded by the Space Telescope Science Institute, which is operated by the Association of Universities for Research in Astronomy, Inc., for NASA, under contract NAS5-26555.
YA is supported by a Research Fellowship for Young Scientists from the Japan Society of the Promotion of Science (JSPS).
MB and GR acknowledge support from the ERC Grant FIRSTLIGHT and from the Slovenian national research agency ARRS through grants N1-0238, P1-0188 and the program HST-GO-16667, provided through a grant from the STScI under NASA contract NAS5-26555.
This research used the Canadian Advanced Network For Astronomy Research (CANFAR) operated in partnership by the Canadian Astronomy Data Centre and The Digital Research Alliance of Canada with support from the National Research Council of Canada the Canadian Space Agency, CANARIE and the Canadian Foundation for Innovation.
\item \textbf{Authors' contributions} Mowla conducted photometry and size measurement, Iyer performed the SED fitting, Asada conducted the spectral analysis, Desprez conducted the lens modeling, and Tan performed the progenitor mass analysis. The above authors also made the figures and wrote the manuscript. Brammer, Willott, Strait, were responsible for the image processing pipeline and generating image mosaics. Martis conducted the BCG subtraction, Sarrouh performed the PSF modelling. All authors made contributions to the manuscript and provided invaluable assistance in data analysis and interpretation.
\end{itemize}







\begin{appendices}






\end{appendices}


\bibliography{sn-bibliography}


\begin{thebibliography}{80}
\ifx \bisbn   \undefined \def \bisbn  #1{ISBN #1}\fi
\ifx \binits  \undefined \def \binits#1{#1}\fi
\ifx \bauthor  \undefined \def \bauthor#1{#1}\fi
\ifx \batitle  \undefined \def \batitle#1{#1}\fi
\ifx \bjtitle  \undefined \def \bjtitle#1{#1}\fi
\ifx \bvolume  \undefined \def \bvolume#1{\textbf{#1}}\fi
\ifx \byear  \undefined \def \byear#1{#1}\fi
\ifx \bissue  \undefined \def \bissue#1{#1}\fi
\ifx \bfpage  \undefined \def \bfpage#1{#1}\fi
\ifx \blpage  \undefined \def \blpage #1{#1}\fi
\ifx \burl  \undefined \def \burl#1{\textsf{#1}}\fi
\ifx \doiurl  \undefined \def \doiurl#1{\url{https://doi.org/#1}}\fi
\ifx \betal  \undefined \def \betal{\textit{et al.}}\fi
\ifx \binstitute  \undefined \def \binstitute#1{#1}\fi
\ifx \binstitutionaled  \undefined \def \binstitutionaled#1{#1}\fi
\ifx \bctitle  \undefined \def \bctitle#1{#1}\fi
\ifx \beditor  \undefined \def \beditor#1{#1}\fi
\ifx \bpublisher  \undefined \def \bpublisher#1{#1}\fi
\ifx \bbtitle  \undefined \def \bbtitle#1{#1}\fi
\ifx \bedition  \undefined \def \bedition#1{#1}\fi
\ifx \bseriesno  \undefined \def \bseriesno#1{#1}\fi
\ifx \blocation  \undefined \def \blocation#1{#1}\fi
\ifx \bsertitle  \undefined \def \bsertitle#1{#1}\fi
\ifx \bsnm \undefined \def \bsnm#1{#1}\fi
\ifx \bsuffix \undefined \def \bsuffix#1{#1}\fi
\ifx \bparticle \undefined \def \bparticle#1{#1}\fi
\ifx \barticle \undefined \def \barticle#1{#1}\fi
\bibcommenthead
\ifx \bconfdate \undefined \def \bconfdate #1{#1}\fi
\ifx \botherref \undefined \def \botherref #1{#1}\fi
\ifx \url \undefined \def \url#1{\textsf{#1}}\fi
\ifx \bchapter \undefined \def \bchapter#1{#1}\fi
\ifx \bbook \undefined \def \bbook#1{#1}\fi
\ifx \bcomment \undefined \def \bcomment#1{#1}\fi
\ifx \oauthor \undefined \def \oauthor#1{#1}\fi
\ifx \citeauthoryear \undefined \def \citeauthoryear#1{#1}\fi
\ifx \endbibitem  \undefined \def \endbibitem {}\fi
\ifx \bconflocation  \undefined \def \bconflocation#1{#1}\fi
\ifx \arxivurl  \undefined \def \arxivurl#1{\textsf{#1}}\fi
\csname PreBibitemsHook\endcsname

\bibitem[\protect\citeauthoryear{{Postman} et~al.}{2012}]{postman12}
\begin{barticle}
\bauthor{\bsnm{{Postman}}, \binits{M.}},
\bauthor{\bsnm{{Coe}}, \binits{D.}},
\bauthor{\bsnm{{Ben{\'\i}tez}}, \binits{N.}},
\bauthor{\bsnm{{Bradley}}, \binits{L.}},
\bauthor{\bsnm{{Broadhurst}}, \binits{T.}},
\bauthor{\bsnm{{Donahue}}, \binits{M.}},
\bauthor{\bsnm{{Ford}}, \binits{H.}},
\bauthor{\bsnm{{Graur}}, \binits{O.}},
\bauthor{\bsnm{{Graves}}, \binits{G.}},
\bauthor{\bsnm{{Jouvel}}, \binits{S.}},
\bauthor{\bsnm{{Koekemoer}}, \binits{A.}},
\bauthor{\bsnm{{Lemze}}, \binits{D.}},
\bauthor{\bsnm{{Medezinski}}, \binits{E.}},
\bauthor{\bsnm{{Molino}}, \binits{A.}},
\bauthor{\bsnm{{Moustakas}}, \binits{L.}},
\bauthor{\bsnm{{Ogaz}}, \binits{S.}},
\bauthor{\bsnm{{Riess}}, \binits{A.}},
\bauthor{\bsnm{{Rodney}}, \binits{S.}},
\bauthor{\bsnm{{Rosati}}, \binits{P.}},
\bauthor{\bsnm{{Umetsu}}, \binits{K.}},
\bauthor{\bsnm{{Zheng}}, \binits{W.}},
\bauthor{\bsnm{{Zitrin}}, \binits{A.}},
\bauthor{\bsnm{{Bartelmann}}, \binits{M.}},
\bauthor{\bsnm{{Bouwens}}, \binits{R.}},
\bauthor{\bsnm{{Czakon}}, \binits{N.}},
\bauthor{\bsnm{{Golwala}}, \binits{S.}},
\bauthor{\bsnm{{Host}}, \binits{O.}},
\bauthor{\bsnm{{Infante}}, \binits{L.}},
\bauthor{\bsnm{{Jha}}, \binits{S.}},
\bauthor{\bsnm{{Jimenez-Teja}}, \binits{Y.}},
\bauthor{\bsnm{{Kelson}}, \binits{D.}},
\bauthor{\bsnm{{Lahav}}, \binits{O.}},
\bauthor{\bsnm{{Lazkoz}}, \binits{R.}},
\bauthor{\bsnm{{Maoz}}, \binits{D.}},
\bauthor{\bsnm{{McCully}}, \binits{C.}},
\bauthor{\bsnm{{Melchior}}, \binits{P.}},
\bauthor{\bsnm{{Meneghetti}}, \binits{M.}},
\bauthor{\bsnm{{Merten}}, \binits{J.}},
\bauthor{\bsnm{{Moustakas}}, \binits{J.}},
\bauthor{\bsnm{{Nonino}}, \binits{M.}},
\bauthor{\bsnm{{Patel}}, \binits{B.}},
\bauthor{\bsnm{{Reg{\"o}s}}, \binits{E.}},
\bauthor{\bsnm{{Sayers}}, \binits{J.}},
\bauthor{\bsnm{{Seitz}}, \binits{S.}},
\bauthor{\bsnm{{Van der Wel}}, \binits{A.}}:
\batitle{{The Cluster Lensing and Supernova Survey with Hubble: An Overview}}.
\bjtitle{\apjs}
\bvolume{199}(\bissue{2}),
\bfpage{25}
(\byear{2012})
\doiurl{10.1088/0067-0049/199/2/25}
{\href{https://arxiv.org/abs/1106.3328}{{arXiv:1106.3328}}}
{[astro-ph.CO]}
\end{barticle}
\endbibitem

\bibitem[\protect\citeauthoryear{{Hoag} et~al.}{2017}]{hoag17}
\begin{barticle}
\bauthor{\bsnm{{Hoag}}, \binits{A.}},
\bauthor{\bsnm{{Brada{\v{c}}}}, \binits{M.}},
\bauthor{\bsnm{{Trenti}}, \binits{M.}},
\bauthor{\bsnm{{Treu}}, \binits{T.}},
\bauthor{\bsnm{{Schmidt}}, \binits{K.B.}},
\bauthor{\bsnm{{Huang}}, \binits{K.-H.}},
\bauthor{\bsnm{{Lemaux}}, \binits{B.C.}},
\bauthor{\bsnm{{He}}, \binits{J.}},
\bauthor{\bsnm{{Bernard}}, \binits{S.R.}},
\bauthor{\bsnm{{Abramson}}, \binits{L.E.}},
\bauthor{\bsnm{{Mason}}, \binits{C.A.}},
\bauthor{\bsnm{{Morishita}}, \binits{T.}},
\bauthor{\bsnm{{Pentericci}}, \binits{L.}},
\bauthor{\bsnm{{Schrabback}}, \binits{T.}}:
\batitle{{Spectroscopic confirmation of an ultra-faint galaxy at the epoch of reionization}}.
\bjtitle{Nature Astronomy}
\bvolume{1},
\bfpage{0091}
(\byear{2017})
\doiurl{10.1038/s41550-017-0091}
{\href{https://arxiv.org/abs/1704.02970}{{arXiv:1704.02970}}}
{[astro-ph.GA]}
\end{barticle}
\endbibitem

\bibitem[\protect\citeauthoryear{{Willott} et~al.}{2022}]{willott22}
\begin{barticle}
\bauthor{\bsnm{{Willott}}, \binits{C.J.}},
\bauthor{\bsnm{{Doyon}}, \binits{R.}},
\bauthor{\bsnm{{Albert}}, \binits{L.}},
\bauthor{\bsnm{{Brammer}}, \binits{G.B.}},
\bauthor{\bsnm{{Dixon}}, \binits{W.V.}},
\bauthor{\bsnm{{Muzic}}, \binits{K.}},
\bauthor{\bsnm{{Ravindranath}}, \binits{S.}},
\bauthor{\bsnm{{Scholz}}, \binits{A.}},
\bauthor{\bsnm{{Abraham}}, \binits{R.}},
\bauthor{\bsnm{{Artigau}}, \binits{{\'E}.}},
\bauthor{\bsnm{{Brada{\v{c}}}}, \binits{M.}},
\bauthor{\bsnm{{Goudfrooij}}, \binits{P.}},
\bauthor{\bsnm{{Hutchings}}, \binits{J.B.}},
\bauthor{\bsnm{{Iyer}}, \binits{K.G.}},
\bauthor{\bsnm{{Jayawardhana}}, \binits{R.}},
\bauthor{\bsnm{{LaMassa}}, \binits{S.}},
\bauthor{\bsnm{{Martis}}, \binits{N.}},
\bauthor{\bsnm{{Meyer}}, \binits{M.R.}},
\bauthor{\bsnm{{Morishita}}, \binits{T.}},
\bauthor{\bsnm{{Mowla}}, \binits{L.}},
\bauthor{\bsnm{{Muzzin}}, \binits{A.}},
\bauthor{\bsnm{{Noirot}}, \binits{G.}},
\bauthor{\bsnm{{Pacifici}}, \binits{C.}},
\bauthor{\bsnm{{Rowlands}}, \binits{N.}},
\bauthor{\bsnm{{Sarrouh}}, \binits{G.}},
\bauthor{\bsnm{{Sawicki}}, \binits{M.}},
\bauthor{\bsnm{{Taylor}}, \binits{J.M.}},
\bauthor{\bsnm{{Volk}}, \binits{K.}},
\bauthor{\bsnm{{Zabl}}, \binits{J.}}:
\batitle{{The Near-infrared Imager and Slitless Spectrograph for the James Webb Space Telescope. II. Wide Field Slitless Spectroscopy}}.
\bjtitle{\pasp}
\bvolume{134}(\bissue{1032}),
\bfpage{025002}
(\byear{2022})
\doiurl{10.1088/1538-3873/ac5158}
{\href{https://arxiv.org/abs/2202.01714}{{arXiv:2202.01714}}}
{[astro-ph.IM]}
\end{barticle}
\endbibitem

\bibitem[\protect\citeauthoryear{Kneib et~al.}{2004}]{Kneib:2004fx}
\begin{barticle}
\bauthor{\bsnm{Kneib}, \binits{J.-P.}},
\bauthor{\bsnm{Ellis}, \binits{R.S.}},
\bauthor{\bsnm{Santos}, \binits{M.R.}},
\bauthor{\bsnm{Richard}, \binits{J.}}:
\batitle{{A Probable z$\sim$7 Galaxy Strongly Lensed by the Rich Cluster A2218: Exploring the Dark Ages}}.
\bjtitle{The Astrophysical Journal}
\bvolume{607}(\bissue{2}),
\bfpage{697}--\blpage{703}
(\byear{2004})
\end{barticle}
\endbibitem

\bibitem[\protect\citeauthoryear{{Jullo} et~al.}{2007}]{jullo07}
\begin{barticle}
\bauthor{\bsnm{{Jullo}}, \binits{E.}},
\bauthor{\bsnm{{Kneib}}, \binits{J.-P.}},
\bauthor{\bsnm{{Limousin}}, \binits{M.}},
\bauthor{\bsnm{{El{\'\i}asd{\'o}ttir}}, \binits{{\'A}.}},
\bauthor{\bsnm{{Marshall}}, \binits{P.J.}},
\bauthor{\bsnm{{Verdugo}}, \binits{T.}}:
\batitle{{A Bayesian approach to strong lensing modelling of galaxy clusters}}.
\bjtitle{New Journal of Physics}
\bvolume{9}(\bissue{12}),
\bfpage{447}
(\byear{2007})
\doiurl{10.1088/1367-2630/9/12/447}
{\href{https://arxiv.org/abs/0706.0048}{{arXiv:0706.0048}}}
{[astro-ph]}
\end{barticle}
\endbibitem

\bibitem[\protect\citeauthoryear{{Labb{\'e}} et~al.}{2023}]{labbe23}
\begin{barticle}
\bauthor{\bsnm{{Labb{\'e}}}, \binits{I.}},
\bauthor{\bsnm{{van Dokkum}}, \binits{P.}},
\bauthor{\bsnm{{Nelson}}, \binits{E.}},
\bauthor{\bsnm{{Bezanson}}, \binits{R.}},
\bauthor{\bsnm{{Suess}}, \binits{K.A.}},
\bauthor{\bsnm{{Leja}}, \binits{J.}},
\bauthor{\bsnm{{Brammer}}, \binits{G.}},
\bauthor{\bsnm{{Whitaker}}, \binits{K.}},
\bauthor{\bsnm{{Mathews}}, \binits{E.}},
\bauthor{\bsnm{{Stefanon}}, \binits{M.}},
\bauthor{\bsnm{{Wang}}, \binits{B.}}:
\batitle{{A population of red candidate massive galaxies 600 Myr after the Big Bang}}.
\bjtitle{\nat}
\bvolume{616}(\bissue{7956}),
\bfpage{266}--\blpage{269}
(\byear{2023})
\doiurl{10.1038/s41586-023-05786-2}
{\href{https://arxiv.org/abs/2207.12446}{{arXiv:2207.12446}}}
{[astro-ph.GA]}
\end{barticle}
\endbibitem

\bibitem[\protect\citeauthoryear{{Desprez} et~al.}{2023}]{desprez23}
\begin{botherref}
\oauthor{\bsnm{{Desprez}}, \binits{G.}},
\oauthor{\bsnm{{Martis}}, \binits{N.S.}},
\oauthor{\bsnm{{Asada}}, \binits{Y.}},
\oauthor{\bsnm{{Sawicki}}, \binits{M.}},
\oauthor{\bsnm{{Willott}}, \binits{C.J.}},
\oauthor{\bsnm{{Muzzin}}, \binits{A.}},
\oauthor{\bsnm{{Abraham}}, \binits{R.G.}},
\oauthor{\bsnm{{Brada{\v{c}}}}, \binits{M.}},
\oauthor{\bsnm{{Brammer}}, \binits{G.}},
\oauthor{\bsnm{{Estrada-Carpenter}}, \binits{V.}},
\oauthor{\bsnm{{Iyer}}, \binits{K.G.}},
\oauthor{\bsnm{{Matharu}}, \binits{J.}},
\oauthor{\bsnm{{Mowla}}, \binits{L.}},
\oauthor{\bsnm{{Noirot}}, \binits{G.}},
\oauthor{\bsnm{{Sarrouh}}, \binits{G.T.E.}},
\oauthor{\bsnm{{Strait}}, \binits{V.}},
\oauthor{\bsnm{{Gledhill}}, \binits{R.}},
\oauthor{\bsnm{{Rihtar{\v{s}}i{\v{c}}}}, \binits{G.}}:
{$\Lambda$CDM not dead yet: massive high-z Balmer break galaxies are less common than previously reported}.
arXiv e-prints,
2310--03063
(2023)
\doiurl{10.48550/arXiv.2310.03063}
{\href{https://arxiv.org/abs/2310.03063}{{arXiv:2310.03063}}}
{[astro-ph.GA]}
\end{botherref}
\endbibitem

\bibitem[\protect\citeauthoryear{{Peng} et~al.}{2010}]{Peng2010}
\begin{barticle}
\bauthor{\bsnm{{Peng}}, \binits{C.Y.}},
\bauthor{\bsnm{{Ho}}, \binits{L.C.}},
\bauthor{\bsnm{{Impey}}, \binits{C.D.}},
\bauthor{\bsnm{{Rix}}, \binits{H.-W.}}:
\batitle{{Detailed Decomposition of Galaxy Images. II. Beyond Axisymmetric Models}}.
\bjtitle{\aj}
\bvolume{139}(\bissue{6}),
\bfpage{2097}--\blpage{2129}
(\byear{2010})
\doiurl{10.1088/0004-6256/139/6/2097}
{\href{https://arxiv.org/abs/0912.0731}{{arXiv:0912.0731}}}
{[astro-ph.CO]}
\end{barticle}
\endbibitem

\bibitem[\protect\citeauthoryear{{Sorba} and {Sawicki}}{2018}]{sorba18}
\begin{barticle}
\bauthor{\bsnm{{Sorba}}, \binits{R.}},
\bauthor{\bsnm{{Sawicki}}, \binits{M.}}:
\batitle{{Spatially unresolved SED fitting can underestimate galaxy masses: a solution to the missing mass problem}}.
\bjtitle{\mnras}
\bvolume{476},
\bfpage{1532}--\blpage{1547}
(\byear{2018})
\doiurl{10.1093/mnras/sty186}
{\href{https://arxiv.org/abs/1801.07368}{{arXiv:1801.07368}}}
\end{barticle}
\endbibitem

\bibitem[\protect\citeauthoryear{{Gim{\'e}nez-Arteaga} et~al.}{2023}]{gimenez-arteaga23}
\begin{barticle}
\bauthor{\bsnm{{Gim{\'e}nez-Arteaga}}, \binits{C.}},
\bauthor{\bsnm{{Oesch}}, \binits{P.A.}},
\bauthor{\bsnm{{Brammer}}, \binits{G.B.}},
\bauthor{\bsnm{{Valentino}}, \binits{F.}},
\bauthor{\bsnm{{Mason}}, \binits{C.A.}},
\bauthor{\bsnm{{Weibel}}, \binits{A.}},
\bauthor{\bsnm{{Barrufet}}, \binits{L.}},
\bauthor{\bsnm{{Fujimoto}}, \binits{S.}},
\bauthor{\bsnm{{Heintz}}, \binits{K.E.}},
\bauthor{\bsnm{{Nelson}}, \binits{E.J.}},
\bauthor{\bsnm{{Strait}}, \binits{V.B.}},
\bauthor{\bsnm{{Suess}}, \binits{K.A.}},
\bauthor{\bsnm{{Gibson}}, \binits{J.}}:
\batitle{{Spatially Resolved Properties of Galaxies at 5 < z < 9 in the SMACS 0723 JWST ERO Field}}.
\bjtitle{\apj}
\bvolume{948}(\bissue{2}),
\bfpage{126}
(\byear{2023})
\doiurl{10.3847/1538-4357/acc5ea}
{\href{https://arxiv.org/abs/2212.08670}{{arXiv:2212.08670}}}
{[astro-ph.GA]}
\end{barticle}
\endbibitem

\bibitem[\protect\citeauthoryear{{Narayanan} et~al.}{2023}]{narayanan23}
\begin{botherref}
\oauthor{\bsnm{{Narayanan}}, \binits{D.}},
\oauthor{\bsnm{{Lower}}, \binits{S.}},
\oauthor{\bsnm{{Torrey}}, \binits{P.}},
\oauthor{\bsnm{{Brammer}}, \binits{G.}},
\oauthor{\bsnm{{Cui}}, \binits{W.}},
\oauthor{\bsnm{{Dave}}, \binits{R.}},
\oauthor{\bsnm{{Iyer}}, \binits{K.}},
\oauthor{\bsnm{{Li}}, \binits{Q.}},
\oauthor{\bsnm{{Lovell}}, \binits{C.}},
\oauthor{\bsnm{{Sales}}, \binits{L.}},
\oauthor{\bsnm{{Stark}}, \binits{D.P.}},
\oauthor{\bsnm{{Marinacci}}, \binits{F.}},
\oauthor{\bsnm{{Vogelsberger}}, \binits{M.}}:
{Outshining by Recent Star Formation Prevents the Accurate Measurement of High-z Galaxy Stellar Masses}.
arXiv e-prints,
2306--10118
(2023)
\doiurl{10.48550/arXiv.2306.10118}
{\href{https://arxiv.org/abs/2306.10118}{{arXiv:2306.10118}}}
{[astro-ph.GA]}
\end{botherref}
\endbibitem

\bibitem[\protect\citeauthoryear{{Pillepich} et~al.}{2023}]{pillepich23}
\begin{botherref}
\oauthor{\bsnm{{Pillepich}}, \binits{A.}},
\oauthor{\bsnm{{Sotillo-Ramos}}, \binits{D.}},
\oauthor{\bsnm{{Ramesh}}, \binits{R.}},
\oauthor{\bsnm{{Nelson}}, \binits{D.}},
\oauthor{\bsnm{{Engler}}, \binits{C.}},
\oauthor{\bsnm{{Rodriguez-Gomez}}, \binits{V.}},
\oauthor{\bsnm{{Fournier}}, \binits{M.}},
\oauthor{\bsnm{{Donnari}}, \binits{M.}},
\oauthor{\bsnm{{Springel}}, \binits{V.}},
\oauthor{\bsnm{{Hernquist}}, \binits{L.}}:
{Milky Way and Andromeda analogs from the TNG50 simulation}.
arXiv e-prints,
2303--16217
(2023)
\doiurl{10.48550/arXiv.2303.16217}
{\href{https://arxiv.org/abs/2303.16217}{{arXiv:2303.16217}}}
{[astro-ph.GA]}
\end{botherref}
\endbibitem

\bibitem[\protect\citeauthoryear{{Behroozi} et~al.}{2013}]{Behroozi:2013}
\begin{barticle}
\bauthor{\bsnm{{Behroozi}}, \binits{P.S.}},
\bauthor{\bsnm{{Marchesini}}, \binits{D.}},
\bauthor{\bsnm{{Wechsler}}, \binits{R.H.}},
\bauthor{\bsnm{{Muzzin}}, \binits{A.}},
\bauthor{\bsnm{{Papovich}}, \binits{C.}},
\bauthor{\bsnm{{Stefanon}}, \binits{M.}}:
\batitle{{Using Cumulative Number Densities to Compare Galaxies across Cosmic Time}}.
\bjtitle{\apjl}
\bvolume{777}(\bissue{1}),
\bfpage{10}
(\byear{2013})
\doiurl{10.1088/2041-8205/777/1/L10}
{\href{https://arxiv.org/abs/1308.3232}{{arXiv:1308.3232}}}
{[astro-ph.CO]}
\end{barticle}
\endbibitem

\bibitem[\protect\citeauthoryear{{Grazian} et~al.}{2015}]{grazian15}
\begin{barticle}
\bauthor{\bsnm{{Grazian}}, \binits{A.}},
\bauthor{\bsnm{{Fontana}}, \binits{A.}},
\bauthor{\bsnm{{Santini}}, \binits{P.}},
\bauthor{\bsnm{{Dunlop}}, \binits{J.S.}},
\bauthor{\bsnm{{Ferguson}}, \binits{H.C.}},
\bauthor{\bsnm{{Castellano}}, \binits{M.}},
\bauthor{\bsnm{{Amorin}}, \binits{R.}},
\bauthor{\bsnm{{Ashby}}, \binits{M.L.N.}},
\bauthor{\bsnm{{Barro}}, \binits{G.}},
\bauthor{\bsnm{{Behroozi}}, \binits{P.}},
\bauthor{\bsnm{{Boutsia}}, \binits{K.}},
\bauthor{\bsnm{{Caputi}}, \binits{K.I.}},
\bauthor{\bsnm{{Chary}}, \binits{R.R.}},
\bauthor{\bsnm{{Dekel}}, \binits{A.}},
\bauthor{\bsnm{{Dickinson}}, \binits{M.E.}},
\bauthor{\bsnm{{Faber}}, \binits{S.M.}},
\bauthor{\bsnm{{Fazio}}, \binits{G.G.}},
\bauthor{\bsnm{{Finkelstein}}, \binits{S.L.}},
\bauthor{\bsnm{{Galametz}}, \binits{A.}},
\bauthor{\bsnm{{Giallongo}}, \binits{E.}},
\bauthor{\bsnm{{Giavalisco}}, \binits{M.}},
\bauthor{\bsnm{{Grogin}}, \binits{N.A.}},
\bauthor{\bsnm{{Guo}}, \binits{Y.}},
\bauthor{\bsnm{{Kocevski}}, \binits{D.}},
\bauthor{\bsnm{{Koekemoer}}, \binits{A.M.}},
\bauthor{\bsnm{{Koo}}, \binits{D.C.}},
\bauthor{\bsnm{{Lee}}, \binits{K.-S.}},
\bauthor{\bsnm{{Lu}}, \binits{Y.}},
\bauthor{\bsnm{{Merlin}}, \binits{E.}},
\bauthor{\bsnm{{Mobasher}}, \binits{B.}},
\bauthor{\bsnm{{Nonino}}, \binits{M.}},
\bauthor{\bsnm{{Papovich}}, \binits{C.}},
\bauthor{\bsnm{{Paris}}, \binits{D.}},
\bauthor{\bsnm{{Pentericci}}, \binits{L.}},
\bauthor{\bsnm{{Reddy}}, \binits{N.}},
\bauthor{\bsnm{{Renzini}}, \binits{A.}},
\bauthor{\bsnm{{Salmon}}, \binits{B.}},
\bauthor{\bsnm{{Salvato}}, \binits{M.}},
\bauthor{\bsnm{{Sommariva}}, \binits{V.}},
\bauthor{\bsnm{{Song}}, \binits{M.}},
\bauthor{\bsnm{{Vanzella}}, \binits{E.}}:
\batitle{{The galaxy stellar mass function at 3.5 {\ensuremath{\leq}}z {\ensuremath{\leq}} 7.5 in the CANDELS/UDS, GOODS-South, and HUDF fields}}.
\bjtitle{\aap}
\bvolume{575},
\bfpage{96}
(\byear{2015})
\doiurl{10.1051/0004-6361/201424750}
{\href{https://arxiv.org/abs/1412.0532}{{arXiv:1412.0532}}}
{[astro-ph.GA]}
\end{barticle}
\endbibitem

\bibitem[\protect\citeauthoryear{{McLeod} et~al.}{2021}]{McLeod:2021}
\begin{barticle}
\bauthor{\bsnm{{McLeod}}, \binits{D.J.}},
\bauthor{\bsnm{{McLure}}, \binits{R.J.}},
\bauthor{\bsnm{{Dunlop}}, \binits{J.S.}},
\bauthor{\bsnm{{Cullen}}, \binits{F.}},
\bauthor{\bsnm{{Carnall}}, \binits{A.C.}},
\bauthor{\bsnm{{Duncan}}, \binits{K.}}:
\batitle{{The evolution of the galaxy stellar-mass function over the last 12 billion years from a combination of ground-based and HST surveys}}.
\bjtitle{\mnras}
\bvolume{503}(\bissue{3}),
\bfpage{4413}--\blpage{4435}
(\byear{2021})
\doiurl{10.1093/mnras/stab731}
{\href{https://arxiv.org/abs/2009.03176}{{arXiv:2009.03176}}}
{[astro-ph.GA]}
\end{barticle}
\endbibitem

\bibitem[\protect\citeauthoryear{{Stefanon} et~al.}{2019}]{stefanon19}
\begin{barticle}
\bauthor{\bsnm{{Stefanon}}, \binits{M.}},
\bauthor{\bsnm{{Labb{\'e}}}, \binits{I.}},
\bauthor{\bsnm{{Bouwens}}, \binits{R.J.}},
\bauthor{\bsnm{{Oesch}}, \binits{P.}},
\bauthor{\bsnm{{Ashby}}, \binits{M.L.N.}},
\bauthor{\bsnm{{Caputi}}, \binits{K.I.}},
\bauthor{\bsnm{{Franx}}, \binits{M.}},
\bauthor{\bsnm{{Fynbo}}, \binits{J.P.U.}},
\bauthor{\bsnm{{Illingworth}}, \binits{G.D.}},
\bauthor{\bsnm{{Le F{\`e}vre}}, \binits{O.}},
\bauthor{\bsnm{{Marchesini}}, \binits{D.}},
\bauthor{\bsnm{{McCracken}}, \binits{H.J.}},
\bauthor{\bsnm{{Milvang-Jensen}}, \binits{B.}},
\bauthor{\bsnm{{Muzzin}}, \binits{A.}},
\bauthor{\bsnm{{van Dokkum}}, \binits{P.}}:
\batitle{{The Brightest z {\ensuremath{\gtrsim}} 8 Galaxies over the COSMOS UltraVISTA Field}}.
\bjtitle{\apj}
\bvolume{883}(\bissue{1}),
\bfpage{99}
(\byear{2019})
\doiurl{10.3847/1538-4357/ab3792}
{\href{https://arxiv.org/abs/1902.10713}{{arXiv:1902.10713}}}
{[astro-ph.GA]}
\end{barticle}
\endbibitem

\bibitem[\protect\citeauthoryear{{Gieles} et~al.}{2011}]{gieles2011}
\begin{barticle}
\bauthor{\bsnm{{Gieles}}, \binits{M.}},
\bauthor{\bsnm{{Heggie}}, \binits{D.C.}},
\bauthor{\bsnm{{Zhao}}, \binits{H.}}:
\batitle{{The life cycle of star clusters in a tidal field}}.
\bjtitle{\mnras}
\bvolume{413}(\bissue{4}),
\bfpage{2509}--\blpage{2524}
(\byear{2011})
\doiurl{10.1111/j.1365-2966.2011.18320.x}
{\href{https://arxiv.org/abs/1101.1821}{{arXiv:1101.1821}}}
{[astro-ph.GA]}
\end{barticle}
\endbibitem

\bibitem[\protect\citeauthoryear{{Wilkins} et~al.}{2024}]{2024MNRAS.527.7965W}
\begin{barticle}
\bauthor{\bsnm{{Wilkins}}, \binits{S.M.}},
\bauthor{\bsnm{{Lovell}}, \binits{C.C.}},
\bauthor{\bsnm{{Irodotou}}, \binits{D.}},
\bauthor{\bsnm{{Vijayan}}, \binits{A.P.}},
\bauthor{\bsnm{{Vikaeus}}, \binits{A.}},
\bauthor{\bsnm{{Zackrisson}}, \binits{E.}},
\bauthor{\bsnm{{Caruana}}, \binits{J.}},
\bauthor{\bsnm{{Stanway}}, \binits{E.R.}},
\bauthor{\bsnm{{Conselice}}, \binits{C.J.}},
\bauthor{\bsnm{{Seeyave}}, \binits{L.T.C.}},
\bauthor{\bsnm{{Roper}}, \binits{W.J.}},
\bauthor{\bsnm{{Chworowsky}}, \binits{K.}},
\bauthor{\bsnm{{Finkelstein}}, \binits{S.L.}}:
\batitle{{First Light and Reionization Epoch Simulations (FLARES) - XIV. The Balmer/4000 {\r{A}} breaks of distant galaxies}}.
\bjtitle{\mnras}
\bvolume{527}(\bissue{3}),
\bfpage{7965}--\blpage{7973}
(\byear{2024})
\doiurl{10.1093/mnras/stad3558}
{\href{https://arxiv.org/abs/2305.18175}{{arXiv:2305.18175}}}
{[astro-ph.GA]}
\end{barticle}
\endbibitem

\bibitem[\protect\citeauthoryear{{Cameron} et~al.}{2023}]{Cameron2023}
\begin{botherref}
\oauthor{\bsnm{{Cameron}}, \binits{A.J.}},
\oauthor{\bsnm{{Katz}}, \binits{H.}},
\oauthor{\bsnm{{Witten}}, \binits{C.}},
\oauthor{\bsnm{{Saxena}}, \binits{A.}},
\oauthor{\bsnm{{Laporte}}, \binits{N.}},
\oauthor{\bsnm{{Bunker}}, \binits{A.J.}}:
{Nebular dominated galaxies in the early Universe with top-heavy stellar initial mass functions}.
arXiv e-prints,
2311--02051
(2023)
\doiurl{10.48550/arXiv.2311.02051}
{\href{https://arxiv.org/abs/2311.02051}{{arXiv:2311.02051}}}
{[astro-ph.GA]}
\end{botherref}
\endbibitem

\bibitem[\protect\citeauthoryear{{Tacchella} et~al.}{2023}]{2023ApJ...952...74T}
\begin{barticle}
\bauthor{\bsnm{{Tacchella}}, \binits{S.}},
\bauthor{\bsnm{{Eisenstein}}, \binits{D.J.}},
\bauthor{\bsnm{{Hainline}}, \binits{K.}},
\bauthor{\bsnm{{Johnson}}, \binits{B.D.}},
\bauthor{\bsnm{{Baker}}, \binits{W.M.}},
\bauthor{\bsnm{{Helton}}, \binits{J.M.}},
\bauthor{\bsnm{{Robertson}}, \binits{B.}},
\bauthor{\bsnm{{Suess}}, \binits{K.A.}},
\bauthor{\bsnm{{Chen}}, \binits{Z.}},
\bauthor{\bsnm{{Nelson}}, \binits{E.}},
\bauthor{\bsnm{{Pusk{\'a}s}}, \binits{D.}},
\bauthor{\bsnm{{Sun}}, \binits{F.}},
\bauthor{\bsnm{{Alberts}}, \binits{S.}},
\bauthor{\bsnm{{Egami}}, \binits{E.}},
\bauthor{\bsnm{{Hausen}}, \binits{R.}},
\bauthor{\bsnm{{Rieke}}, \binits{G.}},
\bauthor{\bsnm{{Rieke}}, \binits{M.}},
\bauthor{\bsnm{{Shivaei}}, \binits{I.}},
\bauthor{\bsnm{{Williams}}, \binits{C.C.}},
\bauthor{\bsnm{{Willmer}}, \binits{C.N.A.}},
\bauthor{\bsnm{{Bunker}}, \binits{A.}},
\bauthor{\bsnm{{Cameron}}, \binits{A.J.}},
\bauthor{\bsnm{{Carniani}}, \binits{S.}},
\bauthor{\bsnm{{Charlot}}, \binits{S.}},
\bauthor{\bsnm{{Curti}}, \binits{M.}},
\bauthor{\bsnm{{Curtis-Lake}}, \binits{E.}},
\bauthor{\bsnm{{Looser}}, \binits{T.J.}},
\bauthor{\bsnm{{Maiolino}}, \binits{R.}},
\bauthor{\bsnm{{Maseda}}, \binits{M.V.}},
\bauthor{\bsnm{{Rawle}}, \binits{T.}},
\bauthor{\bsnm{{Rix}}, \binits{H.-W.}},
\bauthor{\bsnm{{Smit}}, \binits{R.}},
\bauthor{\bsnm{{{\"U}bler}}, \binits{H.}},
\bauthor{\bsnm{{Willott}}, \binits{C.}},
\bauthor{\bsnm{{Witstok}}, \binits{J.}},
\bauthor{\bsnm{{Baum}}, \binits{S.}},
\bauthor{\bsnm{{Bhatawdekar}}, \binits{R.}},
\bauthor{\bsnm{{Boyett}}, \binits{K.}},
\bauthor{\bsnm{{Danhaive}}, \binits{A.L.}},
\bauthor{\bsnm{{de Graaff}}, \binits{A.}},
\bauthor{\bsnm{{Endsley}}, \binits{R.}},
\bauthor{\bsnm{{Ji}}, \binits{Z.}},
\bauthor{\bsnm{{Lyu}}, \binits{J.}},
\bauthor{\bsnm{{Sandles}}, \binits{L.}},
\bauthor{\bsnm{{Saxena}}, \binits{A.}},
\bauthor{\bsnm{{Scholtz}}, \binits{J.}},
\bauthor{\bsnm{{Topping}}, \binits{M.W.}},
\bauthor{\bsnm{{Whitler}}, \binits{L.}}:
\batitle{{JADES Imaging of GN-z11: Revealing the Morphology and Environment of a Luminous Galaxy 430 Myr after the Big Bang}}.
\bjtitle{\apj}
\bvolume{952}(\bissue{1}),
\bfpage{74}
(\byear{2023})
\doiurl{10.3847/1538-4357/acdbc6}
{\href{https://arxiv.org/abs/2302.07234}{{arXiv:2302.07234}}}
{[astro-ph.GA]}
\end{barticle}
\endbibitem

\bibitem[\protect\citeauthoryear{{Carnall} et~al.}{2023}]{2023MNRAS.518L..45C}
\begin{barticle}
\bauthor{\bsnm{{Carnall}}, \binits{A.C.}},
\bauthor{\bsnm{{Begley}}, \binits{R.}},
\bauthor{\bsnm{{McLeod}}, \binits{D.J.}},
\bauthor{\bsnm{{Hamadouche}}, \binits{M.L.}},
\bauthor{\bsnm{{Donnan}}, \binits{C.T.}},
\bauthor{\bsnm{{McLure}}, \binits{R.J.}},
\bauthor{\bsnm{{Dunlop}}, \binits{J.S.}},
\bauthor{\bsnm{{Milvang-Jensen}}, \binits{B.}},
\bauthor{\bsnm{{Bondestam}}, \binits{C.L.}},
\bauthor{\bsnm{{Cullen}}, \binits{F.}},
\bauthor{\bsnm{{Jewell}}, \binits{S.M.}},
\bauthor{\bsnm{{Pollock}}, \binits{C.L.}}:
\batitle{{A first look at the SMACS0723 JWST ERO: spectroscopic redshifts, stellar masses, and star-formation histories}}.
\bjtitle{\mnras}
\bvolume{518}(\bissue{1}),
\bfpage{45}--\blpage{50}
(\byear{2023})
\doiurl{10.1093/mnrasl/slac136}
{\href{https://arxiv.org/abs/2207.08778}{{arXiv:2207.08778}}}
{[astro-ph.GA]}
\end{barticle}
\endbibitem

\bibitem[\protect\citeauthoryear{{Umeda} et~al.}{2023}]{2023arXiv230600487U}
\begin{botherref}
\oauthor{\bsnm{{Umeda}}, \binits{H.}},
\oauthor{\bsnm{{Ouchi}}, \binits{M.}},
\oauthor{\bsnm{{Nakajima}}, \binits{K.}},
\oauthor{\bsnm{{Harikane}}, \binits{Y.}},
\oauthor{\bsnm{{Ono}}, \binits{Y.}},
\oauthor{\bsnm{{Xu}}, \binits{Y.}},
\oauthor{\bsnm{{Isobe}}, \binits{Y.}},
\oauthor{\bsnm{{Zhang}}, \binits{Y.}}:
{JWST Measurements of Neutral Hydrogen Fractions and Ionized Bubble Sizes at $z=7-12$ Obtained with Ly$\alpha$ Damping Wing Absorptions in 26 Bright Continuum Galaxies}.
arXiv e-prints,
2306--00487
(2023)
\doiurl{10.48550/arXiv.2306.00487}
{\href{https://arxiv.org/abs/2306.00487}{{arXiv:2306.00487}}}
{[astro-ph.GA]}
\end{botherref}
\endbibitem

\bibitem[\protect\citeauthoryear{{Heintz} et~al.}{2023}]{2023arXiv230600647H}
\begin{botherref}
\oauthor{\bsnm{{Heintz}}, \binits{K.E.}},
\oauthor{\bsnm{{Watson}}, \binits{D.}},
\oauthor{\bsnm{{Brammer}}, \binits{G.}},
\oauthor{\bsnm{{Vejlgaard}}, \binits{S.}},
\oauthor{\bsnm{{Hutter}}, \binits{A.}},
\oauthor{\bsnm{{Strait}}, \binits{V.B.}},
\oauthor{\bsnm{{Matthee}}, \binits{J.}},
\oauthor{\bsnm{{Oesch}}, \binits{P.A.}},
\oauthor{\bsnm{{Jakobsson}}, \binits{P.}},
\oauthor{\bsnm{{Tanvir}}, \binits{N.R.}},
\oauthor{\bsnm{{Laursen}}, \binits{P.}},
\oauthor{\bsnm{{Naidu}}, \binits{R.P.}},
\oauthor{\bsnm{{Mason}}, \binits{C.A.}},
\oauthor{\bsnm{{Killi}}, \binits{M.}},
\oauthor{\bsnm{{Jung}}, \binits{I.}},
\oauthor{\bsnm{{Hsiao}}, \binits{T.Y.-Y.}},
\oauthor{\bsnm{{Abdurro'uf}}},
\oauthor{\bsnm{{Coe}}, \binits{D.}},
\oauthor{\bsnm{{Arrabal Haro}}, \binits{P.}},
\oauthor{\bsnm{{Finkelstein}}, \binits{S.L.}},
\oauthor{\bsnm{{Toft}}, \binits{S.}}:
{Extreme damped Lyman-$\alpha$ absorption in young star-forming galaxies at $z=9-11$}.
arXiv e-prints,
2306--00647
(2023)
\doiurl{10.48550/arXiv.2306.00647}
{\href{https://arxiv.org/abs/2306.00647}{{arXiv:2306.00647}}}
{[astro-ph.GA]}
\end{botherref}
\endbibitem

\bibitem[\protect\citeauthoryear{{Adamo} et~al.}{2024}]{adamo24}
\begin{botherref}
\oauthor{\bsnm{{Adamo}}, \binits{A.}},
\oauthor{\bsnm{{Bradley}}, \binits{L.D.}},
\oauthor{\bsnm{{Vanzella}}, \binits{E.}},
\oauthor{\bsnm{{Claeyssens}}, \binits{A.}},
\oauthor{\bsnm{{Welch}}, \binits{B.}},
\oauthor{\bsnm{{Diego}}, \binits{J.M.}},
\oauthor{\bsnm{{Mahler}}, \binits{G.}},
\oauthor{\bsnm{{Oguri}}, \binits{M.}},
\oauthor{\bsnm{{Sharon}}, \binits{K.}},
\oauthor{\bsnm{{Abdurro'uf}}},
\oauthor{\bsnm{{Hsiao}}, \binits{T.Y.-Y.}},
\oauthor{\bsnm{{Messa}}, \binits{M.}},
\oauthor{\bsnm{{Zackrisson}}, \binits{E.}},
\oauthor{\bsnm{{Brammer}}, \binits{G.}},
\oauthor{\bsnm{{Coe}}, \binits{D.}},
\oauthor{\bsnm{{Kokorev}}, \binits{V.}},
\oauthor{\bsnm{{Ricotti}}, \binits{M.}},
\oauthor{\bsnm{{Zitrin}}, \binits{A.}},
\oauthor{\bsnm{{Fujimoto}}, \binits{S.}},
\oauthor{\bsnm{{Inoue}}, \binits{A.K.}},
\oauthor{\bsnm{{Resseguier}}, \binits{T.}},
\oauthor{\bsnm{{Rigby}}, \binits{J.R.}},
\oauthor{\bsnm{{Jim{\'e}nez-Teja}}, \binits{Y.}},
\oauthor{\bsnm{{Windhorst}}, \binits{R.A.}},
\oauthor{\bsnm{{Xu}}, \binits{X.}}:
{The discovery of bound star clusters 460 Myr after the Big Bang}.
arXiv e-prints,
2401--03224
(2024)
\doiurl{10.48550/arXiv.2401.03224}
{\href{https://arxiv.org/abs/2401.03224}{{arXiv:2401.03224}}}
{[astro-ph.GA]}
\end{botherref}
\endbibitem

\bibitem[\protect\citeauthoryear{{Harris} and {Racine}}{1979}]{1979ARA&A..17..241H}
\begin{barticle}
\bauthor{\bsnm{{Harris}}, \binits{W.E.}},
\bauthor{\bsnm{{Racine}}, \binits{R.}}:
\batitle{{Globular clusters in galaxies.}}
\bjtitle{\araa}
\bvolume{17},
\bfpage{241}--\blpage{274}
(\byear{1979})
\doiurl{10.1146/annurev.aa.17.090179.001325}
\end{barticle}
\endbibitem

\bibitem[\protect\citeauthoryear{{Chattopadhyay} et~al.}{2015}]{2015ApJ...808...24C}
\begin{barticle}
\bauthor{\bsnm{{Chattopadhyay}}, \binits{T.}},
\bauthor{\bsnm{{De}}, \binits{T.}},
\bauthor{\bsnm{{Warlu}}, \binits{B.}},
\bauthor{\bsnm{{Chattopadhyay}}, \binits{A.K.}}:
\batitle{{Cosmic History of the Integrated Galactic Stellar Initial Mass Function: A Simulation Study}}.
\bjtitle{\apj}
\bvolume{808}(\bissue{1}),
\bfpage{24}
(\byear{2015})
\doiurl{10.1088/0004-637X/808/1/24}
{\href{https://arxiv.org/abs/1411.3848}{{arXiv:1411.3848}}}
{[astro-ph.GA]}
\end{barticle}
\endbibitem

\bibitem[\protect\citeauthoryear{{Carnall} et~al.}{2019}]{carnall19b}
\begin{barticle}
\bauthor{\bsnm{{Carnall}}, \binits{A.C.}},
\bauthor{\bsnm{{McLure}}, \binits{R.J.}},
\bauthor{\bsnm{{Dunlop}}, \binits{J.S.}},
\bauthor{\bsnm{{Cullen}}, \binits{F.}},
\bauthor{\bsnm{{McLeod}}, \binits{D.J.}},
\bauthor{\bsnm{{Wild}}, \binits{V.}},
\bauthor{\bsnm{{Johnson}}, \binits{B.D.}},
\bauthor{\bsnm{{Appleby}}, \binits{S.}},
\bauthor{\bsnm{{Dav{\'e}}}, \binits{R.}},
\bauthor{\bsnm{{Amorin}}, \binits{R.}},
\bauthor{\bsnm{{Bolzonella}}, \binits{M.}},
\bauthor{\bsnm{{Castellano}}, \binits{M.}},
\bauthor{\bsnm{{Cimatti}}, \binits{A.}},
\bauthor{\bsnm{{Cucciati}}, \binits{O.}},
\bauthor{\bsnm{{Gargiulo}}, \binits{A.}},
\bauthor{\bsnm{{Garilli}}, \binits{B.}},
\bauthor{\bsnm{{Marchi}}, \binits{F.}},
\bauthor{\bsnm{{Pentericci}}, \binits{L.}},
\bauthor{\bsnm{{Pozzetti}}, \binits{L.}},
\bauthor{\bsnm{{Schreiber}}, \binits{C.}},
\bauthor{\bsnm{{Talia}}, \binits{M.}},
\bauthor{\bsnm{{Zamorani}}, \binits{G.}}:
\batitle{{The VANDELS survey: the star-formation histories of massive quiescent galaxies at 1.0 < z < 1.3}}.
\bjtitle{\mnras}
\bvolume{490}(\bissue{1}),
\bfpage{417}--\blpage{439}
(\byear{2019})
\doiurl{10.1093/mnras/stz2544}
{\href{https://arxiv.org/abs/1903.11082}{{arXiv:1903.11082}}}
{[astro-ph.GA]}
\end{barticle}
\endbibitem

\bibitem[\protect\citeauthoryear{{Leja} et~al.}{2019}]{leja19}
\begin{barticle}
\bauthor{\bsnm{{Leja}}, \binits{J.}},
\bauthor{\bsnm{{Carnall}}, \binits{A.C.}},
\bauthor{\bsnm{{Johnson}}, \binits{B.D.}},
\bauthor{\bsnm{{Conroy}}, \binits{C.}},
\bauthor{\bsnm{{Speagle}}, \binits{J.S.}}:
\batitle{{How to Measure Galaxy Star Formation Histories. II. Nonparametric Models}}.
\bjtitle{\apj}
\bvolume{876}(\bissue{1}),
\bfpage{3}
(\byear{2019})
\doiurl{10.3847/1538-4357/ab133c}
{\href{https://arxiv.org/abs/1811.03637}{{arXiv:1811.03637}}}
{[astro-ph.GA]}
\end{barticle}
\endbibitem

\bibitem[\protect\citeauthoryear{{Mowla} et~al.}{2022}]{mowlaiyer2022}
\begin{barticle}
\bauthor{\bsnm{{Mowla}}, \binits{L.}},
\bauthor{\bsnm{{Iyer}}, \binits{K.G.}},
\bauthor{\bsnm{{Desprez}}, \binits{G.}},
\bauthor{\bsnm{{Estrada-Carpenter}}, \binits{V.}},
\bauthor{\bsnm{{Martis}}, \binits{N.S.}},
\bauthor{\bsnm{{Noirot}}, \binits{G.}},
\bauthor{\bsnm{{Sarrouh}}, \binits{G.T.}},
\bauthor{\bsnm{{Strait}}, \binits{V.}},
\bauthor{\bsnm{{Asada}}, \binits{Y.}},
\bauthor{\bsnm{{Abraham}}, \binits{R.G.}},
\bauthor{\bsnm{{Brammer}}, \binits{G.}},
\bauthor{\bsnm{{Sawicki}}, \binits{M.}},
\bauthor{\bsnm{{Willott}}, \binits{C.J.}},
\bauthor{\bsnm{{Bradac}}, \binits{M.}},
\bauthor{\bsnm{{Doyon}}, \binits{R.}},
\bauthor{\bsnm{{Muzzin}}, \binits{A.}},
\bauthor{\bsnm{{Pacifici}}, \binits{C.}},
\bauthor{\bsnm{{Ravindranath}}, \binits{S.}},
\bauthor{\bsnm{{Zabl}}, \binits{J.}}:
\batitle{{The Sparkler: Evolved High-redshift Globular Cluster Candidates Captured by JWST}}.
\bjtitle{\apjl}
\bvolume{937}(\bissue{2}),
\bfpage{35}
(\byear{2022})
\doiurl{10.3847/2041-8213/ac90ca}
{\href{https://arxiv.org/abs/2208.02233}{{arXiv:2208.02233}}}
{[astro-ph.GA]}
\end{barticle}
\endbibitem

\bibitem[\protect\citeauthoryear{{Papovich} et~al.}{2015}]{papovich15}
\begin{barticle}
\bauthor{\bsnm{{Papovich}}, \binits{C.}},
\bauthor{\bsnm{{Labb{\'e}}}, \binits{I.}},
\bauthor{\bsnm{{Quadri}}, \binits{R.}},
\bauthor{\bsnm{{Tilvi}}, \binits{V.}},
\bauthor{\bsnm{{Behroozi}}, \binits{P.}},
\bauthor{\bsnm{{Bell}}, \binits{E.F.}},
\bauthor{\bsnm{{Glazebrook}}, \binits{K.}},
\bauthor{\bsnm{{Spitler}}, \binits{L.}},
\bauthor{\bsnm{{Straatman}}, \binits{C.M.S.}},
\bauthor{\bsnm{{Tran}}, \binits{K.-V.}},
\bauthor{\bsnm{{Cowley}}, \binits{M.}},
\bauthor{\bsnm{{Dav{\'e}}}, \binits{R.}},
\bauthor{\bsnm{{Dekel}}, \binits{A.}},
\bauthor{\bsnm{{Dickinson}}, \binits{M.}},
\bauthor{\bsnm{{Ferguson}}, \binits{H.C.}},
\bauthor{\bsnm{{Finkelstein}}, \binits{S.L.}},
\bauthor{\bsnm{{Gawiser}}, \binits{E.}},
\bauthor{\bsnm{{Inami}}, \binits{H.}},
\bauthor{\bsnm{{Faber}}, \binits{S.M.}},
\bauthor{\bsnm{{Kacprzak}}, \binits{G.G.}},
\bauthor{\bsnm{{Kawinwanichakij}}, \binits{L.}},
\bauthor{\bsnm{{Kocevski}}, \binits{D.}},
\bauthor{\bsnm{{Koekemoer}}, \binits{A.}},
\bauthor{\bsnm{{Koo}}, \binits{D.C.}},
\bauthor{\bsnm{{Kurczynski}}, \binits{P.}},
\bauthor{\bsnm{{Lotz}}, \binits{J.M.}},
\bauthor{\bsnm{{Lu}}, \binits{Y.}},
\bauthor{\bsnm{{Lucas}}, \binits{R.A.}},
\bauthor{\bsnm{{McIntosh}}, \binits{D.}},
\bauthor{\bsnm{{Mehrtens}}, \binits{N.}},
\bauthor{\bsnm{{Mobasher}}, \binits{B.}},
\bauthor{\bsnm{{Monson}}, \binits{A.}},
\bauthor{\bsnm{{Morrison}}, \binits{G.}},
\bauthor{\bsnm{{Nanayakkara}}, \binits{T.}},
\bauthor{\bsnm{{Persson}}, \binits{S.E.}},
\bauthor{\bsnm{{Salmon}}, \binits{B.}},
\bauthor{\bsnm{{Simons}}, \binits{R.}},
\bauthor{\bsnm{{Tomczak}}, \binits{A.}},
\bauthor{\bsnm{{van Dokkum}}, \binits{P.}},
\bauthor{\bsnm{{Weiner}}, \binits{B.}},
\bauthor{\bsnm{{Willner}}, \binits{S.P.}}:
\batitle{{ZFOURGE/CANDELS: On the Evolution of M* Galaxy Progenitors from z = 3 to 0.5}}.
\bjtitle{\apj}
\bvolume{803}(\bissue{1}),
\bfpage{26}
(\byear{2015})
\doiurl{10.1088/0004-637X/803/1/26}
{\href{https://arxiv.org/abs/1412.3806}{{arXiv:1412.3806}}}
{[astro-ph.GA]}
\end{barticle}
\endbibitem

\bibitem[\protect\citeauthoryear{{Graziani} et~al.}{2017}]{graziani17}
\begin{barticle}
\bauthor{\bsnm{{Graziani}}, \binits{L.}},
\bauthor{\bsnm{{de Bennassuti}}, \binits{M.}},
\bauthor{\bsnm{{Schneider}}, \binits{R.}},
\bauthor{\bsnm{{Kawata}}, \binits{D.}},
\bauthor{\bsnm{{Salvadori}}, \binits{S.}}:
\batitle{{The history of the dark and luminous side of Milky Way-like progenitors}}.
\bjtitle{\mnras}
\bvolume{469}(\bissue{1}),
\bfpage{1101}--\blpage{1116}
(\byear{2017})
\doiurl{10.1093/mnras/stx900}
{\href{https://arxiv.org/abs/1704.02983}{{arXiv:1704.02983}}}
{[astro-ph.GA]}
\end{barticle}
\endbibitem

\bibitem[\protect\citeauthoryear{{Papovich} et~al.}{2016}]{papovich16}
\begin{barticle}
\bauthor{\bsnm{{Papovich}}, \binits{C.}},
\bauthor{\bsnm{{Labb{\'e}}}, \binits{I.}},
\bauthor{\bsnm{{Glazebrook}}, \binits{K.}},
\bauthor{\bsnm{{Quadri}}, \binits{R.}},
\bauthor{\bsnm{{Bekiaris}}, \binits{G.}},
\bauthor{\bsnm{{Dickinson}}, \binits{M.}},
\bauthor{\bsnm{{Finkelstein}}, \binits{S.L.}},
\bauthor{\bsnm{{Fisher}}, \binits{D.}},
\bauthor{\bsnm{{Inami}}, \binits{H.}},
\bauthor{\bsnm{{Livermore}}, \binits{R.C.}},
\bauthor{\bsnm{{Spitler}}, \binits{L.}},
\bauthor{\bsnm{{Straatman}}, \binits{C.}},
\bauthor{\bsnm{{Tran}}, \binits{K.-V.}}:
\batitle{{Large molecular gas reservoirs in ancestors of Milky Way-mass galaxies nine billion years ago}}.
\bjtitle{Nature Astronomy}
\bvolume{1},
\bfpage{0003}
(\byear{2016})
\doiurl{10.1038/s41550-016-0003}
{\href{https://arxiv.org/abs/1610.05313}{{arXiv:1610.05313}}}
{[astro-ph.GA]}
\end{barticle}
\endbibitem

\bibitem[\protect\citeauthoryear{{Iza} et~al.}{2023}]{2023BAAA...64..143I}
\begin{barticle}
\bauthor{\bsnm{{Iza}}, \binits{F.G.}},
\bauthor{\bsnm{{Nuza}}, \binits{S.E.}},
\bauthor{\bsnm{{Scannapieco}}, \binits{C.}}:
\batitle{{The temporal evolution of gas accretion onto the discs of simulated Milky Way-mass galaxies}}.
\bjtitle{Boletin de la Asociacion Argentina de Astronomia La Plata Argentina}
\bvolume{64},
\bfpage{143}--\blpage{145}
(\byear{2023})
\doiurl{10.48550/arXiv.2306.14760}
{\href{https://arxiv.org/abs/2306.14760}{{arXiv:2306.14760}}}
{[astro-ph.GA]}
\end{barticle}
\endbibitem

\bibitem[\protect\citeauthoryear{{Feng} et~al.}{2015}]{2015ApJ...808L..17F}
\begin{barticle}
\bauthor{\bsnm{{Feng}}, \binits{Y.}},
\bauthor{\bsnm{{Di Matteo}}, \binits{T.}},
\bauthor{\bsnm{{Croft}}, \binits{R.}},
\bauthor{\bsnm{{Tenneti}}, \binits{A.}},
\bauthor{\bsnm{{Bird}}, \binits{S.}},
\bauthor{\bsnm{{Battaglia}}, \binits{N.}},
\bauthor{\bsnm{{Wilkins}}, \binits{S.}}:
\batitle{{The Formation of Milky Way-mass Disk Galaxies in the First 500 Million Years of a Cold Dark Matter Universe}}.
\bjtitle{\apjl}
\bvolume{808}(\bissue{1}),
\bfpage{17}
(\byear{2015})
\doiurl{10.1088/2041-8205/808/1/L17}
{\href{https://arxiv.org/abs/1504.06618}{{arXiv:1504.06618}}}
{[astro-ph.GA]}
\end{barticle}
\endbibitem

\bibitem[\protect\citeauthoryear{{Nepal} et~al.}{2024}]{2024arXiv240200561N}
\begin{botherref}
\oauthor{\bsnm{{Nepal}}, \binits{S.}},
\oauthor{\bsnm{{Chiappini}}, \binits{C.}},
\oauthor{\bsnm{{Queiroz}}, \binits{A.B.A.}},
\oauthor{\bsnm{{Guiglion}}, \binits{G.}},
\oauthor{\bsnm{{Montalb{\'a}n}}, \binits{J.}},
\oauthor{\bsnm{{Steinmetz}}, \binits{M.}},
\oauthor{\bsnm{{Miglio}}, \binits{A.}},
\oauthor{\bsnm{{Khalatyan}}, \binits{A.}}:
{Discovery of the local counterpart of disc galaxies at z > 4: The oldest thin disc of Milky Way using Gaia-RVS}.
arXiv e-prints,
2402--00561
(2024)
\doiurl{10.48550/arXiv.2402.00561}
{\href{https://arxiv.org/abs/2402.00561}{{arXiv:2402.00561}}}
{[astro-ph.GA]}
\end{botherref}
\endbibitem

\bibitem[\protect\citeauthoryear{{Bournaud}}{2016}]{2016ASSL..418..355B}
\begin{bchapter}
\bauthor{\bsnm{{Bournaud}}, \binits{F.}}:
\bctitle{{Bulge Growth Through Disc Instabilities in High-Redshift Galaxies}}.
In: \beditor{\bsnm{{Laurikainen}}, \binits{E.}},
\beditor{\bsnm{{Peletier}}, \binits{R.}},
\beditor{\bsnm{{Gadotti}}, \binits{D.}} (eds.)
\bbtitle{Galactic Bulges}.
\bsertitle{Astrophysics and Space Science Library},
vol. \bseriesno{418},
p. \bfpage{355}
(\byear{2016}).
\doiurl{10.1007/978-3-319-19378-6_13}
\end{bchapter}
\endbibitem

\bibitem[\protect\citeauthoryear{{Mandelker} et~al.}{2017}]{2017MNRAS.464..635M}
\begin{barticle}
\bauthor{\bsnm{{Mandelker}}, \binits{N.}},
\bauthor{\bsnm{{Dekel}}, \binits{A.}},
\bauthor{\bsnm{{Ceverino}}, \binits{D.}},
\bauthor{\bsnm{{DeGraf}}, \binits{C.}},
\bauthor{\bsnm{{Guo}}, \binits{Y.}},
\bauthor{\bsnm{{Primack}}, \binits{J.}}:
\batitle{{Giant clumps in simulated high- z Galaxies: properties, evolution and dependence on feedback}}.
\bjtitle{\mnras}
\bvolume{464}(\bissue{1}),
\bfpage{635}--\blpage{665}
(\byear{2017})
\doiurl{10.1093/mnras/stw2358}
{\href{https://arxiv.org/abs/1512.08791}{{arXiv:1512.08791}}}
{[astro-ph.GA]}
\end{barticle}
\endbibitem

\bibitem[\protect\citeauthoryear{{Mandelker} et~al.}{2018}]{2018ApJ...861..148M}
\begin{barticle}
\bauthor{\bsnm{{Mandelker}}, \binits{N.}},
\bauthor{\bsnm{{van Dokkum}}, \binits{P.G.}},
\bauthor{\bsnm{{Brodie}}, \binits{J.P.}},
\bauthor{\bsnm{{van den Bosch}}, \binits{F.C.}},
\bauthor{\bsnm{{Ceverino}}, \binits{D.}}:
\batitle{{Cold Filamentary Accretion and the Formation of Metal-poor Globular Clusters and Halo Stars}}.
\bjtitle{\apj}
\bvolume{861}(\bissue{2}),
\bfpage{148}
(\byear{2018})
\doiurl{10.3847/1538-4357/aaca98}
{\href{https://arxiv.org/abs/1711.09108}{{arXiv:1711.09108}}}
{[astro-ph.GA]}
\end{barticle}
\endbibitem

\bibitem[\protect\citeauthoryear{{van Donkelaar} et~al.}{2022}]{2022MNRAS.512.3806V}
\begin{barticle}
\bauthor{\bsnm{{van Donkelaar}}, \binits{F.}},
\bauthor{\bsnm{{Agertz}}, \binits{O.}},
\bauthor{\bsnm{{Renaud}}, \binits{F.}}:
\batitle{{From giant clumps to clouds - II. The emergence of thick disc kinematics from the conditions of star formation in high redshift gas rich galaxies}}.
\bjtitle{\mnras}
\bvolume{512}(\bissue{3}),
\bfpage{3806}--\blpage{3814}
(\byear{2022})
\doiurl{10.1093/mnras/stac692}
{\href{https://arxiv.org/abs/2110.13165}{{arXiv:2110.13165}}}
{[astro-ph.GA]}
\end{barticle}
\endbibitem

\bibitem[\protect\citeauthoryear{{Dekel} et~al.}{2023}]{2023MNRAS.523.3201D}
\begin{barticle}
\bauthor{\bsnm{{Dekel}}, \binits{A.}},
\bauthor{\bsnm{{Sarkar}}, \binits{K.C.}},
\bauthor{\bsnm{{Birnboim}}, \binits{Y.}},
\bauthor{\bsnm{{Mandelker}}, \binits{N.}},
\bauthor{\bsnm{{Li}}, \binits{Z.}}:
\batitle{{Efficient formation of massive galaxies at cosmic dawn by feedback-free starbursts}}.
\bjtitle{\mnras}
\bvolume{523}(\bissue{3}),
\bfpage{3201}--\blpage{3218}
(\byear{2023})
\doiurl{10.1093/mnras/stad1557}
{\href{https://arxiv.org/abs/2303.04827}{{arXiv:2303.04827}}}
{[astro-ph.GA]}
\end{barticle}
\endbibitem

\bibitem[\protect\citeauthoryear{{Yung} et~al.}{2024}]{2024MNRAS.527.5929Y}
\begin{barticle}
\bauthor{\bsnm{{Yung}}, \binits{L.Y.A.}},
\bauthor{\bsnm{{Somerville}}, \binits{R.S.}},
\bauthor{\bsnm{{Finkelstein}}, \binits{S.L.}},
\bauthor{\bsnm{{Wilkins}}, \binits{S.M.}},
\bauthor{\bsnm{{Gardner}}, \binits{J.P.}}:
\batitle{{Are the ultra-high-redshift galaxies at z > 10 surprising in the context of standard galaxy formation models?}}
\bjtitle{\mnras}
\bvolume{527}(\bissue{3}),
\bfpage{5929}--\blpage{5948}
(\byear{2024})
\doiurl{10.1093/mnras/stad3484}
{\href{https://arxiv.org/abs/2304.04348}{{arXiv:2304.04348}}}
{[astro-ph.GA]}
\end{barticle}
\endbibitem

\bibitem[\protect\citeauthoryear{{Renzini}}{2023}]{2023MNRAS.525L.117R}
\begin{barticle}
\bauthor{\bsnm{{Renzini}}, \binits{A.}}:
\batitle{{A transient overcooling in the early Universe? Clues from globular clusters formation}}.
\bjtitle{\mnras}
\bvolume{525}(\bissue{1}),
\bfpage{117}--\blpage{120}
(\byear{2023})
\doiurl{10.1093/mnrasl/slad091}
{\href{https://arxiv.org/abs/2305.14476}{{arXiv:2305.14476}}}
{[astro-ph.GA]}
\end{barticle}
\endbibitem

\bibitem[\protect\citeauthoryear{{Lancaster} et~al.}{2021}]{lancaster21}
\begin{barticle}
\bauthor{\bsnm{{Lancaster}}, \binits{L.}},
\bauthor{\bsnm{{Ostriker}}, \binits{E.C.}},
\bauthor{\bsnm{{Kim}}, \binits{J.-G.}},
\bauthor{\bsnm{{Kim}}, \binits{C.-G.}}:
\batitle{{Efficiently Cooled Stellar Wind Bubbles in Turbulent Clouds. II. Validation of Theory with Hydrodynamic Simulations}}.
\bjtitle{\apj}
\bvolume{914}(\bissue{2}),
\bfpage{90}
(\byear{2021})
\doiurl{10.3847/1538-4357/abf8ac}
{\href{https://arxiv.org/abs/2104.07722}{{arXiv:2104.07722}}}
{[astro-ph.GA]}
\end{barticle}
\endbibitem

\bibitem[\protect\citeauthoryear{{Tamfal} et~al.}{2022}]{2022ApJ...928..106T}
\begin{barticle}
\bauthor{\bsnm{{Tamfal}}, \binits{T.}},
\bauthor{\bsnm{{Mayer}}, \binits{L.}},
\bauthor{\bsnm{{Quinn}}, \binits{T.R.}},
\bauthor{\bsnm{{Babul}}, \binits{A.}},
\bauthor{\bsnm{{Madau}}, \binits{P.}},
\bauthor{\bsnm{{Capelo}}, \binits{P.R.}},
\bauthor{\bsnm{{Shen}}, \binits{S.}},
\bauthor{\bsnm{{Staub}}, \binits{M.}}:
\batitle{{The Dawn of Disk Formation in a Milky Way-sized Galaxy Halo: Thin Stellar Disks at z > 4}}.
\bjtitle{\apj}
\bvolume{928}(\bissue{2}),
\bfpage{106}
(\byear{2022})
\doiurl{10.3847/1538-4357/ac558e}
{\href{https://arxiv.org/abs/2106.11981}{{arXiv:2106.11981}}}
{[astro-ph.GA]}
\end{barticle}
\endbibitem

\bibitem[\protect\citeauthoryear{{Bournaud} et~al.}{2014}]{bournaud14}
\begin{barticle}
\bauthor{\bsnm{{Bournaud}}, \binits{F.}},
\bauthor{\bsnm{{Perret}}, \binits{V.}},
\bauthor{\bsnm{{Renaud}}, \binits{F.}},
\bauthor{\bsnm{{Dekel}}, \binits{A.}},
\bauthor{\bsnm{{Elmegreen}}, \binits{B.G.}},
\bauthor{\bsnm{{Elmegreen}}, \binits{D.M.}},
\bauthor{\bsnm{{Teyssier}}, \binits{R.}},
\bauthor{\bsnm{{Amram}}, \binits{P.}},
\bauthor{\bsnm{{Daddi}}, \binits{E.}},
\bauthor{\bsnm{{Duc}}, \binits{P.-A.}},
\bauthor{\bsnm{{Elbaz}}, \binits{D.}},
\bauthor{\bsnm{{Epinat}}, \binits{B.}},
\bauthor{\bsnm{{Gabor}}, \binits{J.M.}},
\bauthor{\bsnm{{Juneau}}, \binits{S.}},
\bauthor{\bsnm{{Kraljic}}, \binits{K.}},
\bauthor{\bsnm{{Le Floch'}}, \binits{E.}}:
\batitle{{The Long Lives of Giant Clumps and the Birth of Outflows in Gas-rich Galaxies at High Redshift}}.
\bjtitle{\apj}
\bvolume{780}(\bissue{1}),
\bfpage{57}
(\byear{2014})
\doiurl{10.1088/0004-637X/780/1/57}
{\href{https://arxiv.org/abs/1307.7136}{{arXiv:1307.7136}}}
{[astro-ph.CO]}
\end{barticle}
\endbibitem

\bibitem[\protect\citeauthoryear{{Gnedin} et~al.}{2014}]{2014ApJ...785...71G}
\begin{barticle}
\bauthor{\bsnm{{Gnedin}}, \binits{O.Y.}},
\bauthor{\bsnm{{Ostriker}}, \binits{J.P.}},
\bauthor{\bsnm{{Tremaine}}, \binits{S.}}:
\batitle{{Co-evolution of Galactic Nuclei and Globular Cluster Systems}}.
\bjtitle{\apj}
\bvolume{785}(\bissue{1}),
\bfpage{71}
(\byear{2014})
\doiurl{10.1088/0004-637X/785/1/71}
{\href{https://arxiv.org/abs/1308.0021}{{arXiv:1308.0021}}}
{[astro-ph.CO]}
\end{barticle}
\endbibitem

\bibitem[\protect\citeauthoryear{{Mandelker} et~al.}{2014}]{2014MNRAS.443.3675M}
\begin{barticle}
\bauthor{\bsnm{{Mandelker}}, \binits{N.}},
\bauthor{\bsnm{{Dekel}}, \binits{A.}},
\bauthor{\bsnm{{Ceverino}}, \binits{D.}},
\bauthor{\bsnm{{Tweed}}, \binits{D.}},
\bauthor{\bsnm{{Moody}}, \binits{C.E.}},
\bauthor{\bsnm{{Primack}}, \binits{J.}}:
\batitle{{The population of giant clumps in simulated high-z galaxies: in situ and ex situ migration and survival}}.
\bjtitle{\mnras}
\bvolume{443}(\bissue{4}),
\bfpage{3675}--\blpage{3702}
(\byear{2014})
\doiurl{10.1093/mnras/stu1340}
{\href{https://arxiv.org/abs/1311.0013}{{arXiv:1311.0013}}}
{[astro-ph.CO]}
\end{barticle}
\endbibitem

\bibitem[\protect\citeauthoryear{{Elmegreen}}{2015}]{2015llg..book..477E}
\begin{bchapter}
\bauthor{\bsnm{{Elmegreen}}, \binits{B.G.}}:
\bctitle{{Formation of Stars and Clusters over Cosmological Time}}.
In: \bbtitle{Lessons from the Local Group: A Conference in Honor of David Block and Bruce Elmegreen},
p. \bfpage{477}
(\byear{2015}).
\doiurl{10.1007/978-3-319-10614-4_39}
\end{bchapter}
\endbibitem

\bibitem[\protect\citeauthoryear{{Forbes} et~al.}{2018}]{2018RSPSA.47470616F}
\begin{barticle}
\bauthor{\bsnm{{Forbes}}, \binits{D.A.}},
\bauthor{\bsnm{{Bastian}}, \binits{N.}},
\bauthor{\bsnm{{Gieles}}, \binits{M.}},
\bauthor{\bsnm{{Crain}}, \binits{R.A.}},
\bauthor{\bsnm{{Kruijssen}}, \binits{J.M.D.}},
\bauthor{\bsnm{{Larsen}}, \binits{S.S.}},
\bauthor{\bsnm{{Ploeckinger}}, \binits{S.}},
\bauthor{\bsnm{{Agertz}}, \binits{O.}},
\bauthor{\bsnm{{Trenti}}, \binits{M.}},
\bauthor{\bsnm{{Ferguson}}, \binits{A.M.N.}},
\bauthor{\bsnm{{Pfeffer}}, \binits{J.}},
\bauthor{\bsnm{{Gnedin}}, \binits{O.Y.}}:
\batitle{{Globular cluster formation and evolution in the context of cosmological galaxy assembly: open questions}}.
\bjtitle{Proceedings of the Royal Society of London Series A}
\bvolume{474}(\bissue{2210}),
\bfpage{20170616}
(\byear{2018})
\doiurl{10.1098/rspa.2017.0616}
{\href{https://arxiv.org/abs/1801.05818}{{arXiv:1801.05818}}}
{[astro-ph.GA]}
\end{barticle}
\endbibitem

\bibitem[\protect\citeauthoryear{{Asada} et~al.}{2024}]{asada24}
\begin{barticle}
\bauthor{\bsnm{{Asada}}, \binits{Y.}},
\bauthor{\bsnm{{Sawicki}}, \binits{M.}},
\bauthor{\bsnm{{Abraham}}, \binits{R.}},
\bauthor{\bsnm{{Brada{\v{c}}}}, \binits{M.}},
\bauthor{\bsnm{{Brammer}}, \binits{G.}},
\bauthor{\bsnm{{Desprez}}, \binits{G.}},
\bauthor{\bsnm{{Estrada-Carpenter}}, \binits{V.}},
\bauthor{\bsnm{{Iyer}}, \binits{K.}},
\bauthor{\bsnm{{Martis}}, \binits{N.}},
\bauthor{\bsnm{{Matharu}}, \binits{J.}},
\bauthor{\bsnm{{Mowla}}, \binits{L.}},
\bauthor{\bsnm{{Muzzin}}, \binits{A.}},
\bauthor{\bsnm{{Noirot}}, \binits{G.}},
\bauthor{\bsnm{{Sarrouh}}, \binits{G.T.E.}},
\bauthor{\bsnm{{Strait}}, \binits{V.}},
\bauthor{\bsnm{{Willott}}, \binits{C.J.}},
\bauthor{\bsnm{{Harshan}}, \binits{A.}}:
\batitle{{Bursty star formation and galaxy-galaxy interactions in low-mass galaxies 1 Gyr after the Big Bang}}.
\bjtitle{\mnras}
\bvolume{527}(\bissue{4}),
\bfpage{11372}--\blpage{11392}
(\byear{2024})
\doiurl{10.1093/mnras/stad3902}
{\href{https://arxiv.org/abs/2310.02314}{{arXiv:2310.02314}}}
{[astro-ph.GA]}
\end{barticle}
\endbibitem

\bibitem[\protect\citeauthoryear{{Brammer}}{2023}]{grizli23}
\begin{botherref}
\oauthor{\bsnm{{Brammer}}, \binits{G.}}:
{grizli}.
Zenodo
(2023).
\doiurl{10.5281/zenodo.1146904}
\end{botherref}
\endbibitem

\bibitem[\protect\citeauthoryear{{Gaia Collaboration} et~al.}{2023}]{gaia2023}
\begin{barticle}
\bauthor{\bsnm{{Gaia Collaboration}}},
\bauthor{\bsnm{{Vallenari}}, \binits{A.}},
\bauthor{\bsnm{{Brown}}, \binits{A.G.A.}},
\bauthor{\bsnm{{Prusti}}, \binits{T.}},
\bauthor{\bsnm{{de Bruijne}}, \binits{J.H.J.}},
\bauthor{\bsnm{{Arenou}}, \binits{F.}},
\bauthor{\bsnm{{Babusiaux}}, \binits{C.}},
\bauthor{\bsnm{{Biermann}}, \binits{M.}},
\bauthor{\bsnm{{Creevey}}, \binits{O.L.}},
\bauthor{\bsnm{{Ducourant}}, \binits{C.}},
\bauthor{\bsnm{{Evans}}, \binits{D.W.}},
\bauthor{\bsnm{{Eyer}}, \binits{L.}},
\bauthor{\bsnm{{Guerra}}, \binits{R.}},
\bauthor{\bsnm{{Hutton}}, \binits{A.}},
\bauthor{\bsnm{{Jordi}}, \binits{C.}},
\bauthor{\bsnm{{Klioner}}, \binits{S.A.}},
\bauthor{\bsnm{{Lammers}}, \binits{U.L.}},
\bauthor{\bsnm{{Lindegren}}, \binits{L.}},
\bauthor{\bsnm{{Luri}}, \binits{X.}},
\bauthor{\bsnm{{Mignard}}, \binits{F.}},
\bauthor{\bsnm{{Panem}}, \binits{C.}},
\bauthor{\bsnm{{Pourbaix}}, \binits{D.}},
\bauthor{\bsnm{{Randich}}, \binits{S.}},
\bauthor{\bsnm{{Sartoretti}}, \binits{P.}},
\bauthor{\bsnm{{Soubiran}}, \binits{C.}},
\bauthor{\bsnm{{Tanga}}, \binits{P.}},
\bauthor{\bsnm{{Walton}}, \binits{N.A.}},
\bauthor{\bsnm{{Bailer-Jones}}, \binits{C.A.L.}},
\bauthor{\bsnm{{Bastian}}, \binits{U.}},
\bauthor{\bsnm{{Drimmel}}, \binits{R.}},
\bauthor{\bsnm{{Jansen}}, \binits{F.}},
\bauthor{\bsnm{{Katz}}, \binits{D.}},
\bauthor{\bsnm{{Lattanzi}}, \binits{M.G.}},
\bauthor{\bsnm{{van Leeuwen}}, \binits{F.}},
\bauthor{\bsnm{{Bakker}}, \binits{J.}},
\bauthor{\bsnm{{Cacciari}}, \binits{C.}},
\bauthor{\bsnm{{Casta{\~n}eda}}, \binits{J.}},
\bauthor{\bsnm{{De Angeli}}, \binits{F.}},
\bauthor{\bsnm{{Fabricius}}, \binits{C.}},
\bauthor{\bsnm{{Fouesneau}}, \binits{M.}},
\bauthor{\bsnm{{Fr{\'e}mat}}, \binits{Y.}},
\bauthor{\bsnm{{Galluccio}}, \binits{L.}},
\bauthor{\bsnm{{Guerrier}}, \binits{A.}},
\bauthor{\bsnm{{Heiter}}, \binits{U.}},
\bauthor{\bsnm{{Masana}}, \binits{E.}},
\bauthor{\bsnm{{Messineo}}, \binits{R.}},
\bauthor{\bsnm{{Mowlavi}}, \binits{N.}},
\bauthor{\bsnm{{Nicolas}}, \binits{C.}},
\bauthor{\bsnm{{Nienartowicz}}, \binits{K.}},
\bauthor{\bsnm{{Pailler}}, \binits{F.}},
\bauthor{\bsnm{{Panuzzo}}, \binits{P.}},
\bauthor{\bsnm{{Riclet}}, \binits{F.}},
\bauthor{\bsnm{{Roux}}, \binits{W.}},
\bauthor{\bsnm{{Seabroke}}, \binits{G.M.}},
\bauthor{\bsnm{{Sordo}}, \binits{R.}},
\bauthor{\bsnm{{Th{\'e}venin}}, \binits{F.}},
\bauthor{\bsnm{{Gracia-Abril}}, \binits{G.}},
\bauthor{\bsnm{{Portell}}, \binits{J.}},
\bauthor{\bsnm{{Teyssier}}, \binits{D.}},
\bauthor{\bsnm{{Altmann}}, \binits{M.}},
\bauthor{\bsnm{{Andrae}}, \binits{R.}},
\bauthor{\bsnm{{Audard}}, \binits{M.}},
\bauthor{\bsnm{{Bellas-Velidis}}, \binits{I.}},
\bauthor{\bsnm{{Benson}}, \binits{K.}},
\bauthor{\bsnm{{Berthier}}, \binits{J.}},
\bauthor{\bsnm{{Blomme}}, \binits{R.}},
\bauthor{\bsnm{{Burgess}}, \binits{P.W.}},
\bauthor{\bsnm{{Busonero}}, \binits{D.}},
\bauthor{\bsnm{{Busso}}, \binits{G.}},
\bauthor{\bsnm{{C{\'a}novas}}, \binits{H.}},
\bauthor{\bsnm{{Carry}}, \binits{B.}},
\bauthor{\bsnm{{Cellino}}, \binits{A.}},
\bauthor{\bsnm{{Cheek}}, \binits{N.}},
\bauthor{\bsnm{{Clementini}}, \binits{G.}},
\bauthor{\bsnm{{Damerdji}}, \binits{Y.}},
\bauthor{\bsnm{{Davidson}}, \binits{M.}},
\bauthor{\bsnm{{de Teodoro}}, \binits{P.}},
\bauthor{\bsnm{{Nu{\~n}ez Campos}}, \binits{M.}},
\bauthor{\bsnm{{Delchambre}}, \binits{L.}},
\bauthor{\bsnm{{Dell'Oro}}, \binits{A.}},
\bauthor{\bsnm{{Esquej}}, \binits{P.}},
\bauthor{\bsnm{{Fern{\'a}ndez-Hern{\'a}ndez}}, \binits{J.}},
\bauthor{\bsnm{{Fraile}}, \binits{E.}},
\bauthor{\bsnm{{Garabato}}, \binits{D.}},
\bauthor{\bsnm{{Garc{\'\i}a-Lario}}, \binits{P.}},
\bauthor{\bsnm{{Gosset}}, \binits{E.}},
\bauthor{\bsnm{{Haigron}}, \binits{R.}},
\bauthor{\bsnm{{Halbwachs}}, \binits{J.-L.}},
\bauthor{\bsnm{{Hambly}}, \binits{N.C.}},
\bauthor{\bsnm{{Harrison}}, \binits{D.L.}},
\bauthor{\bsnm{{Hern{\'a}ndez}}, \binits{J.}},
\bauthor{\bsnm{{Hestroffer}}, \binits{D.}},
\bauthor{\bsnm{{Hodgkin}}, \binits{S.T.}},
\bauthor{\bsnm{{Holl}}, \binits{B.}},
\bauthor{\bsnm{{Jan{\ss}en}}, \binits{K.}},
\bauthor{\bsnm{{Jevardat de Fombelle}}, \binits{G.}},
\bauthor{\bsnm{{Jordan}}, \binits{S.}},
\bauthor{\bsnm{{Krone-Martins}}, \binits{A.}},
\bauthor{\bsnm{{Lanzafame}}, \binits{A.C.}},
\bauthor{\bsnm{{L{\"o}ffler}}, \binits{W.}},
\bauthor{\bsnm{{Marchal}}, \binits{O.}},
\bauthor{\bsnm{{Marrese}}, \binits{P.M.}},
\bauthor{\bsnm{{Moitinho}}, \binits{A.}},
\bauthor{\bsnm{{Muinonen}}, \binits{K.}},
\bauthor{\bsnm{{Osborne}}, \binits{P.}},
\bauthor{\bsnm{{Pancino}}, \binits{E.}},
\bauthor{\bsnm{{Pauwels}}, \binits{T.}},
\bauthor{\bsnm{{Recio-Blanco}}, \binits{A.}},
\bauthor{\bsnm{{Reyl{\'e}}}, \binits{C.}},
\bauthor{\bsnm{{Riello}}, \binits{M.}},
\bauthor{\bsnm{{Rimoldini}}, \binits{L.}},
\bauthor{\bsnm{{Roegiers}}, \binits{T.}},
\bauthor{\bsnm{{Rybizki}}, \binits{J.}},
\bauthor{\bsnm{{Sarro}}, \binits{L.M.}},
\bauthor{\bsnm{{Siopis}}, \binits{C.}},
\bauthor{\bsnm{{Smith}}, \binits{M.}},
\bauthor{\bsnm{{Sozzetti}}, \binits{A.}},
\bauthor{\bsnm{{Utrilla}}, \binits{E.}},
\bauthor{\bsnm{{van Leeuwen}}, \binits{M.}},
\bauthor{\bsnm{{Abbas}}, \binits{U.}},
\bauthor{\bsnm{{{\'A}brah{\'a}m}}, \binits{P.}},
\bauthor{\bsnm{{Abreu Aramburu}}, \binits{A.}},
\bauthor{\bsnm{{Aerts}}, \binits{C.}},
\bauthor{\bsnm{{Aguado}}, \binits{J.J.}},
\bauthor{\bsnm{{Ajaj}}, \binits{M.}},
\bauthor{\bsnm{{Aldea-Montero}}, \binits{F.}},
\bauthor{\bsnm{{Altavilla}}, \binits{G.}},
\bauthor{\bsnm{{{\'A}lvarez}}, \binits{M.A.}},
\bauthor{\bsnm{{Alves}}, \binits{J.}},
\bauthor{\bsnm{{Anders}}, \binits{F.}},
\bauthor{\bsnm{{Anderson}}, \binits{R.I.}},
\bauthor{\bsnm{{Anglada Varela}}, \binits{E.}},
\bauthor{\bsnm{{Antoja}}, \binits{T.}},
\bauthor{\bsnm{{Baines}}, \binits{D.}},
\bauthor{\bsnm{{Baker}}, \binits{S.G.}},
\bauthor{\bsnm{{Balaguer-N{\'u}{\~n}ez}}, \binits{L.}},
\bauthor{\bsnm{{Balbinot}}, \binits{E.}},
\bauthor{\bsnm{{Balog}}, \binits{Z.}},
\bauthor{\bsnm{{Barache}}, \binits{C.}},
\bauthor{\bsnm{{Barbato}}, \binits{D.}},
\bauthor{\bsnm{{Barros}}, \binits{M.}},
\bauthor{\bsnm{{Barstow}}, \binits{M.A.}},
\bauthor{\bsnm{{Bartolom{\'e}}}, \binits{S.}},
\bauthor{\bsnm{{Bassilana}}, \binits{J.-L.}},
\bauthor{\bsnm{{Bauchet}}, \binits{N.}},
\bauthor{\bsnm{{Becciani}}, \binits{U.}},
\bauthor{\bsnm{{Bellazzini}}, \binits{M.}},
\bauthor{\bsnm{{Berihuete}}, \binits{A.}},
\bauthor{\bsnm{{Bernet}}, \binits{M.}},
\bauthor{\bsnm{{Bertone}}, \binits{S.}},
\bauthor{\bsnm{{Bianchi}}, \binits{L.}},
\bauthor{\bsnm{{Binnenfeld}}, \binits{A.}},
\bauthor{\bsnm{{Blanco-Cuaresma}}, \binits{S.}},
\bauthor{\bsnm{{Blazere}}, \binits{A.}},
\bauthor{\bsnm{{Boch}}, \binits{T.}},
\bauthor{\bsnm{{Bombrun}}, \binits{A.}},
\bauthor{\bsnm{{Bossini}}, \binits{D.}},
\bauthor{\bsnm{{Bouquillon}}, \binits{S.}},
\bauthor{\bsnm{{Bragaglia}}, \binits{A.}},
\bauthor{\bsnm{{Bramante}}, \binits{L.}},
\bauthor{\bsnm{{Breedt}}, \binits{E.}},
\bauthor{\bsnm{{Bressan}}, \binits{A.}},
\bauthor{\bsnm{{Brouillet}}, \binits{N.}},
\bauthor{\bsnm{{Brugaletta}}, \binits{E.}},
\bauthor{\bsnm{{Bucciarelli}}, \binits{B.}},
\bauthor{\bsnm{{Burlacu}}, \binits{A.}},
\bauthor{\bsnm{{Butkevich}}, \binits{A.G.}},
\bauthor{\bsnm{{Buzzi}}, \binits{R.}},
\bauthor{\bsnm{{Caffau}}, \binits{E.}},
\bauthor{\bsnm{{Cancelliere}}, \binits{R.}},
\bauthor{\bsnm{{Cantat-Gaudin}}, \binits{T.}},
\bauthor{\bsnm{{Carballo}}, \binits{R.}},
\bauthor{\bsnm{{Carlucci}}, \binits{T.}},
\bauthor{\bsnm{{Carnerero}}, \binits{M.I.}},
\bauthor{\bsnm{{Carrasco}}, \binits{J.M.}},
\bauthor{\bsnm{{Casamiquela}}, \binits{L.}},
\bauthor{\bsnm{{Castellani}}, \binits{M.}},
\bauthor{\bsnm{{Castro-Ginard}}, \binits{A.}},
\bauthor{\bsnm{{Chaoul}}, \binits{L.}},
\bauthor{\bsnm{{Charlot}}, \binits{P.}},
\bauthor{\bsnm{{Chemin}}, \binits{L.}},
\bauthor{\bsnm{{Chiaramida}}, \binits{V.}},
\bauthor{\bsnm{{Chiavassa}}, \binits{A.}},
\bauthor{\bsnm{{Chornay}}, \binits{N.}},
\bauthor{\bsnm{{Comoretto}}, \binits{G.}},
\bauthor{\bsnm{{Contursi}}, \binits{G.}},
\bauthor{\bsnm{{Cooper}}, \binits{W.J.}},
\bauthor{\bsnm{{Cornez}}, \binits{T.}},
\bauthor{\bsnm{{Cowell}}, \binits{S.}},
\bauthor{\bsnm{{Crifo}}, \binits{F.}},
\bauthor{\bsnm{{Cropper}}, \binits{M.}},
\bauthor{\bsnm{{Crosta}}, \binits{M.}},
\bauthor{\bsnm{{Crowley}}, \binits{C.}},
\bauthor{\bsnm{{Dafonte}}, \binits{C.}},
\bauthor{\bsnm{{Dapergolas}}, \binits{A.}},
\bauthor{\bsnm{{David}}, \binits{M.}},
\bauthor{\bsnm{{David}}, \binits{P.}},
\bauthor{\bsnm{{de Laverny}}, \binits{P.}},
\bauthor{\bsnm{{De Luise}}, \binits{F.}},
\bauthor{\bsnm{{De March}}, \binits{R.}},
\bauthor{\bsnm{{De Ridder}}, \binits{J.}},
\bauthor{\bsnm{{de Souza}}, \binits{R.}},
\bauthor{\bsnm{{de Torres}}, \binits{A.}},
\bauthor{\bsnm{{del Peloso}}, \binits{E.F.}},
\bauthor{\bsnm{{del Pozo}}, \binits{E.}},
\bauthor{\bsnm{{Delbo}}, \binits{M.}},
\bauthor{\bsnm{{Delgado}}, \binits{A.}},
\bauthor{\bsnm{{Delisle}}, \binits{J.-B.}},
\bauthor{\bsnm{{Demouchy}}, \binits{C.}},
\bauthor{\bsnm{{Dharmawardena}}, \binits{T.E.}},
\bauthor{\bsnm{{Di Matteo}}, \binits{P.}},
\bauthor{\bsnm{{Diakite}}, \binits{S.}},
\bauthor{\bsnm{{Diener}}, \binits{C.}},
\bauthor{\bsnm{{Distefano}}, \binits{E.}},
\bauthor{\bsnm{{Dolding}}, \binits{C.}},
\bauthor{\bsnm{{Edvardsson}}, \binits{B.}},
\bauthor{\bsnm{{Enke}}, \binits{H.}},
\bauthor{\bsnm{{Fabre}}, \binits{C.}},
\bauthor{\bsnm{{Fabrizio}}, \binits{M.}},
\bauthor{\bsnm{{Faigler}}, \binits{S.}},
\bauthor{\bsnm{{Fedorets}}, \binits{G.}},
\bauthor{\bsnm{{Fernique}}, \binits{P.}},
\bauthor{\bsnm{{Fienga}}, \binits{A.}},
\bauthor{\bsnm{{Figueras}}, \binits{F.}},
\bauthor{\bsnm{{Fournier}}, \binits{Y.}},
\bauthor{\bsnm{{Fouron}}, \binits{C.}},
\bauthor{\bsnm{{Fragkoudi}}, \binits{F.}},
\bauthor{\bsnm{{Gai}}, \binits{M.}},
\bauthor{\bsnm{{Garcia-Gutierrez}}, \binits{A.}},
\bauthor{\bsnm{{Garcia-Reinaldos}}, \binits{M.}},
\bauthor{\bsnm{{Garc{\'\i}a-Torres}}, \binits{M.}},
\bauthor{\bsnm{{Garofalo}}, \binits{A.}},
\bauthor{\bsnm{{Gavel}}, \binits{A.}},
\bauthor{\bsnm{{Gavras}}, \binits{P.}},
\bauthor{\bsnm{{Gerlach}}, \binits{E.}},
\bauthor{\bsnm{{Geyer}}, \binits{R.}},
\bauthor{\bsnm{{Giacobbe}}, \binits{P.}},
\bauthor{\bsnm{{Gilmore}}, \binits{G.}},
\bauthor{\bsnm{{Girona}}, \binits{S.}},
\bauthor{\bsnm{{Giuffrida}}, \binits{G.}},
\bauthor{\bsnm{{Gomel}}, \binits{R.}},
\bauthor{\bsnm{{Gomez}}, \binits{A.}},
\bauthor{\bsnm{{Gonz{\'a}lez-N{\'u}{\~n}ez}}, \binits{J.}},
\bauthor{\bsnm{{Gonz{\'a}lez-Santamar{\'\i}a}}, \binits{I.}},
\bauthor{\bsnm{{Gonz{\'a}lez-Vidal}}, \binits{J.J.}},
\bauthor{\bsnm{{Granvik}}, \binits{M.}},
\bauthor{\bsnm{{Guillout}}, \binits{P.}},
\bauthor{\bsnm{{Guiraud}}, \binits{J.}},
\bauthor{\bsnm{{Guti{\'e}rrez-S{\'a}nchez}}, \binits{R.}},
\bauthor{\bsnm{{Guy}}, \binits{L.P.}},
\bauthor{\bsnm{{Hatzidimitriou}}, \binits{D.}},
\bauthor{\bsnm{{Hauser}}, \binits{M.}},
\bauthor{\bsnm{{Haywood}}, \binits{M.}},
\bauthor{\bsnm{{Helmer}}, \binits{A.}},
\bauthor{\bsnm{{Helmi}}, \binits{A.}},
\bauthor{\bsnm{{Sarmiento}}, \binits{M.H.}},
\bauthor{\bsnm{{Hidalgo}}, \binits{S.L.}},
\bauthor{\bsnm{{Hilger}}, \binits{T.}},
\bauthor{\bsnm{{H{\l}adczuk}}, \binits{N.}},
\bauthor{\bsnm{{Hobbs}}, \binits{D.}},
\bauthor{\bsnm{{Holland}}, \binits{G.}},
\bauthor{\bsnm{{Huckle}}, \binits{H.E.}},
\bauthor{\bsnm{{Jardine}}, \binits{K.}},
\bauthor{\bsnm{{Jasniewicz}}, \binits{G.}},
\bauthor{\bsnm{{Jean-Antoine Piccolo}}, \binits{A.}},
\bauthor{\bsnm{{Jim{\'e}nez-Arranz}}, \binits{{\'O}.}},
\bauthor{\bsnm{{Jorissen}}, \binits{A.}},
\bauthor{\bsnm{{Juaristi Campillo}}, \binits{J.}},
\bauthor{\bsnm{{Julbe}}, \binits{F.}},
\bauthor{\bsnm{{Karbevska}}, \binits{L.}},
\bauthor{\bsnm{{Kervella}}, \binits{P.}},
\bauthor{\bsnm{{Khanna}}, \binits{S.}},
\bauthor{\bsnm{{Kontizas}}, \binits{M.}},
\bauthor{\bsnm{{Kordopatis}}, \binits{G.}},
\bauthor{\bsnm{{Korn}}, \binits{A.J.}},
\bauthor{\bsnm{{K{\'o}sp{\'a}l}}, \binits{{\'A}.}},
\bauthor{\bsnm{{Kostrzewa-Rutkowska}}, \binits{Z.}},
\bauthor{\bsnm{{Kruszy{\'n}ska}}, \binits{K.}},
\bauthor{\bsnm{{Kun}}, \binits{M.}},
\bauthor{\bsnm{{Laizeau}}, \binits{P.}},
\bauthor{\bsnm{{Lambert}}, \binits{S.}},
\bauthor{\bsnm{{Lanza}}, \binits{A.F.}},
\bauthor{\bsnm{{Lasne}}, \binits{Y.}},
\bauthor{\bsnm{{Le Campion}}, \binits{J.-F.}},
\bauthor{\bsnm{{Lebreton}}, \binits{Y.}},
\bauthor{\bsnm{{Lebzelter}}, \binits{T.}},
\bauthor{\bsnm{{Leccia}}, \binits{S.}},
\bauthor{\bsnm{{Leclerc}}, \binits{N.}},
\bauthor{\bsnm{{Lecoeur-Taibi}}, \binits{I.}},
\bauthor{\bsnm{{Liao}}, \binits{S.}},
\bauthor{\bsnm{{Licata}}, \binits{E.L.}},
\bauthor{\bsnm{{Lindstr{\o}m}}, \binits{H.E.P.}},
\bauthor{\bsnm{{Lister}}, \binits{T.A.}},
\bauthor{\bsnm{{Livanou}}, \binits{E.}},
\bauthor{\bsnm{{Lobel}}, \binits{A.}},
\bauthor{\bsnm{{Lorca}}, \binits{A.}},
\bauthor{\bsnm{{Loup}}, \binits{C.}},
\bauthor{\bsnm{{Madrero Pardo}}, \binits{P.}},
\bauthor{\bsnm{{Magdaleno Romeo}}, \binits{A.}},
\bauthor{\bsnm{{Managau}}, \binits{S.}},
\bauthor{\bsnm{{Mann}}, \binits{R.G.}},
\bauthor{\bsnm{{Manteiga}}, \binits{M.}},
\bauthor{\bsnm{{Marchant}}, \binits{J.M.}},
\bauthor{\bsnm{{Marconi}}, \binits{M.}},
\bauthor{\bsnm{{Marcos}}, \binits{J.}},
\bauthor{\bsnm{{Marcos Santos}}, \binits{M.M.S.}},
\bauthor{\bsnm{{Mar{\'\i}n Pina}}, \binits{D.}},
\bauthor{\bsnm{{Marinoni}}, \binits{S.}},
\bauthor{\bsnm{{Marocco}}, \binits{F.}},
\bauthor{\bsnm{{Marshall}}, \binits{D.J.}},
\bauthor{\bsnm{{Martin Polo}}, \binits{L.}},
\bauthor{\bsnm{{Mart{\'\i}n-Fleitas}}, \binits{J.M.}},
\bauthor{\bsnm{{Marton}}, \binits{G.}},
\bauthor{\bsnm{{Mary}}, \binits{N.}},
\bauthor{\bsnm{{Masip}}, \binits{A.}},
\bauthor{\bsnm{{Massari}}, \binits{D.}},
\bauthor{\bsnm{{Mastrobuono-Battisti}}, \binits{A.}},
\bauthor{\bsnm{{Mazeh}}, \binits{T.}},
\bauthor{\bsnm{{McMillan}}, \binits{P.J.}},
\bauthor{\bsnm{{Messina}}, \binits{S.}},
\bauthor{\bsnm{{Michalik}}, \binits{D.}},
\bauthor{\bsnm{{Millar}}, \binits{N.R.}},
\bauthor{\bsnm{{Mints}}, \binits{A.}},
\bauthor{\bsnm{{Molina}}, \binits{D.}},
\bauthor{\bsnm{{Molinaro}}, \binits{R.}},
\bauthor{\bsnm{{Moln{\'a}r}}, \binits{L.}},
\bauthor{\bsnm{{Monari}}, \binits{G.}},
\bauthor{\bsnm{{Mongui{\'o}}}, \binits{M.}},
\bauthor{\bsnm{{Montegriffo}}, \binits{P.}},
\bauthor{\bsnm{{Montero}}, \binits{A.}},
\bauthor{\bsnm{{Mor}}, \binits{R.}},
\bauthor{\bsnm{{Mora}}, \binits{A.}},
\bauthor{\bsnm{{Morbidelli}}, \binits{R.}},
\bauthor{\bsnm{{Morel}}, \binits{T.}},
\bauthor{\bsnm{{Morris}}, \binits{D.}},
\bauthor{\bsnm{{Muraveva}}, \binits{T.}},
\bauthor{\bsnm{{Murphy}}, \binits{C.P.}},
\bauthor{\bsnm{{Musella}}, \binits{I.}},
\bauthor{\bsnm{{Nagy}}, \binits{Z.}},
\bauthor{\bsnm{{Noval}}, \binits{L.}},
\bauthor{\bsnm{{Oca{\~n}a}}, \binits{F.}},
\bauthor{\bsnm{{Ogden}}, \binits{A.}},
\bauthor{\bsnm{{Ordenovic}}, \binits{C.}},
\bauthor{\bsnm{{Osinde}}, \binits{J.O.}},
\bauthor{\bsnm{{Pagani}}, \binits{C.}},
\bauthor{\bsnm{{Pagano}}, \binits{I.}},
\bauthor{\bsnm{{Palaversa}}, \binits{L.}},
\bauthor{\bsnm{{Palicio}}, \binits{P.A.}},
\bauthor{\bsnm{{Pallas-Quintela}}, \binits{L.}},
\bauthor{\bsnm{{Panahi}}, \binits{A.}},
\bauthor{\bsnm{{Payne-Wardenaar}}, \binits{S.}},
\bauthor{\bsnm{{Pe{\~n}alosa Esteller}}, \binits{X.}},
\bauthor{\bsnm{{Penttil{\"a}}}, \binits{A.}},
\bauthor{\bsnm{{Pichon}}, \binits{B.}},
\bauthor{\bsnm{{Piersimoni}}, \binits{A.M.}},
\bauthor{\bsnm{{Pineau}}, \binits{F.-X.}},
\bauthor{\bsnm{{Plachy}}, \binits{E.}},
\bauthor{\bsnm{{Plum}}, \binits{G.}},
\bauthor{\bsnm{{Poggio}}, \binits{E.}},
\bauthor{\bsnm{{Pr{\v{s}}a}}, \binits{A.}},
\bauthor{\bsnm{{Pulone}}, \binits{L.}},
\bauthor{\bsnm{{Racero}}, \binits{E.}},
\bauthor{\bsnm{{Ragaini}}, \binits{S.}},
\bauthor{\bsnm{{Rainer}}, \binits{M.}},
\bauthor{\bsnm{{Raiteri}}, \binits{C.M.}},
\bauthor{\bsnm{{Rambaux}}, \binits{N.}},
\bauthor{\bsnm{{Ramos}}, \binits{P.}},
\bauthor{\bsnm{{Ramos-Lerate}}, \binits{M.}},
\bauthor{\bsnm{{Re Fiorentin}}, \binits{P.}},
\bauthor{\bsnm{{Regibo}}, \binits{S.}},
\bauthor{\bsnm{{Richards}}, \binits{P.J.}},
\bauthor{\bsnm{{Rios Diaz}}, \binits{C.}},
\bauthor{\bsnm{{Ripepi}}, \binits{V.}},
\bauthor{\bsnm{{Riva}}, \binits{A.}},
\bauthor{\bsnm{{Rix}}, \binits{H.-W.}},
\bauthor{\bsnm{{Rixon}}, \binits{G.}},
\bauthor{\bsnm{{Robichon}}, \binits{N.}},
\bauthor{\bsnm{{Robin}}, \binits{A.C.}},
\bauthor{\bsnm{{Robin}}, \binits{C.}},
\bauthor{\bsnm{{Roelens}}, \binits{M.}},
\bauthor{\bsnm{{Rogues}}, \binits{H.R.O.}},
\bauthor{\bsnm{{Rohrbasser}}, \binits{L.}},
\bauthor{\bsnm{{Romero-G{\'o}mez}}, \binits{M.}},
\bauthor{\bsnm{{Rowell}}, \binits{N.}},
\bauthor{\bsnm{{Royer}}, \binits{F.}},
\bauthor{\bsnm{{Ruz Mieres}}, \binits{D.}},
\bauthor{\bsnm{{Rybicki}}, \binits{K.A.}},
\bauthor{\bsnm{{Sadowski}}, \binits{G.}},
\bauthor{\bsnm{{S{\'a}ez N{\'u}{\~n}ez}}, \binits{A.}},
\bauthor{\bsnm{{Sagrist{\`a} Sell{\'e}s}}, \binits{A.}},
\bauthor{\bsnm{{Sahlmann}}, \binits{J.}},
\bauthor{\bsnm{{Salguero}}, \binits{E.}},
\bauthor{\bsnm{{Samaras}}, \binits{N.}},
\bauthor{\bsnm{{Sanchez Gimenez}}, \binits{V.}},
\bauthor{\bsnm{{Sanna}}, \binits{N.}},
\bauthor{\bsnm{{Santove{\~n}a}}, \binits{R.}},
\bauthor{\bsnm{{Sarasso}}, \binits{M.}},
\bauthor{\bsnm{{Schultheis}}, \binits{M.}},
\bauthor{\bsnm{{Sciacca}}, \binits{E.}},
\bauthor{\bsnm{{Segol}}, \binits{M.}},
\bauthor{\bsnm{{Segovia}}, \binits{J.C.}},
\bauthor{\bsnm{{S{\'e}gransan}}, \binits{D.}},
\bauthor{\bsnm{{Semeux}}, \binits{D.}},
\bauthor{\bsnm{{Shahaf}}, \binits{S.}},
\bauthor{\bsnm{{Siddiqui}}, \binits{H.I.}},
\bauthor{\bsnm{{Siebert}}, \binits{A.}},
\bauthor{\bsnm{{Siltala}}, \binits{L.}},
\bauthor{\bsnm{{Silvelo}}, \binits{A.}},
\bauthor{\bsnm{{Slezak}}, \binits{E.}},
\bauthor{\bsnm{{Slezak}}, \binits{I.}},
\bauthor{\bsnm{{Smart}}, \binits{R.L.}},
\bauthor{\bsnm{{Snaith}}, \binits{O.N.}},
\bauthor{\bsnm{{Solano}}, \binits{E.}},
\bauthor{\bsnm{{Solitro}}, \binits{F.}},
\bauthor{\bsnm{{Souami}}, \binits{D.}},
\bauthor{\bsnm{{Souchay}}, \binits{J.}},
\bauthor{\bsnm{{Spagna}}, \binits{A.}},
\bauthor{\bsnm{{Spina}}, \binits{L.}},
\bauthor{\bsnm{{Spoto}}, \binits{F.}},
\bauthor{\bsnm{{Steele}}, \binits{I.A.}},
\bauthor{\bsnm{{Steidelm{\"u}ller}}, \binits{H.}},
\bauthor{\bsnm{{Stephenson}}, \binits{C.A.}},
\bauthor{\bsnm{{S{\"u}veges}}, \binits{M.}},
\bauthor{\bsnm{{Surdej}}, \binits{J.}},
\bauthor{\bsnm{{Szabados}}, \binits{L.}},
\bauthor{\bsnm{{Szegedi-Elek}}, \binits{E.}},
\bauthor{\bsnm{{Taris}}, \binits{F.}},
\bauthor{\bsnm{{Taylor}}, \binits{M.B.}},
\bauthor{\bsnm{{Teixeira}}, \binits{R.}},
\bauthor{\bsnm{{Tolomei}}, \binits{L.}},
\bauthor{\bsnm{{Tonello}}, \binits{N.}},
\bauthor{\bsnm{{Torra}}, \binits{F.}},
\bauthor{\bsnm{{Torra}}, \binits{J.}},
\bauthor{\bsnm{{Torralba Elipe}}, \binits{G.}},
\bauthor{\bsnm{{Trabucchi}}, \binits{M.}},
\bauthor{\bsnm{{Tsounis}}, \binits{A.T.}},
\bauthor{\bsnm{{Turon}}, \binits{C.}},
\bauthor{\bsnm{{Ulla}}, \binits{A.}},
\bauthor{\bsnm{{Unger}}, \binits{N.}},
\bauthor{\bsnm{{Vaillant}}, \binits{M.V.}},
\bauthor{\bsnm{{van Dillen}}, \binits{E.}},
\bauthor{\bsnm{{van Reeven}}, \binits{W.}},
\bauthor{\bsnm{{Vanel}}, \binits{O.}},
\bauthor{\bsnm{{Vecchiato}}, \binits{A.}},
\bauthor{\bsnm{{Viala}}, \binits{Y.}},
\bauthor{\bsnm{{Vicente}}, \binits{D.}},
\bauthor{\bsnm{{Voutsinas}}, \binits{S.}},
\bauthor{\bsnm{{Weiler}}, \binits{M.}},
\bauthor{\bsnm{{Wevers}}, \binits{T.}},
\bauthor{\bsnm{{Wyrzykowski}}, \binits{{\L}.}},
\bauthor{\bsnm{{Yoldas}}, \binits{A.}},
\bauthor{\bsnm{{Yvard}}, \binits{P.}},
\bauthor{\bsnm{{Zhao}}, \binits{H.}},
\bauthor{\bsnm{{Zorec}}, \binits{J.}},
\bauthor{\bsnm{{Zucker}}, \binits{S.}},
\bauthor{\bsnm{{Zwitter}}, \binits{T.}}:
\batitle{{Gaia Data Release 3. Summary of the content and survey properties}}.
\bjtitle{\aap}
\bvolume{674},
\bfpage{1}
(\byear{2023})
\doiurl{10.1051/0004-6361/202243940}
{\href{https://arxiv.org/abs/2208.00211}{{arXiv:2208.00211}}}
{[astro-ph.GA]}
\end{barticle}
\endbibitem

\bibitem[\protect\citeauthoryear{Brammer}{2022}]{grizliphot}
\begin{botherref}
\oauthor{\bsnm{Brammer}, \binits{G.}}:
{Preliminary updates to the NIRCam photometric calibration}.
Zenodo
(2022).
\doiurl{10.5281/zenodo.7143382} .
\url{https://doi.org/10.5281/zenodo.7143382}
\end{botherref}
\endbibitem

\bibitem[\protect\citeauthoryear{{Martis} et~al.}{2024}]{martis2024}
\begin{botherref}
\oauthor{\bsnm{{Martis}}, \binits{N.S.}},
\oauthor{\bsnm{{Sarrouh}}, \binits{G.T.E.}},
\oauthor{\bsnm{{Willott}}, \binits{C.J.}},
\oauthor{\bsnm{{Abraham}}, \binits{R.}},
\oauthor{\bsnm{{Asada}}, \binits{Y.}},
\oauthor{\bsnm{{Brada{\v{c}}}}, \binits{M.}},
\oauthor{\bsnm{{Brammer}}, \binits{G.}},
\oauthor{\bsnm{{Harshan}}, \binits{A.}},
\oauthor{\bsnm{{Muzzin}}, \binits{A.}},
\oauthor{\bsnm{{Noirot}}, \binits{G.}},
\oauthor{\bsnm{{Sawicki}}, \binits{M.}},
\oauthor{\bsnm{{Rihtar{\v{s}}i{\v{c}}}}, \binits{G.}}:
{Modelling and Subtracting Diffuse Cluster Light in JWST Images: A Relation between the Spatial Distribution of Globular Clusters, Dwarf Galaxies, and Intracluster Light in the Lensing Cluster SMACS 0723}.
arXiv e-prints,
2401--01945
(2024)
\doiurl{10.48550/arXiv.2401.01945}
{\href{https://arxiv.org/abs/2401.01945}{{arXiv:2401.01945}}}
{[astro-ph.GA]}
\end{botherref}
\endbibitem

\bibitem[\protect\citeauthoryear{{Willott} et~al.}{2023}]{willott23}
\begin{botherref}
\oauthor{\bsnm{{Willott}}, \binits{C.J.}},
\oauthor{\bsnm{{Desprez}}, \binits{G.}},
\oauthor{\bsnm{{Asada}}, \binits{Y.}},
\oauthor{\bsnm{{Sarrouh}}, \binits{G.T.E.}},
\oauthor{\bsnm{{Abraham}}, \binits{R.}},
\oauthor{\bsnm{{Brada{\v{c}}}}, \binits{M.}},
\oauthor{\bsnm{{Brammer}}, \binits{G.}},
\oauthor{\bsnm{{Estrada-Carpenter}}, \binits{V.}},
\oauthor{\bsnm{{Iyer}}, \binits{K.G.}},
\oauthor{\bsnm{{Martis}}, \binits{N.S.}},
\oauthor{\bsnm{{Matharu}}, \binits{J.}},
\oauthor{\bsnm{{Mowla}}, \binits{L.}},
\oauthor{\bsnm{{Muzzin}}, \binits{A.}},
\oauthor{\bsnm{{Noirot}}, \binits{G.}},
\oauthor{\bsnm{{Sawicki}}, \binits{M.}},
\oauthor{\bsnm{{Strait}}, \binits{V.}},
\oauthor{\bsnm{{Rihtar{\v{s}}i{\v{c}}}}, \binits{G.}},
\oauthor{\bsnm{{Withers}}, \binits{S.}}:
{A Steep Decline in the Galaxy Space Density Beyond Redshift 9 in the CANUCS UV Luminosity Function}.
arXiv e-prints,
2311--12234
(2023)
\doiurl{10.48550/arXiv.2311.12234}
{\href{https://arxiv.org/abs/2311.12234}{{arXiv:2311.12234}}}
{[astro-ph.GA]}
\end{botherref}
\endbibitem

\bibitem[\protect\citeauthoryear{{Peng} et~al.}{2010}]{peng10}
\begin{barticle}
\bauthor{\bsnm{{Peng}}, \binits{Y.-j.}},
\bauthor{\bsnm{{Lilly}}, \binits{S.J.}},
\bauthor{\bsnm{{Kova{\v c}}}, \binits{K.}},
\bauthor{\bsnm{{Bolzonella}}, \binits{M.}},
\bauthor{\bsnm{{Pozzetti}}, \binits{L.}},
\bauthor{\bsnm{{Renzini}}, \binits{A.}},
\bauthor{\bsnm{{Zamorani}}, \binits{G.}},
\bauthor{\bsnm{{Ilbert}}, \binits{O.}},
\bauthor{\bsnm{{Knobel}}, \binits{C.}},
\bauthor{\bsnm{{Iovino}}, \binits{A.}},
\bauthor{\bsnm{{Maier}}, \binits{C.}},
\bauthor{\bsnm{{Cucciati}}, \binits{O.}},
\bauthor{\bsnm{{Tasca}}, \binits{L.}},
\bauthor{\bsnm{{Carollo}}, \binits{C.M.}},
\bauthor{\bsnm{{Silverman}}, \binits{J.}},
\bauthor{\bsnm{{Kampczyk}}, \binits{P.}},
\bauthor{\bsnm{{de Ravel}}, \binits{L.}},
\bauthor{\bsnm{{Sanders}}, \binits{D.}},
\bauthor{\bsnm{{Scoville}}, \binits{N.}},
\bauthor{\bsnm{{Contini}}, \binits{T.}},
\bauthor{\bsnm{{Mainieri}}, \binits{V.}},
\bauthor{\bsnm{{Scodeggio}}, \binits{M.}},
\bauthor{\bsnm{{Kneib}}, \binits{J.-P.}},
\bauthor{\bsnm{{Le F{\`e}vre}}, \binits{O.}},
\bauthor{\bsnm{{Bardelli}}, \binits{S.}},
\bauthor{\bsnm{{Bongiorno}}, \binits{A.}},
\bauthor{\bsnm{{Caputi}}, \binits{K.}},
\bauthor{\bsnm{{Coppa}}, \binits{G.}},
\bauthor{\bsnm{{de la Torre}}, \binits{S.}},
\bauthor{\bsnm{{Franzetti}}, \binits{P.}},
\bauthor{\bsnm{{Garilli}}, \binits{B.}},
\bauthor{\bsnm{{Lamareille}}, \binits{F.}},
\bauthor{\bsnm{{Le Borgne}}, \binits{J.-F.}},
\bauthor{\bsnm{{Le Brun}}, \binits{V.}},
\bauthor{\bsnm{{Mignoli}}, \binits{M.}},
\bauthor{\bsnm{{Perez Montero}}, \binits{E.}},
\bauthor{\bsnm{{Pello}}, \binits{R.}},
\bauthor{\bsnm{{Ricciardelli}}, \binits{E.}},
\bauthor{\bsnm{{Tanaka}}, \binits{M.}},
\bauthor{\bsnm{{Tresse}}, \binits{L.}},
\bauthor{\bsnm{{Vergani}}, \binits{D.}},
\bauthor{\bsnm{{Welikala}}, \binits{N.}},
\bauthor{\bsnm{{Zucca}}, \binits{E.}},
\bauthor{\bsnm{{Oesch}}, \binits{P.}},
\bauthor{\bsnm{{Abbas}}, \binits{U.}},
\bauthor{\bsnm{{Barnes}}, \binits{L.}},
\bauthor{\bsnm{{Bordoloi}}, \binits{R.}},
\bauthor{\bsnm{{Bottini}}, \binits{D.}},
\bauthor{\bsnm{{Cappi}}, \binits{A.}},
\bauthor{\bsnm{{Cassata}}, \binits{P.}},
\bauthor{\bsnm{{Cimatti}}, \binits{A.}},
\bauthor{\bsnm{{Fumana}}, \binits{M.}},
\bauthor{\bsnm{{Hasinger}}, \binits{G.}},
\bauthor{\bsnm{{Koekemoer}}, \binits{A.}},
\bauthor{\bsnm{{Leauthaud}}, \binits{A.}},
\bauthor{\bsnm{{Maccagni}}, \binits{D.}},
\bauthor{\bsnm{{Marinoni}}, \binits{C.}},
\bauthor{\bsnm{{McCracken}}, \binits{H.}},
\bauthor{\bsnm{{Memeo}}, \binits{P.}},
\bauthor{\bsnm{{Meneux}}, \binits{B.}},
\bauthor{\bsnm{{Nair}}, \binits{P.}},
\bauthor{\bsnm{{Porciani}}, \binits{C.}},
\bauthor{\bsnm{{Presotto}}, \binits{V.}},
\bauthor{\bsnm{{Scaramella}}, \binits{R.}}:
\batitle{{Mass and Environment as Drivers of Galaxy Evolution in SDSS and zCOSMOS and the Origin of the Schechter Function}}.
\bjtitle{\apj}
\bvolume{721},
\bfpage{193}--\blpage{221}
(\byear{2010})
\doiurl{10.1088/0004-637X/721/1/193}
{\href{https://arxiv.org/abs/1003.4747}{{arXiv:1003.4747}}}
{[astro-ph.CO]}
\end{barticle}
\endbibitem

\bibitem[\protect\citeauthoryear{{Schlafly} and {Finkbeiner}}{2011}]{schlafly11}
\begin{barticle}
\bauthor{\bsnm{{Schlafly}}, \binits{E.F.}},
\bauthor{\bsnm{{Finkbeiner}}, \binits{D.P.}}:
\batitle{{Measuring Reddening with Sloan Digital Sky Survey Stellar Spectra and Recalibrating SFD}}.
\bjtitle{\apj}
\bvolume{737}(\bissue{2}),
\bfpage{103}
(\byear{2011})
\doiurl{10.1088/0004-637X/737/2/103}
{\href{https://arxiv.org/abs/1012.4804}{{arXiv:1012.4804}}}
{[astro-ph.GA]}
\end{barticle}
\endbibitem

\bibitem[\protect\citeauthoryear{{Fitzpatrick}}{1999}]{fitzpatrick99}
\begin{barticle}
\bauthor{\bsnm{{Fitzpatrick}}, \binits{E.L.}}:
\batitle{{Correcting for the Effects of Interstellar Extinction}}.
\bjtitle{\pasp}
\bvolume{111}(\bissue{755}),
\bfpage{63}--\blpage{75}
(\byear{1999})
\doiurl{10.1086/316293}
{\href{https://arxiv.org/abs/astro-ph/9809387}{{arXiv:astro-ph/9809387}}}
{[astro-ph]}
\end{barticle}
\endbibitem

\bibitem[\protect\citeauthoryear{Brammer}{2022}]{brammer2022msaexp}
\begin{botherref}
\oauthor{\bsnm{Brammer}, \binits{G.}}:
msaexp: Nirspec analyis tools.
Zenodo
(2022)
\end{botherref}
\endbibitem

\bibitem[\protect\citeauthoryear{{Horne}}{1986}]{Horne_1986}
\begin{barticle}
\bauthor{\bsnm{{Horne}}, \binits{K.}}:
\batitle{{An optimal extraction algorithm for CCD spectroscopy.}}
\bjtitle{\pasp}
\bvolume{98},
\bfpage{609}--\blpage{617}
(\byear{1986})
\doiurl{10.1086/131801}
\end{barticle}
\endbibitem

\bibitem[\protect\citeauthoryear{{Chatzikos} et~al.}{2023}]{Chatzikos2023}
\begin{barticle}
\bauthor{\bsnm{{Chatzikos}}, \binits{M.}},
\bauthor{\bsnm{{Bianchi}}, \binits{S.}},
\bauthor{\bsnm{{Camilloni}}, \binits{F.}},
\bauthor{\bsnm{{Chakraborty}}, \binits{P.}},
\bauthor{\bsnm{{Gunasekera}}, \binits{C.M.}},
\bauthor{\bsnm{{Guzm{\'a}n}}, \binits{F.}},
\bauthor{\bsnm{{Milby}}, \binits{J.S.}},
\bauthor{\bsnm{{Sarkar}}, \binits{A.}},
\bauthor{\bsnm{{Shaw}}, \binits{G.}},
\bauthor{\bsnm{{van Hoof}}, \binits{P.A.M.}},
\bauthor{\bsnm{{Ferland}}, \binits{G.J.}}:
\batitle{{The 2023 Release of Cloudy}}.
\bjtitle{\rmxaa}
\bvolume{59},
\bfpage{327}--\blpage{343}
(\byear{2023})
\doiurl{10.22201/ia.01851101p.2023.59.02.12}
{\href{https://arxiv.org/abs/2308.06396}{{arXiv:2308.06396}}}
{[astro-ph.GA]}
\end{barticle}
\endbibitem

\bibitem[\protect\citeauthoryear{{Osterbrock} and {Ferland}}{2006}]{Osterbrock2006}
\begin{bbook}
\bauthor{\bsnm{{Osterbrock}}, \binits{D.E.}},
\bauthor{\bsnm{{Ferland}}, \binits{G.J.}}:
\bbtitle{{Astrophysics of Gaseous Nebulae and Active Galactic Nuclei}},
(\byear{2006})
\end{bbook}
\endbibitem

\bibitem[\protect\citeauthoryear{{Izotov} et~al.}{2006}]{Izotov2006}
\begin{barticle}
\bauthor{\bsnm{{Izotov}}, \binits{Y.I.}},
\bauthor{\bsnm{{Stasi{\'n}ska}}, \binits{G.}},
\bauthor{\bsnm{{Meynet}}, \binits{G.}},
\bauthor{\bsnm{{Guseva}}, \binits{N.G.}},
\bauthor{\bsnm{{Thuan}}, \binits{T.X.}}:
\batitle{{The chemical composition of metal-poor emission-line galaxies in the Data Release 3 of the Sloan Digital Sky Survey}}.
\bjtitle{\aap}
\bvolume{448}(\bissue{3}),
\bfpage{955}--\blpage{970}
(\byear{2006})
\doiurl{10.1051/0004-6361:20053763}
{\href{https://arxiv.org/abs/astro-ph/0511644}{{arXiv:astro-ph/0511644}}}
{[astro-ph]}
\end{barticle}
\endbibitem

\bibitem[\protect\citeauthoryear{{Berg} et~al.}{2021}]{Berg2021}
\begin{barticle}
\bauthor{\bsnm{{Berg}}, \binits{D.A.}},
\bauthor{\bsnm{{Chisholm}}, \binits{J.}},
\bauthor{\bsnm{{Erb}}, \binits{D.K.}},
\bauthor{\bsnm{{Skillman}}, \binits{E.D.}},
\bauthor{\bsnm{{Pogge}}, \binits{R.W.}},
\bauthor{\bsnm{{Olivier}}, \binits{G.M.}}:
\batitle{{Characterizing Extreme Emission-line Galaxies. I. A Four-zone Ionization Model for Very High-ionization Emission}}.
\bjtitle{\apj}
\bvolume{922}(\bissue{2}),
\bfpage{170}
(\byear{2021})
\doiurl{10.3847/1538-4357/ac141b}
{\href{https://arxiv.org/abs/2105.12765}{{arXiv:2105.12765}}}
{[astro-ph.GA]}
\end{barticle}
\endbibitem

\bibitem[\protect\citeauthoryear{{Kewley} and {Dopita}}{2002}]{Kewley2002}
\begin{barticle}
\bauthor{\bsnm{{Kewley}}, \binits{L.J.}},
\bauthor{\bsnm{{Dopita}}, \binits{M.A.}}:
\batitle{{Using Strong Lines to Estimate Abundances in Extragalactic H II Regions and Starburst Galaxies}}.
\bjtitle{\apjs}
\bvolume{142}(\bissue{1}),
\bfpage{35}--\blpage{52}
(\byear{2002})
\doiurl{10.1086/341326}
{\href{https://arxiv.org/abs/astro-ph/0206495}{{arXiv:astro-ph/0206495}}}
{[astro-ph]}
\end{barticle}
\endbibitem

\bibitem[\protect\citeauthoryear{{Levesque} and {Richardson}}{2014}]{Levesque2014}
\begin{barticle}
\bauthor{\bsnm{{Levesque}}, \binits{E.M.}},
\bauthor{\bsnm{{Richardson}}, \binits{M.L.A.}}:
\batitle{{[Ne III]/[O II] as an Ionization Parameter Diagnostic in Star-Forming Galaxies}}.
\bjtitle{\apj}
\bvolume{780}(\bissue{1}),
\bfpage{100}
(\byear{2014})
\doiurl{10.1088/0004-637X/780/1/100}
{\href{https://arxiv.org/abs/1309.0513}{{arXiv:1309.0513}}}
{[astro-ph.GA]}
\end{barticle}
\endbibitem

\bibitem[\protect\citeauthoryear{{Sanders} et~al.}{2023}]{Sanders2023}
\begin{botherref}
\oauthor{\bsnm{{Sanders}}, \binits{R.L.}},
\oauthor{\bsnm{{Shapley}}, \binits{A.E.}},
\oauthor{\bsnm{{Topping}}, \binits{M.W.}},
\oauthor{\bsnm{{Reddy}}, \binits{N.A.}},
\oauthor{\bsnm{{Brammer}}, \binits{G.B.}}:
{Direct T\_e-based Metallicities of z=2-9 Galaxies with JWST/NIRSpec: Empirical Metallicity Calibrations Applicable from Reionization to Cosmic Noon}.
arXiv e-prints,
2303--08149
(2023)
\doiurl{10.48550/arXiv.2303.08149}
{\href{https://arxiv.org/abs/2303.08149}{{arXiv:2303.08149}}}
{[astro-ph.GA]}
\end{botherref}
\endbibitem

\bibitem[\protect\citeauthoryear{{Ahn} et~al.}{2014}]{Ahn2014}
\begin{barticle}
\bauthor{\bsnm{{Ahn}}, \binits{C.P.}},
\bauthor{\bsnm{{Alexandroff}}, \binits{R.}},
\bauthor{\bsnm{{Allende Prieto}}, \binits{C.}},
\bauthor{\bsnm{{Anders}}, \binits{F.}},
\bauthor{\bsnm{{Anderson}}, \binits{S.F.}},
\bauthor{\bsnm{{Anderton}}, \binits{T.}},
\bauthor{\bsnm{{Andrews}}, \binits{B.H.}},
\bauthor{\bsnm{{Aubourg}}, \binits{{\'E}.}},
\bauthor{\bsnm{{Bailey}}, \binits{S.}},
\bauthor{\bsnm{{Bastien}}, \binits{F.A.}},
\bauthor{\bsnm{{Bautista}}, \binits{J.E.}},
\bauthor{\bsnm{{Beers}}, \binits{T.C.}},
\bauthor{\bsnm{{Beifiori}}, \binits{A.}},
\bauthor{\bsnm{{Bender}}, \binits{C.F.}},
\bauthor{\bsnm{{Berlind}}, \binits{A.A.}},
\bauthor{\bsnm{{Beutler}}, \binits{F.}},
\bauthor{\bsnm{{Bhardwaj}}, \binits{V.}},
\bauthor{\bsnm{{Bird}}, \binits{J.C.}},
\bauthor{\bsnm{{Bizyaev}}, \binits{D.}},
\bauthor{\bsnm{{Blake}}, \binits{C.H.}},
\bauthor{\bsnm{{Blanton}}, \binits{M.R.}},
\bauthor{\bsnm{{Blomqvist}}, \binits{M.}},
\bauthor{\bsnm{{Bochanski}}, \binits{J.J.}},
\bauthor{\bsnm{{Bolton}}, \binits{A.S.}},
\bauthor{\bsnm{{Borde}}, \binits{A.}},
\bauthor{\bsnm{{Bovy}}, \binits{J.}},
\bauthor{\bsnm{{Shelden Bradley}}, \binits{A.}},
\bauthor{\bsnm{{Brandt}}, \binits{W.N.}},
\bauthor{\bsnm{{Brauer}}, \binits{D.}},
\bauthor{\bsnm{{Brinkmann}}, \binits{J.}},
\bauthor{\bsnm{{Brownstein}}, \binits{J.R.}},
\bauthor{\bsnm{{Busca}}, \binits{N.G.}},
\bauthor{\bsnm{{Carithers}}, \binits{W.}},
\bauthor{\bsnm{{Carlberg}}, \binits{J.K.}},
\bauthor{\bsnm{{Carnero}}, \binits{A.R.}},
\bauthor{\bsnm{{Carr}}, \binits{M.A.}},
\bauthor{\bsnm{{Chiappini}}, \binits{C.}},
\bauthor{\bsnm{{Chojnowski}}, \binits{S.D.}},
\bauthor{\bsnm{{Chuang}}, \binits{C.-H.}},
\bauthor{\bsnm{{Comparat}}, \binits{J.}},
\bauthor{\bsnm{{Crepp}}, \binits{J.R.}},
\bauthor{\bsnm{{Cristiani}}, \binits{S.}},
\bauthor{\bsnm{{Croft}}, \binits{R.A.C.}},
\bauthor{\bsnm{{Cuesta}}, \binits{A.J.}},
\bauthor{\bsnm{{Cunha}}, \binits{K.}},
\bauthor{\bsnm{{da Costa}}, \binits{L.N.}},
\bauthor{\bsnm{{Dawson}}, \binits{K.S.}},
\bauthor{\bsnm{{De Lee}}, \binits{N.}},
\bauthor{\bsnm{{Dean}}, \binits{J.D.R.}},
\bauthor{\bsnm{{Delubac}}, \binits{T.}},
\bauthor{\bsnm{{Deshpande}}, \binits{R.}},
\bauthor{\bsnm{{Dhital}}, \binits{S.}},
\bauthor{\bsnm{{Ealet}}, \binits{A.}},
\bauthor{\bsnm{{Ebelke}}, \binits{G.L.}},
\bauthor{\bsnm{{Edmondson}}, \binits{E.M.}},
\bauthor{\bsnm{{Eisenstein}}, \binits{D.J.}},
\bauthor{\bsnm{{Epstein}}, \binits{C.R.}},
\bauthor{\bsnm{{Escoffier}}, \binits{S.}},
\bauthor{\bsnm{{Esposito}}, \binits{M.}},
\bauthor{\bsnm{{Evans}}, \binits{M.L.}},
\bauthor{\bsnm{{Fabbian}}, \binits{D.}},
\bauthor{\bsnm{{Fan}}, \binits{X.}},
\bauthor{\bsnm{{Favole}}, \binits{G.}},
\bauthor{\bsnm{{Femen{\'\i}a Castell{\'a}}}, \binits{B.}},
\bauthor{\bsnm{{Fern{\'a}ndez Alvar}}, \binits{E.}},
\bauthor{\bsnm{{Feuillet}}, \binits{D.}},
\bauthor{\bsnm{{Filiz Ak}}, \binits{N.}},
\bauthor{\bsnm{{Finley}}, \binits{H.}},
\bauthor{\bsnm{{Fleming}}, \binits{S.W.}},
\bauthor{\bsnm{{Font-Ribera}}, \binits{A.}},
\bauthor{\bsnm{{Frinchaboy}}, \binits{P.M.}},
\bauthor{\bsnm{{Galbraith-Frew}}, \binits{J.G.}},
\bauthor{\bsnm{{Garc{\'\i}a-Hern{\'a}ndez}}, \binits{D.A.}},
\bauthor{\bsnm{{Garc{\'\i}a P{\'e}rez}}, \binits{A.E.}},
\bauthor{\bsnm{{Ge}}, \binits{J.}},
\bauthor{\bsnm{{G{\'e}nova-Santos}}, \binits{R.}},
\bauthor{\bsnm{{Gillespie}}, \binits{B.A.}},
\bauthor{\bsnm{{Girardi}}, \binits{L.}},
\bauthor{\bsnm{{Gonz{\'a}lez Hern{\'a}ndez}}, \binits{J.I.}},
\bauthor{\bsnm{{Gott}}, \binits{I.} \bsuffix{J.~Richard}},
\bauthor{\bsnm{{Gunn}}, \binits{J.E.}},
\bauthor{\bsnm{{Guo}}, \binits{H.}},
\bauthor{\bsnm{{Halverson}}, \binits{S.}},
\bauthor{\bsnm{{Harding}}, \binits{P.}},
\bauthor{\bsnm{{Harris}}, \binits{D.W.}},
\bauthor{\bsnm{{Hasselquist}}, \binits{S.}},
\bauthor{\bsnm{{Hawley}}, \binits{S.L.}},
\bauthor{\bsnm{{Hayden}}, \binits{M.}},
\bauthor{\bsnm{{Hearty}}, \binits{F.R.}},
\bauthor{\bsnm{{Herrero Dav{\'o}}}, \binits{A.}},
\bauthor{\bsnm{{Ho}}, \binits{S.}},
\bauthor{\bsnm{{Hogg}}, \binits{D.W.}},
\bauthor{\bsnm{{Holtzman}}, \binits{J.A.}},
\bauthor{\bsnm{{Honscheid}}, \binits{K.}},
\bauthor{\bsnm{{Huehnerhoff}}, \binits{J.}},
\bauthor{\bsnm{{Ivans}}, \binits{I.I.}},
\bauthor{\bsnm{{Jackson}}, \binits{K.M.}},
\bauthor{\bsnm{{Jiang}}, \binits{P.}},
\bauthor{\bsnm{{Johnson}}, \binits{J.A.}},
\bauthor{\bsnm{{Kinemuchi}}, \binits{K.}},
\bauthor{\bsnm{{Kirkby}}, \binits{D.}},
\bauthor{\bsnm{{Klaene}}, \binits{M.A.}},
\bauthor{\bsnm{{Kneib}}, \binits{J.-P.}},
\bauthor{\bsnm{{Koesterke}}, \binits{L.}},
\bauthor{\bsnm{{Lan}}, \binits{T.-W.}},
\bauthor{\bsnm{{Lang}}, \binits{D.}},
\bauthor{\bsnm{{Le Goff}}, \binits{J.-M.}},
\bauthor{\bsnm{{Leauthaud}}, \binits{A.}},
\bauthor{\bsnm{{Lee}}, \binits{K.-G.}},
\bauthor{\bsnm{{Lee}}, \binits{Y.S.}},
\bauthor{\bsnm{{Long}}, \binits{D.C.}},
\bauthor{\bsnm{{Loomis}}, \binits{C.P.}},
\bauthor{\bsnm{{Lucatello}}, \binits{S.}},
\bauthor{\bsnm{{Lupton}}, \binits{R.H.}},
\bauthor{\bsnm{{Ma}}, \binits{B.}},
\bauthor{\bsnm{{Mack}}, \binits{I.} \bsuffix{Claude~E.}},
\bauthor{\bsnm{{Mahadevan}}, \binits{S.}},
\bauthor{\bsnm{{Maia}}, \binits{M.A.G.}},
\bauthor{\bsnm{{Majewski}}, \binits{S.R.}},
\bauthor{\bsnm{{Malanushenko}}, \binits{E.}},
\bauthor{\bsnm{{Malanushenko}}, \binits{V.}},
\bauthor{\bsnm{{Manchado}}, \binits{A.}},
\bauthor{\bsnm{{Manera}}, \binits{M.}},
\bauthor{\bsnm{{Maraston}}, \binits{C.}},
\bauthor{\bsnm{{Margala}}, \binits{D.}},
\bauthor{\bsnm{{Martell}}, \binits{S.L.}},
\bauthor{\bsnm{{Masters}}, \binits{K.L.}},
\bauthor{\bsnm{{McBride}}, \binits{C.K.}},
\bauthor{\bsnm{{McGreer}}, \binits{I.D.}},
\bauthor{\bsnm{{McMahon}}, \binits{R.G.}},
\bauthor{\bsnm{{M{\'e}nard}}, \binits{B.}},
\bauthor{\bsnm{{M{\'e}sz{\'a}ros}}, \binits{S.}},
\bauthor{\bsnm{{Miralda-Escud{\'e}}}, \binits{J.}},
\bauthor{\bsnm{{Miyatake}}, \binits{H.}},
\bauthor{\bsnm{{Montero-Dorta}}, \binits{A.D.}},
\bauthor{\bsnm{{Montesano}}, \binits{F.}},
\bauthor{\bsnm{{More}}, \binits{S.}},
\bauthor{\bsnm{{Morrison}}, \binits{H.L.}},
\bauthor{\bsnm{{Muna}}, \binits{D.}},
\bauthor{\bsnm{{Munn}}, \binits{J.A.}},
\bauthor{\bsnm{{Myers}}, \binits{A.D.}},
\bauthor{\bsnm{{Nguyen}}, \binits{D.C.}},
\bauthor{\bsnm{{Nichol}}, \binits{R.C.}},
\bauthor{\bsnm{{Nidever}}, \binits{D.L.}},
\bauthor{\bsnm{{Noterdaeme}}, \binits{P.}},
\bauthor{\bsnm{{Nuza}}, \binits{S.E.}},
\bauthor{\bsnm{{O'Connell}}, \binits{J.E.}},
\bauthor{\bsnm{{O'Connell}}, \binits{R.W.}},
\bauthor{\bsnm{{O'Connell}}, \binits{R.}},
\bauthor{\bsnm{{Olmstead}}, \binits{M.D.}},
\bauthor{\bsnm{{Oravetz}}, \binits{D.J.}},
\bauthor{\bsnm{{Owen}}, \binits{R.}},
\bauthor{\bsnm{{Padmanabhan}}, \binits{N.}},
\bauthor{\bsnm{{Palanque-Delabrouille}}, \binits{N.}},
\bauthor{\bsnm{{Pan}}, \binits{K.}},
\bauthor{\bsnm{{Parejko}}, \binits{J.K.}},
\bauthor{\bsnm{{Parihar}}, \binits{P.}},
\bauthor{\bsnm{{P{\^a}ris}}, \binits{I.}},
\bauthor{\bsnm{{Pepper}}, \binits{J.}},
\bauthor{\bsnm{{Percival}}, \binits{W.J.}},
\bauthor{\bsnm{{P{\'e}rez-R{\`a}fols}}, \binits{I.}},
\bauthor{\bsnm{{Dotto Perottoni}}, \binits{H.}},
\bauthor{\bsnm{{Petitjean}}, \binits{P.}},
\bauthor{\bsnm{{Pieri}}, \binits{M.M.}},
\bauthor{\bsnm{{Pinsonneault}}, \binits{M.H.}},
\bauthor{\bsnm{{Prada}}, \binits{F.}},
\bauthor{\bsnm{{Price-Whelan}}, \binits{A.M.}},
\bauthor{\bsnm{{Raddick}}, \binits{M.J.}},
\bauthor{\bsnm{{Rahman}}, \binits{M.}},
\bauthor{\bsnm{{Rebolo}}, \binits{R.}},
\bauthor{\bsnm{{Reid}}, \binits{B.A.}},
\bauthor{\bsnm{{Richards}}, \binits{J.C.}},
\bauthor{\bsnm{{Riffel}}, \binits{R.}},
\bauthor{\bsnm{{Robin}}, \binits{A.C.}},
\bauthor{\bsnm{{Rocha-Pinto}}, \binits{H.J.}},
\bauthor{\bsnm{{Rockosi}}, \binits{C.M.}},
\bauthor{\bsnm{{Roe}}, \binits{N.A.}},
\bauthor{\bsnm{{Ross}}, \binits{A.J.}},
\bauthor{\bsnm{{Ross}}, \binits{N.P.}},
\bauthor{\bsnm{{Rossi}}, \binits{G.}},
\bauthor{\bsnm{{Roy}}, \binits{A.}},
\bauthor{\bsnm{{Rubi{\~n}o-Martin}}, \binits{J.A.}},
\bauthor{\bsnm{{Sabiu}}, \binits{C.G.}},
\bauthor{\bsnm{{S{\'a}nchez}}, \binits{A.G.}},
\bauthor{\bsnm{{Santiago}}, \binits{B.}},
\bauthor{\bsnm{{Sayres}}, \binits{C.}},
\bauthor{\bsnm{{Schiavon}}, \binits{R.P.}},
\bauthor{\bsnm{{Schlegel}}, \binits{D.J.}},
\bauthor{\bsnm{{Schlesinger}}, \binits{K.J.}},
\bauthor{\bsnm{{Schmidt}}, \binits{S.J.}},
\bauthor{\bsnm{{Schneider}}, \binits{D.P.}},
\bauthor{\bsnm{{Schultheis}}, \binits{M.}},
\bauthor{\bsnm{{Sellgren}}, \binits{K.}},
\bauthor{\bsnm{{Seo}}, \binits{H.-J.}},
\bauthor{\bsnm{{Shen}}, \binits{Y.}},
\bauthor{\bsnm{{Shetrone}}, \binits{M.}},
\bauthor{\bsnm{{Shu}}, \binits{Y.}},
\bauthor{\bsnm{{Simmons}}, \binits{A.E.}},
\bauthor{\bsnm{{Skrutskie}}, \binits{M.F.}},
\bauthor{\bsnm{{Slosar}}, \binits{A.}},
\bauthor{\bsnm{{Smith}}, \binits{V.V.}},
\bauthor{\bsnm{{Snedden}}, \binits{S.A.}},
\bauthor{\bsnm{{Sobeck}}, \binits{J.S.}},
\bauthor{\bsnm{{Sobreira}}, \binits{F.}},
\bauthor{\bsnm{{Stassun}}, \binits{K.G.}},
\bauthor{\bsnm{{Steinmetz}}, \binits{M.}},
\bauthor{\bsnm{{Strauss}}, \binits{M.A.}},
\bauthor{\bsnm{{Streblyanska}}, \binits{A.}},
\bauthor{\bsnm{{Suzuki}}, \binits{N.}},
\bauthor{\bsnm{{Swanson}}, \binits{M.E.C.}},
\bauthor{\bsnm{{Terrien}}, \binits{R.C.}},
\bauthor{\bsnm{{Thakar}}, \binits{A.R.}},
\bauthor{\bsnm{{Thomas}}, \binits{D.}},
\bauthor{\bsnm{{Thompson}}, \binits{B.A.}},
\bauthor{\bsnm{{Tinker}}, \binits{J.L.}},
\bauthor{\bsnm{{Tojeiro}}, \binits{R.}},
\bauthor{\bsnm{{Troup}}, \binits{N.W.}},
\bauthor{\bsnm{{Vandenberg}}, \binits{J.}},
\bauthor{\bsnm{{Vargas Maga{\~n}a}}, \binits{M.}},
\bauthor{\bsnm{{Viel}}, \binits{M.}},
\bauthor{\bsnm{{Vogt}}, \binits{N.P.}},
\bauthor{\bsnm{{Wake}}, \binits{D.A.}},
\bauthor{\bsnm{{Weaver}}, \binits{B.A.}},
\bauthor{\bsnm{{Weinberg}}, \binits{D.H.}},
\bauthor{\bsnm{{Weiner}}, \binits{B.J.}},
\bauthor{\bsnm{{White}}, \binits{M.}},
\bauthor{\bsnm{{White}}, \binits{S.D.M.}},
\bauthor{\bsnm{{Wilson}}, \binits{J.C.}},
\bauthor{\bsnm{{Wisniewski}}, \binits{J.P.}},
\bauthor{\bsnm{{Wood-Vasey}}, \binits{W.M.}},
\bauthor{\bsnm{{Y{\`e}che}}, \binits{C.}},
\bauthor{\bsnm{{York}}, \binits{D.G.}},
\bauthor{\bsnm{{Zamora}}, \binits{O.}},
\bauthor{\bsnm{{Zasowski}}, \binits{G.}},
\bauthor{\bsnm{{Zehavi}}, \binits{I.}},
\bauthor{\bsnm{{Zhao}}, \binits{G.-B.}},
\bauthor{\bsnm{{Zheng}}, \binits{Z.}},
\bauthor{\bsnm{{Zhu}}, \binits{G.}}:
\batitle{{The Tenth Data Release of the Sloan Digital Sky Survey: First Spectroscopic Data from the SDSS-III Apache Point Observatory Galactic Evolution Experiment}}.
\bjtitle{\apjs}
\bvolume{211}(\bissue{2}),
\bfpage{17}
(\byear{2014})
\doiurl{10.1088/0067-0049/211/2/17}
{\href{https://arxiv.org/abs/1307.7735}{{arXiv:1307.7735}}}
{[astro-ph.IM]}
\end{barticle}
\endbibitem

\bibitem[\protect\citeauthoryear{{Iyer} and {Gawiser}}{2017}]{iyer17}
\begin{barticle}
\bauthor{\bsnm{{Iyer}}, \binits{K.}},
\bauthor{\bsnm{{Gawiser}}, \binits{E.}}:
\batitle{{Reconstruction of Galaxy Star Formation Histories through SED Fitting:The Dense Basis Approach}}.
\bjtitle{\apj}
\bvolume{838}(\bissue{2}),
\bfpage{127}
(\byear{2017})
\doiurl{10.3847/1538-4357/aa63f0}
{\href{https://arxiv.org/abs/1702.04371}{{arXiv:1702.04371}}}
{[astro-ph.GA]}
\end{barticle}
\endbibitem

\bibitem[\protect\citeauthoryear{{Iyer} et~al.}{2019}]{iyer19}
\begin{barticle}
\bauthor{\bsnm{{Iyer}}, \binits{K.G.}},
\bauthor{\bsnm{{Gawiser}}, \binits{E.}},
\bauthor{\bsnm{{Faber}}, \binits{S.M.}},
\bauthor{\bsnm{{Ferguson}}, \binits{H.C.}},
\bauthor{\bsnm{{Kartaltepe}}, \binits{J.}},
\bauthor{\bsnm{{Koekemoer}}, \binits{A.M.}},
\bauthor{\bsnm{{Pacifici}}, \binits{C.}},
\bauthor{\bsnm{{Somerville}}, \binits{R.S.}}:
\batitle{{Nonparametric Star Formation History Reconstruction with Gaussian Processes. I. Counting Major Episodes of Star Formation}}.
\bjtitle{\apj}
\bvolume{879}(\bissue{2}),
\bfpage{116}
(\byear{2019})
\doiurl{10.3847/1538-4357/ab2052}
{\href{https://arxiv.org/abs/1901.02877}{{arXiv:1901.02877}}}
{[astro-ph.GA]}
\end{barticle}
\endbibitem

\bibitem[\protect\citeauthoryear{{Calzetti}}{2001}]{calzetti01}
\begin{barticle}
\bauthor{\bsnm{{Calzetti}}, \binits{D.}}:
\batitle{{The Dust Opacity of Star-forming Galaxies}}.
\bjtitle{\pasp}
\bvolume{113}(\bissue{790}),
\bfpage{1449}--\blpage{1485}
(\byear{2001})
\doiurl{10.1086/324269}
{\href{https://arxiv.org/abs/astro-ph/0109035}{{arXiv:astro-ph/0109035}}}
{[astro-ph]}
\end{barticle}
\endbibitem

\bibitem[\protect\citeauthoryear{{Kroupa}}{2001}]{Kroupa2001}
\begin{barticle}
\bauthor{\bsnm{{Kroupa}}, \binits{P.}}:
\batitle{{On the variation of the initial mass function}}.
\bjtitle{\mnras}
\bvolume{322}(\bissue{2}),
\bfpage{231}--\blpage{246}
(\byear{2001})
\doiurl{10.1046/j.1365-8711.2001.04022.x}
{\href{https://arxiv.org/abs/astro-ph/0009005}{{arXiv:astro-ph/0009005}}}
{[astro-ph]}
\end{barticle}
\endbibitem

\bibitem[\protect\citeauthoryear{{Tacchella} et~al.}{2023}]{tacchella23}
\begin{barticle}
\bauthor{\bsnm{{Tacchella}}, \binits{S.}},
\bauthor{\bsnm{{Johnson}}, \binits{B.D.}},
\bauthor{\bsnm{{Robertson}}, \binits{B.E.}},
\bauthor{\bsnm{{Carniani}}, \binits{S.}},
\bauthor{\bsnm{{D'Eugenio}}, \binits{F.}},
\bauthor{\bsnm{{Kumari}}, \binits{N.}},
\bauthor{\bsnm{{Maiolino}}, \binits{R.}},
\bauthor{\bsnm{{Nelson}}, \binits{E.J.}},
\bauthor{\bsnm{{Suess}}, \binits{K.A.}},
\bauthor{\bsnm{{{\"U}bler}}, \binits{H.}},
\bauthor{\bsnm{{Williams}}, \binits{C.C.}},
\bauthor{\bsnm{{Adebusola}}, \binits{A.}},
\bauthor{\bsnm{{Alberts}}, \binits{S.}},
\bauthor{\bsnm{{Arribas}}, \binits{S.}},
\bauthor{\bsnm{{Bhatawdekar}}, \binits{R.}},
\bauthor{\bsnm{{Bonaventura}}, \binits{N.}},
\bauthor{\bsnm{{Bowler}}, \binits{R.A.A.}},
\bauthor{\bsnm{{Bunker}}, \binits{A.J.}},
\bauthor{\bsnm{{Cameron}}, \binits{A.J.}},
\bauthor{\bsnm{{Curti}}, \binits{M.}},
\bauthor{\bsnm{{Egami}}, \binits{E.}},
\bauthor{\bsnm{{Eisenstein}}, \binits{D.J.}},
\bauthor{\bsnm{{Frye}}, \binits{B.}},
\bauthor{\bsnm{{Hainline}}, \binits{K.}},
\bauthor{\bsnm{{Helton}}, \binits{J.M.}},
\bauthor{\bsnm{{Ji}}, \binits{Z.}},
\bauthor{\bsnm{{Looser}}, \binits{T.J.}},
\bauthor{\bsnm{{Lyu}}, \binits{J.}},
\bauthor{\bsnm{{Perna}}, \binits{M.}},
\bauthor{\bsnm{{Rawle}}, \binits{T.}},
\bauthor{\bsnm{{Rieke}}, \binits{G.}},
\bauthor{\bsnm{{Rieke}}, \binits{M.}},
\bauthor{\bsnm{{Saxena}}, \binits{A.}},
\bauthor{\bsnm{{Sandles}}, \binits{L.}},
\bauthor{\bsnm{{Shivaei}}, \binits{I.}},
\bauthor{\bsnm{{Simmonds}}, \binits{C.}},
\bauthor{\bsnm{{Sun}}, \binits{F.}},
\bauthor{\bsnm{{Willmer}}, \binits{C.N.A.}},
\bauthor{\bsnm{{Willott}}, \binits{C.J.}},
\bauthor{\bsnm{{Witstok}}, \binits{J.}}:
\batitle{{JWST NIRCam + NIRSpec: interstellar medium and stellar populations of young galaxies with rising star formation and evolving gas reservoirs}}.
\bjtitle{\mnras}
\bvolume{522}(\bissue{4}),
\bfpage{6236}--\blpage{6249}
(\byear{2023})
\doiurl{10.1093/mnras/stad1408}
{\href{https://arxiv.org/abs/2208.03281}{{arXiv:2208.03281}}}
\end{barticle}
\endbibitem

\bibitem[\protect\citeauthoryear{{Lower} et~al.}{2020}]{lower20}
\begin{barticle}
\bauthor{\bsnm{{Lower}}, \binits{S.}},
\bauthor{\bsnm{{Narayanan}}, \binits{D.}},
\bauthor{\bsnm{{Leja}}, \binits{J.}},
\bauthor{\bsnm{{Johnson}}, \binits{B.D.}},
\bauthor{\bsnm{{Conroy}}, \binits{C.}},
\bauthor{\bsnm{{Dav{\'e}}}, \binits{R.}}:
\batitle{{How Well Can We Measure the Stellar Mass of a Galaxy: The Impact of the Assumed Star Formation History Model in SED Fitting}}.
\bjtitle{\apj}
\bvolume{904}(\bissue{1}),
\bfpage{33}
(\byear{2020})
\doiurl{10.3847/1538-4357/abbfa7}
{\href{https://arxiv.org/abs/2006.03599}{{arXiv:2006.03599}}}
{[astro-ph.GA]}
\end{barticle}
\endbibitem

\bibitem[\protect\citeauthoryear{{Limousin} et~al.}{2010}]{limousin10}
\begin{barticle}
\bauthor{\bsnm{{Limousin}}, \binits{M.}},
\bauthor{\bsnm{{Ebeling}}, \binits{H.}},
\bauthor{\bsnm{{Ma}}, \binits{C.-J.}},
\bauthor{\bsnm{{Swinbank}}, \binits{A.M.}},
\bauthor{\bsnm{{Smith}}, \binits{G.P.}},
\bauthor{\bsnm{{Richard}}, \binits{J.}},
\bauthor{\bsnm{{Edge}}, \binits{A.C.}},
\bauthor{\bsnm{{Jauzac}}, \binits{M.}},
\bauthor{\bsnm{{Kneib}}, \binits{J.-P.}},
\bauthor{\bsnm{{Marshall}}, \binits{P.}},
\bauthor{\bsnm{{Schrabback}}, \binits{T.}}:
\batitle{{MACS J1423.8+2404: gravitational lensing by a massive, relaxed cluster of galaxies at z = 0.54}}.
\bjtitle{\mnras}
\bvolume{405}(\bissue{2}),
\bfpage{777}--\blpage{782}
(\byear{2010})
\doiurl{10.1111/j.1365-2966.2010.16518.x}
{\href{https://arxiv.org/abs/0911.4125}{{arXiv:0911.4125}}}
{[astro-ph.CO]}
\end{barticle}
\endbibitem

\bibitem[\protect\citeauthoryear{{Zitrin} et~al.}{2015}]{zitrin15}
\begin{barticle}
\bauthor{\bsnm{{Zitrin}}, \binits{A.}},
\bauthor{\bsnm{{Fabris}}, \binits{A.}},
\bauthor{\bsnm{{Merten}}, \binits{J.}},
\bauthor{\bsnm{{Melchior}}, \binits{P.}},
\bauthor{\bsnm{{Meneghetti}}, \binits{M.}},
\bauthor{\bsnm{{Koekemoer}}, \binits{A.}},
\bauthor{\bsnm{{Coe}}, \binits{D.}},
\bauthor{\bsnm{{Maturi}}, \binits{M.}},
\bauthor{\bsnm{{Bartelmann}}, \binits{M.}},
\bauthor{\bsnm{{Postman}}, \binits{M.}},
\bauthor{\bsnm{{Umetsu}}, \binits{K.}},
\bauthor{\bsnm{{Seidel}}, \binits{G.}},
\bauthor{\bsnm{{Sendra}}, \binits{I.}},
\bauthor{\bsnm{{Broadhurst}}, \binits{T.}},
\bauthor{\bsnm{{Balestra}}, \binits{I.}},
\bauthor{\bsnm{{Biviano}}, \binits{A.}},
\bauthor{\bsnm{{Grillo}}, \binits{C.}},
\bauthor{\bsnm{{Mercurio}}, \binits{A.}},
\bauthor{\bsnm{{Nonino}}, \binits{M.}},
\bauthor{\bsnm{{Rosati}}, \binits{P.}},
\bauthor{\bsnm{{Bradley}}, \binits{L.}},
\bauthor{\bsnm{{Carrasco}}, \binits{M.}},
\bauthor{\bsnm{{Donahue}}, \binits{M.}},
\bauthor{\bsnm{{Ford}}, \binits{H.}},
\bauthor{\bsnm{{Frye}}, \binits{B.L.}},
\bauthor{\bsnm{{Moustakas}}, \binits{J.}}:
\batitle{{Hubble Space Telescope Combined Strong and Weak Lensing Analysis of the CLASH Sample: Mass and Magnification Models and Systematic Uncertainties}}.
\bjtitle{\apj}
\bvolume{801}(\bissue{1}),
\bfpage{44}
(\byear{2015})
\doiurl{10.1088/0004-637X/801/1/44}
{\href{https://arxiv.org/abs/1411.1414}{{arXiv:1411.1414}}}
{[astro-ph.CO]}
\end{barticle}
\endbibitem

\bibitem[\protect\citeauthoryear{{El{\'\i}asd{\'o}ttir} et~al.}{2007}]{eliasdottir07}
\begin{botherref}
\oauthor{\bsnm{{El{\'\i}asd{\'o}ttir}}, \binits{{\'A}.}},
\oauthor{\bsnm{{Limousin}}, \binits{M.}},
\oauthor{\bsnm{{Richard}}, \binits{J.}},
\oauthor{\bsnm{{Hjorth}}, \binits{J.}},
\oauthor{\bsnm{{Kneib}}, \binits{J.-P.}},
\oauthor{\bsnm{{Natarajan}}, \binits{P.}},
\oauthor{\bsnm{{Pedersen}}, \binits{K.}},
\oauthor{\bsnm{{Jullo}}, \binits{E.}},
\oauthor{\bsnm{{Paraficz}}, \binits{D.}}:
{Where is the matter in the Merging Cluster Abell 2218?}
arXiv e-prints,
0710--5636
(2007)
\doiurl{10.48550/arXiv.0710.5636}
{\href{https://arxiv.org/abs/0710.5636}{{arXiv:0710.5636}}}
{[astro-ph]}
\end{botherref}
\endbibitem

\bibitem[\protect\citeauthoryear{{Richard} et~al.}{2010}]{richard10}
\begin{barticle}
\bauthor{\bsnm{{Richard}}, \binits{J.}},
\bauthor{\bsnm{{Smith}}, \binits{G.P.}},
\bauthor{\bsnm{{Kneib}}, \binits{J.-P.}},
\bauthor{\bsnm{{Ellis}}, \binits{R.S.}},
\bauthor{\bsnm{{Sanderson}}, \binits{A.J.R.}},
\bauthor{\bsnm{{Pei}}, \binits{L.}},
\bauthor{\bsnm{{Targett}}, \binits{T.A.}},
\bauthor{\bsnm{{Sand}}, \binits{D.J.}},
\bauthor{\bsnm{{Swinbank}}, \binits{A.M.}},
\bauthor{\bsnm{{Dannerbauer}}, \binits{H.}},
\bauthor{\bsnm{{Mazzotta}}, \binits{P.}},
\bauthor{\bsnm{{Limousin}}, \binits{M.}},
\bauthor{\bsnm{{Egami}}, \binits{E.}},
\bauthor{\bsnm{{Jullo}}, \binits{E.}},
\bauthor{\bsnm{{Hamilton-Morris}}, \binits{V.}},
\bauthor{\bsnm{{Moran}}, \binits{S.M.}}:
\batitle{{LoCuSS: first results from strong-lensing analysis of 20 massive galaxy clusters at z = 0.2}}.
\bjtitle{\mnras}
\bvolume{404},
\bfpage{325}--\blpage{349}
(\byear{2010})
\doiurl{10.1111/j.1365-2966.2009.16274.x}
{\href{https://arxiv.org/abs/0911.3302}{{arXiv:0911.3302}}}
\end{barticle}
\endbibitem

\bibitem[\protect\citeauthoryear{{Grazian} et~al.}{2015}]{Grazian:2015}
\begin{barticle}
\bauthor{\bsnm{{Grazian}}, \binits{A.}},
\bauthor{\bsnm{{Fontana}}, \binits{A.}},
\bauthor{\bsnm{{Santini}}, \binits{P.}},
\bauthor{\bsnm{{Dunlop}}, \binits{J.S.}},
\bauthor{\bsnm{{Ferguson}}, \binits{H.C.}},
\bauthor{\bsnm{{Castellano}}, \binits{M.}},
\bauthor{\bsnm{{Amorin}}, \binits{R.}},
\bauthor{\bsnm{{Ashby}}, \binits{M.L.N.}},
\bauthor{\bsnm{{Barro}}, \binits{G.}},
\bauthor{\bsnm{{Behroozi}}, \binits{P.}},
\bauthor{\bsnm{{Boutsia}}, \binits{K.}},
\bauthor{\bsnm{{Caputi}}, \binits{K.I.}},
\bauthor{\bsnm{{Chary}}, \binits{R.R.}},
\bauthor{\bsnm{{Dekel}}, \binits{A.}},
\bauthor{\bsnm{{Dickinson}}, \binits{M.E.}},
\bauthor{\bsnm{{Faber}}, \binits{S.M.}},
\bauthor{\bsnm{{Fazio}}, \binits{G.G.}},
\bauthor{\bsnm{{Finkelstein}}, \binits{S.L.}},
\bauthor{\bsnm{{Galametz}}, \binits{A.}},
\bauthor{\bsnm{{Giallongo}}, \binits{E.}},
\bauthor{\bsnm{{Giavalisco}}, \binits{M.}},
\bauthor{\bsnm{{Grogin}}, \binits{N.A.}},
\bauthor{\bsnm{{Guo}}, \binits{Y.}},
\bauthor{\bsnm{{Kocevski}}, \binits{D.}},
\bauthor{\bsnm{{Koekemoer}}, \binits{A.M.}},
\bauthor{\bsnm{{Koo}}, \binits{D.C.}},
\bauthor{\bsnm{{Lee}}, \binits{K.-S.}},
\bauthor{\bsnm{{Lu}}, \binits{Y.}},
\bauthor{\bsnm{{Merlin}}, \binits{E.}},
\bauthor{\bsnm{{Mobasher}}, \binits{B.}},
\bauthor{\bsnm{{Nonino}}, \binits{M.}},
\bauthor{\bsnm{{Papovich}}, \binits{C.}},
\bauthor{\bsnm{{Paris}}, \binits{D.}},
\bauthor{\bsnm{{Pentericci}}, \binits{L.}},
\bauthor{\bsnm{{Reddy}}, \binits{N.}},
\bauthor{\bsnm{{Renzini}}, \binits{A.}},
\bauthor{\bsnm{{Salmon}}, \binits{B.}},
\bauthor{\bsnm{{Salvato}}, \binits{M.}},
\bauthor{\bsnm{{Sommariva}}, \binits{V.}},
\bauthor{\bsnm{{Song}}, \binits{M.}},
\bauthor{\bsnm{{Vanzella}}, \binits{E.}}:
\batitle{{The galaxy stellar mass function at 3.5 {\ensuremath{\leq}}z {\ensuremath{\leq}} 7.5 in the CANDELS/UDS, GOODS-South, and HUDF fields}}.
\bjtitle{\aap}
\bvolume{575},
\bfpage{96}
(\byear{2015})
\doiurl{10.1051/0004-6361/201424750}
{\href{https://arxiv.org/abs/1412.0532}{{arXiv:1412.0532}}}
{[astro-ph.GA]}
\end{barticle}
\endbibitem

\bibitem[\protect\citeauthoryear{{Stefanon} et~al.}{2021}]{Stefanon:2021}
\begin{barticle}
\bauthor{\bsnm{{Stefanon}}, \binits{M.}},
\bauthor{\bsnm{{Bouwens}}, \binits{R.J.}},
\bauthor{\bsnm{{Labb{\'e}}}, \binits{I.}},
\bauthor{\bsnm{{Illingworth}}, \binits{G.D.}},
\bauthor{\bsnm{{Gonzalez}}, \binits{V.}},
\bauthor{\bsnm{{Oesch}}, \binits{P.A.}}:
\batitle{{Galaxy Stellar Mass Functions from z 10 to z 6 using the Deepest Spitzer/Infrared Array Camera Data: No Significant Evolution in the Stellar-to-halo Mass Ratio of Galaxies in the First Gigayear of Cosmic Time}}.
\bjtitle{\apj}
\bvolume{922}(\bissue{1}),
\bfpage{29}
(\byear{2021})
\doiurl{10.3847/1538-4357/ac1bb6}
{\href{https://arxiv.org/abs/2103.16571}{{arXiv:2103.16571}}}
{[astro-ph.GA]}
\end{barticle}
\endbibitem

\end{thebibliography}

\end{document}